\documentclass[numberedappendix]{emulateapj}
\usepackage{apjfonts}

\newcommand{\etal}{et al.\,}

\slugcomment{Accepted for publication in the Astrophysical Journal}

\shorttitle{Kpc-scale rest-optical imaging of $z \sim 2$ galaxies}
\shortauthors{F\"orster Schreiber \etal}

\begin{document}

\title{Constraints on the assembly and dynamics of galaxies:~
   I.~Detailed rest-frame optical morphologies on kiloparsec-scale
   of $z \sim 2$ star-forming galaxies \,\altaffilmark{1}}

\author{N. M. F\"orster Schreiber\altaffilmark{2},
        A. E. Shapley\altaffilmark{3},
        D. K. Erb\altaffilmark{4,5},
        R. Genzel\altaffilmark{2,6},
        C. C. Steidel\altaffilmark{7},
        N. Bouch\'e\altaffilmark{4,8,9},
        G. Cresci\altaffilmark{10},
        R. Davies\altaffilmark{2}}

\altaffiltext{1}{Based on observations made with the NASA/ESA
  {\em Hubble Space Telescope\/} ({\em HST\/}), obtained at the
  Space Telescope Science Institute, which is operated by AURA, Inc.,
  under NASA contract NAS 5$-$26555,
  and at the Very Large Telescope of the European Southern Observatory,
  Paranal, Chile (ESO Programme IDs 073.B-9018, 074.A-9011, 075.A-0466,
  076.A-0527, 077.A-0576, 078.A-0600, 079.A-0341, 080.A-0330, and 080.A-0339).}
\altaffiltext{2}{Max-Planck-Institut f\"ur extraterrestrische Physik,
                 Giessenbachstrasse, D-85748 Garching, Germany}
\altaffiltext{3}{Department of Physics and Astronomy,
                 University of California, Los Angeles, CA 90095-1547}
\altaffiltext{4}{Department of Physics, University of California Santa Barbara,
                 Santa Barbara, California, CA 93106-9530; Marie Curie Fellow}
\altaffiltext{5}{Present address: Department of Physics,
                 University of Wisconsin Milwaukee,
                 Milwaukee, Wisconsin, WI 53211}
\altaffiltext{6}{Department of Physics, University of California,
                 Berkeley, CA 94720}
\altaffiltext{7}{California Institute of Technology, MS 105-24,
                 Pasadena, CA 91125}
\altaffiltext{8}{CNRS,
                 Institut de Recherche en Astrophysique et Plan\'etologie 
                 de Toulouse, 14 Avenue E. Belin, F-31400 Toulouse, France}
\altaffiltext{9}{Universit\'e de Toulouse; UPS-OMP; IRAP;
                 F-31400 Toulouse, France}
\altaffiltext{10}{INAF-Osservatorio Astrofisico di Arcetri,
                 Largo E. Fermi 5, I-50125 Firenze, Italy}

\begin{abstract}

We present deep and high-resolution {\em Hubble Space Telescope\/} NIC2
F160W imaging at 1.6\,\micron\ of six $z \sim 2$ star-forming galaxies
with existing near-infrared integral field spectroscopy from SINFONI at
the Very Large Telescope.  The unique combination of rest-frame optical
imaging and nebular emission-line maps provides simultaneous insight
into morphologies and dynamical properties.  The overall rest-frame
optical emission of the galaxies is characterized by shallow profiles
in general (S\'ersic index $n < 1$), with median effective radii of
$R_{\rm e} \sim 5~{\rm kpc}$.  The morphologies are significantly clumpy
and irregular, which we quantify through a non-parametric morphological
approach, estimating the Gini ($G$), Multiplicity ($\Psi$), and $M_{20}$
coefficients.  The estimated strength of the rest-frame optical emission
lines in the F160W bandpass indicates that the observed structure is not
dominated by the morphology of line-emitting gas, and must reflect the
underlying stellar mass distribution of the galaxies.
The sizes and structural parameters in the rest-frame optical continuum
and H$\alpha$ emission reveal no significant differences, suggesting 
similar global distributions of the on-going star formation and more
evolved stellar population.
While no strong correlations are observed between stellar population
parameters and morphology within the NIC2/SINFONI sample itself, a
consideration of the sample in the context of a broader range of
$z \sim 2$ galaxy types ($K$-selected quiescent, AGN, and star-forming;
24\,\micron -selected dusty, infrared-luminous) indicates that these
galaxies probe the high specific star formation rate and low stellar
mass surface density part of the massive $z \sim 2$ galaxy population,
with correspondingly large effective radii, low S\'ersic indices, low $G$,
and high $\Psi$ and $M_{20}$.  The combined NIC2 and SINFONI dataset yields
insights of unprecedented detail into the nature of mass accretion at high
redshift.

\end{abstract}


\section{INTRODUCTION}    \label{Sect-intro}

A fundamental but still unsolved issue is the origin of the Hubble
sequence and the formation of the spheroidal and disk components of
present-day galaxies.  Observations indicate that the Hubble sequence
emerges around $z \sim 1 - 2$, along with the closely connected bimodality
in integrated rest-frame optical colors characterized by a ``red sequence''
of mostly elliptical/bulge-dominated early-types and a ``blue cloud'' of
generally disk-dominated and irregular late-types
\citep[e.g.,][]
  {Bel04,vdB96,vdB01,Lil98,Sta04,Rav04,Pap05,Kri08b,Kri09,Wil09}.
Although colors are easier to measure, morphologies potentially yield more
information about the processes that shape galaxies over time.  Morphological
studies reveal a progressively larger proportion of irregular and disk-like
types at higher redshift compared to the local galaxy population
\citep*[e.g.,][]{Lot04, Lot06, Law07a, Sca07a, Sca07b, Elm07, Pan09},
hinting at the diversity of processes driving early galaxy evolution.

Morphologies, however, can be difficult to interpret.
Bandshifting can plausibly explain part of the observed changes
in morphologies; high resolution imaging conducted so far mostly
in the optical regime traces rest-UV wavelengths at $z \ga 1$, which 
are more affected by dust obscuration and on-going star formation
\citep[e.g.,][]{Tof07, Ove10, Cam10}.
Star-forming sites within galaxies may
also evolve in size and spatial distribution as a result of different
dominant triggering mechanisms, and different physical and dynamical
conditions prevailing at earlier epochs.  Kiloparsec-sized ``clumps''
are in fact ubiquitous among high-$z$ star-forming galaxies and are
believed to be at least in part due to fragmentation in Toomre-unstable
gas-rich disks \citep[e.g.,][and references therein]{Elm07, Elm09}.
There is now direct kinematic evidence that high-$z$ disks are
significantly more turbulent (and geometrically thicker) than their
$z \sim 0$ counterparts, implying accordingly larger Jeans length
setting the characteristic size scale of self-gravitating star-forming
complexes \citep[e.g.,][]{Bour07, Bour08, Gen08, Gen10}.  High gas 
mass fractions at $z \sim 1 - 2.5$ have also been confirmed through
observations of CO rotational transitions \citep[e.g.,][]{Tac10, Dad10}.
Higher merger rates in the young universe may further contribute to the
irregular appearance of galaxies.  Merger-induced morphological disturbances
can however last over a wide range of timescales, from $\rm < 100~Myr$ up to
$\rm > 1~Gyr$ depending on progenitors baryonic mass ratio, gas fraction,
and orbital parameters \citep[e.g.,][]{Lot08, Lot10a, Lot10b}.

Information on the galaxy kinematics is therefore crucial to establish the
nature and dynamical state of high $z$ galaxies.  Kinematics provide
powerful diagnostics to distinguish reliably between disks and mergers as a
function of the degree of symmetry of velocity fields and velocity dispersion
maps \citep[e.g.,][]{Kra06, Flo06, Sha08, Epi10}, and to determine the main
source of dynamical support from the ratio of rotation velocity to local
intrinsic velocity dispersion \citep[e.g.][]{Pue07, Cre09, Epi09}.
Spatially-resolved mapping of the two-dimensional kinematics at $z \ga 1$
have become possible during the past decade, with the newest generation of
sensitive near-IR integral field spectrographs mounted on ground-based
10\,m-class telescopes.

Using the near-IR integral field spectrometer SINFONI at the ESO Very
Large Telescope (VLT), we carried out the first and largest survey to
date of the gas kinematics of 80 $z \sim 1 - 3$ galaxies, ``SINS'' 
\citep{FS06b, FS09, Gen06, Gen08, Bou07, Sha08, Cre09}.  The survey
includes 62 rest-UV/optically selected objects at $1.3 < z < 2.6$ for which
we targeted primarily the H$\alpha$ and [\ion{N}{2}] emission lines.  This
``SINS H$\alpha$ sample'' covers roughly two orders of magnitude in stellar 
mass ($M_{\star} \approx 3 \times 10^{9} - 3 \times 10^{11}~{\rm M_{\odot}}$)
and star formation rate ($\rm SFR \approx 10 - 800~M_{\odot}\,yr^{-1}$).
The line emission is resolved on typical spatial scales of $\rm 4 - 5~kpc$
for the seeing-limited data and down to $\rm 1 - 2~kpc$ for the subset of
11 sources followed-up with adaptive optics (AO).  Kinematically, about
one-third of the SINS galaxies are rotation-dominated yet turbulent disks,
another third comprises typically more compact dispersion-dominated objects,
and the remaining sources are interacting/merging systems.  Concurring
results on the kinematic mix at $z \sim 1 - 3$ have been found by other
groups \citep[e.g.,][]
       {Erb03,Erb04,Erb06,Wri07,Wri09,Law07b,Law09,Sta08,Epi09,Jon10}.
If the relative proportions vary somewhat between different studies,
possibly reflecting in part differences in samples and mass/redshift
ranges, the existence of disks in the early universe and their typically
high intrinsic local velocity dispersion now seem fairly well established.
Under these conditions, and in the absence of major mergers, internal
dynamical processes can drive the secular evolution of disks and
formation of bulges/spheroids on timescales $\rm \sim 1~Gyr$, about
an order of magnitude faster than in present-day disks
\citep*[e.g.][]{Nog99, Imm04a, Imm04b, Car07, Bour07, Gen08, Elm09}.

While H$\alpha$ emission provides an excellent probe of the sites
of ongoing star formation activity and the current dynamical state
of a galaxy, broad-band rest-frame optical imaging with high spatial
resolution is required to characterize the structural properties of the
underlying stellar component.  Dynamical processes at play in the past
are imprinted in the stellar mass distribution and the stellar population
contains a record of earlier star formation.  Unique insights into early
galaxy evolution can therefore be gained from the powerful combination of
ionized gas kinematics and spatial distribution from H$\alpha$ with detailed
morphologies from high resolution broad-band imaging.

In this paper, we present the results of deep high resolution imaging
at 1.6~\micron\ with the NICMOS/NIC2 camera on-board the Hubble Space
Telescope ({\em HST\/}) of six of the SINS galaxies at $2.1 < z < 2.5$.
The NICMOS imaging probes the rest-frame 5000\,\AA\ continuum of the 
galaxies, redwards of the Balmer/4000\,\AA\ break.  In this range,
light from longer-lived, lower-mass stars that make up the bulk of the
stellar mass becomes more important compared to that of young hot stars
shining more brightly at shorter wavelengths and the effects of dust
extinction are smaller than in the rest-UV.  Five of the targets exhibit
compelling signatures of disk-like kinematics.  The kinematics of
the sixth source unambiguously reveal a major merger, although this is
not obvious from the surface brightness distribution of H$\alpha$ at
seeing-limited resolution.  The original motivation for our NIC2
follow-up was to determine whether the stellar light/mass distribution
of the kinematically-identified disks also follows a disk-like profile,
and to unveil the components of the merger system.

With the deep $\rm \sim 1~kpc$ resolution NIC2 1.6\,\micron\ imaging,
our primary goals in the present study are to derive the global structural
parameters and the detailed morphologies of the galaxies as traced by the
stellar light.  We further explore the connection with stellar population
properties and, taking advantage of the high quality SINFONI data of our
targets, we test if and how the structure relates to the dynamical properties.
Our targets extend up to the high-mass end of the rest-UV selected $z \sim 2$
population, and overlap in mass and redshift range with other massive galaxy
samples with deep and high resolution NIC2 1.6\,\micron\ data studied by
\citet{Dok08}, \citet{Das08}, and \citet{Kri09}.  The similar NIC2 data
sets between these samples and ours allow consistent comparisons of the
structural and morphological properties of massive $z \approx 2 - 2.5$
galaxies selected through different criteria and, anticipating our
results, show how galaxies of similar masses can be very different in
those respects, reflecting a wide range in star formation activity.
In a companion paper \citep[][hereafter Paper~II]{FS11}, we focus on
the properties of the kpc-scale clumps identified in the NIC2 images
of our targets, and derive constraints on the nature and evolution of
these clumps.

The paper is organized as follows.
We introduce the sample and the SINFONI data sets, and describe
the NIC2 observations and data reduction in \S~\ref{Sect-data}.
In \S~\ref{Sect-global}, we revisit the global photometric and stellar
populations properties of the galaxies with the inclusion of the new
NIC2 1.6\,\micron\ data, and place them in the broader context of
$z \sim 2$ populations in terms of stellar mass, star formation rate,
and kinematics.
We present the structural and morphological analysis in \S~\ref{Sect-res},
and combine the NIC2 broad-band imaging together with the emission line
data from SINFONI in \S~\ref{Sect-sinf}.
We compare our sample to other massive galaxy samples at $z \sim 2 - 2.6$
in terms of structural and morphological properties, and address the
connection with stellar populations in \S~\ref{Sect-comp}.
In \S~\ref{Sect-disc}, we constrain the spatial distribution of stellar
mass in one of our targets with additional high resolution data from
{\em HST\/}/ACS, compare the rest-frame optical and H$\alpha$ sizes,
and discuss the relation between morphologies and H$\alpha$ kinematics.
We summarize the paper in \S~\ref{Sect-conclu}.
Throughout the paper, we assume a $\Lambda$-dominated cosmology
with $H_{0} = 70\,h_{70}~{\rm km\,s^{-1}\,Mpc^{-1}}$,
$\Omega_{\rm m} = 0.3$, and $\Omega_{\Lambda} = 0.7$.
For this cosmology, 1\arcsec\ corresponds to $\rm \approx 8.2~kpc$
at $z = 2.2$.  Magnitudes are given in the AB photometric system
unless otherwise specified.

\section{SAMPLE, OBSERVATIONS, DATA REDUCTION}    \label{Sect-data}

\subsection{Choice of NICMOS/NIC2 Targets}   \label{Sub-sample}

Table~\ref{tab-targets} lists the galaxies observed along with their
photometric properties.  They were drawn from the initial sample of
17 rest-UV selected objects at $z \sim 2$ from the SINS survey
\citep{FS06b, FS09} carried out with SINFONI \citep{Eis03, Bon04} at
the ESO VLT.  The galaxies were originally part of the large optical
spectroscopic survey of $z \sim 1.5 - 2.5$ candidates selected by
their $U_{n}G\mathcal{R}$ colors described by
\citet[][see also \citealt{Ade04}]{Ste04}.
Additional multi-wavelength data include ground-based near-IR $J$ and
$K_{\rm s}$ imaging, and space-based {\em Spitzer\/} mid-IR photometry at
$\rm 3 - 8~\mu m$ with IRAC and at 24\,\micron\ with MIPS for the majority
of our targets
\citep[][and \citealt{FS09} for the $K_{\rm s}$ band photometry of MD\,41]
        {Erb06, Red10}.
Prior to the SINFONI observations, near-IR long-slit spectroscopy
was obtained for all of our NIC2 targets with NIRSPEC at the Keck~II
telescope and ISAAC at the VLT by \citet{Erb03, Erb06} and \citet{Sha04}.

The choice of our NIC2 targets was primarily driven by their kinematic
nature along with their high signal-to-noise ratio (S/N), high quality
SINFONI data mapping the spatial distribution and relative gas motions
from H$\alpha$ out to radii $\rm \ga 10~kpc$ \citep{FS06b}.
They include five rotating disks (BX\,663, MD\,41, BX\,389, BX\,610,
and BX\,482), and one major merger (BX\,528).  The quantitative kinematic
classification performed through application of kinemetry is described by
\citet{Sha08}, and detailed dynamical modeling of the disks is presented by
\citet{Gen08} and \citet{Cre09}.  The disks show the expected signatures of
a fairly ordered two-dimensional velocity field, with a monotonic gradient
across the source and flattening of the velocity curve at large radii,
alignment of the H$\alpha$ kinematic and morphological major axes, and
peak of the velocity dispersion close to the dynamical and geometrical
center.  In contrast, BX\,528 exhibits a reversal in velocity gradient
along the major axis, as observed in nearby counter-rotating binary
mergers.  Interestingly, the H$\alpha$ morphology of this merger from
the seeing-limited SINFONI data does not show features that would
distinguish it from the disks; all sources have extended, more or less
irregular/clumpy H$\alpha$ distributions at a typical seeing-limited
resolution of $\approx 0\farcs 5$.

Two of our targets (BX\,663 and BX\,610) are candidate AGN hosts
on the basis of their observed mid-IR spectral energy distributions
from {\em Spitzer\/} IRAC and MIPS photometry
\citep[][and N. Reddy, private communication]{Red10}.
BX\,663 also exhibits spectral signatures of Type 2 AGN in its
integrated rest-UV and rest-optical spectrum \citep{Sha04, Erb06}.
In our spatially-resolved SINFONI data (as well as the new NIC2
imaging presented in this paper), the host galaxy is well detected
and can be distinguished from the central compact emission peak with
higher [\ion{N}{2}]/H$\alpha$ and broad H$\alpha$ velocity component
associated with the AGN, underneath the narrower component dominated 
by star formation.
In contrast, BX\,610 exhibits no sign of AGN in the rest-frame UV and
optical, including the spatially- and spectrally-resolved SINFONI data,
so that its AGN is likely to be very obscured and will not affect any
aspect of our analysis.  Therefore, in the context of this paper, we
will only consider BX\,663 explicitly as an AGN source.

\subsection{SINFONI Data Sets}   \label{Sub-sinfdata}

The SINFONI data sets of our NIC2 targets are fully described by
\citet{FS09}.  They are among the deepest of the SINS survey, with
on-source integration times in the $K$-band (targeting the H$\alpha$ 
and [\ion{N}{2}]\,$\lambda\lambda 6548,6584$ emission lines) ranging
from 3 to 7.3 hours.  All galaxies were first observed using the largest
pixel scale ($\rm 0\farcs 125\,pixel^{-1}$), five of them in seeing-limited
mode under very good seeing conditions, and one of them (BX\,663) with
the aid of AO using a nearby natural guide star.
The typical effective angular resolution of these SINFONI $K$-band data
has a $\rm FWHM \approx 0\farcs 5$, corresponding to $\rm \approx 4.1~kpc$
at the redshift of the sources ($\rm FWHM = 0\farcs 39$ or 3.2~kpc for
the NGS-AO data of BX\,663).  Seeing-limited $H$ band data were taken
for all but BX\,528 to map the [\ion{O}{3}]\,$\lambda\lambda 4959,5007$
and H$\beta$ emission lines, also under very good near-IR seeing of
$\approx 0\farcs 55$ and using the $\rm 0\farcs 125\,pixel^{-1}$ scale.
In addition, Laser Guide Star (LGS) AO-assisted $K$ band observations
were obtained for BX\,482 using the intermediate pixel scale of
$\rm 0\farcs 05\,pixel^{-1}$, with total on-source integration time of
6.8\,hours and effective resolution of $0\farcs 17$ (1.4~kpc), very
similar to that of the NIC2 1.6\,\micron\ imaging.

The central averaged H$\alpha$ surface brightness (within the
H$\alpha$ half-light radius) of the NIC2 targets range from
$\rm \approx 1 \times 10^{-17}$ to
$\rm \approx 1 \times 10^{-16}~erg\,s^{-1}\,cm^{-2}\,arcsec^{-2}$.
Using the conversion factor of \citet{Ken98} between H$\alpha$ luminosity
and star formation rate, divided by 1.7 to scale from a \citet{Sal55} to
a \citet{Chab03} IMF, and accounting for beam-smearing in computing the
intrinsic physical area, we calculate star formation rates per unit area of
$\rm \approx 0.15 - 0.4~M_{\odot}\,yr^{-1}\,kpc^{-2}$ (uncorrected for dust
extinction).  The median S/N per pixel of the H$\alpha$ maps over the regions
where line emission is detected (i.e., $\rm S/N > 3$) are $\approx 10 - 15$
for the six galaxies.  From the velocity-integrated line maps and from the
noise properties of the data cubes \citep[see details in][]{FS09}, the
$3\,\sigma$ H$\alpha$ surface brightness sensitivities are typically
$\rm \approx 10^{-17}~erg\,s^{-1}\,cm^{-2}\,arcsec^{-2}$, corresponding
to a limiting star formation rate of
$\rm \approx 0.03~M_{\odot}\,yr^{-1}\,kpc^{-2}$.

\subsection{NICMOS/NIC2 Observations}   \label{Sub-obs}

The NICMOS observations were carried out between 2007 April and 2007
September with the NIC2 camera onboard the {\em HST\/} and using the F160W
filter (hereafter referred to as the $H_{160}$ bandpass).  The $H_{160}$
bandpass, centered at $\rm 1.6~\mu m$ ($\rm FWHM = 0.4~\mu m$), probes
the longest wavelengths at which the {\em HST\/} thermal emission is
unimportant, taking full advantage of the lack of sky background that
limits the sensitivity of ground-based near-IR observations.  NIC2 has
a pixel scale of $0\farcs 075$, critically sampling the {\em HST\/}
point spread function (PSF) at 1.6\,\micron, and a field of view of 
$19\farcs 2 \times 19\farcs 2$.
While the NIC3 camera has greater surface brightness sensitivity,
NIC2 offers the best combination of resolution and sensitivity for our
science goals, and also allows more consistent comparisons with the
available {\em HST\/}/ACS imaging of MD\,41 and SINFONI$+$AO observations
of BX\,482, which have comparable or better resolution than NIC2.

Each target was observed for four orbits, with each orbit split into four
exposures with a subpixel dither pattern to ensure good sampling of the PSF
and minimize the impact of hot/cold bad pixels and other such artifacts
(e.g., ``grot,'' electronic bars).  The individual exposure time was 640\,s,
giving a total on-source integration time of 10240\,s.  The read-out mode
employed was the SPARS-64 MULTIACCUM sequence with $\rm NSAMP = 12$.
We used a square four-point dither pattern with step size of $1\farcs 5375$
(20.5 pixels).  We defined the starting positions of each dither sequence 
as the mid-points of the sides of a $0\farcs 9 \times 0\farcs 45$ rectangle
($\rm 12 \times 6$ pixels).
Table~\ref{tab-obs} summarizes the log of the observations.

\subsection{NICMOS/NIC2 Data reduction}   \label{Sub-red}

We started with the pipeline-reduced data products, processed with the STSDAS
task {\em calnica\/}, version 4.1.1, of the {\bf nicmos} package within the
IRAF environment.  We refer to the NICMOS Data Handbook\,\footnote{
 http://www.stsci.edu/hst/nicmos/documents/handbooks/DataHandbookv8}
for details and the description of specific instrumental features of NICMOS.
The steps performed in this stage include bias, zero read signal, and dark
current subtraction, data linearization, correction for so-called
``electronic bars,'' flat-fielding, and conversion to count rates.
The task {\em calnica\/} also identifies cosmic ray hits and various
bad pixels.  These identifications missed on a significant number of
obvious bad pixels present in our data, so we extended the bad pixel
lists for each exposure as described below.

We then used the task {\em pedsky\/} in the same IRAF package to subtract
a first-order estimate of the background and the quadrant-dependent residual
bias (or ``pedestal'') in each individual exposure.  We subtracted the median
value along each column to correct for bias jumps.
We constructed a static bad pixel mask for each galaxy based on deviant
pixels in the normalized flat field and those identified by {\em calnica\/}
with bad data quality flags.  To identify remaining bad pixels and cosmic
rays, we first drizzled the images for each galaxy onto a common finer
pixel grid accounting for their relative shifts, the (small) NIC2 distortion,
and the static bad pixel masks, using the IRAF task {\em drizzle\/} in the
STSDAS {\bf analysis} package.  A clean reference image was generated by
median-combining the registered drizzled images with ``minmax'' rejection.
This image was inverse-drizzled to the original pixel scale and coordinate
system of each individual exposure with the task {\em blot\/} in the STSDAS
{\bf analysis} package.  We compared the ``blotted'' reference images with
their corresponding input image and updated the static bad pixel mask to
make a mask for each exposure.  The background-subtracted exposures were
again drizzled, now using the updated individual bad pixel masks, to create
the combined drizzled image of each galaxy.  For the drizzling, we chose an
output pixel scale of $0\farcs 05$, corresponding to that of the SINFONI$+$AO
data of BX\,482, and used a square kernel with 0.7 times the size of the
input pixels (i.e., a ``dropsize'' of 0.7).  We found from experimentation that
this combination of parameters gave the best results in terms of preserving
the angular resolution and minimizing the impact on the pixel-to-pixel
rms noise level of our NIC2 data.

The combined drizzled images showed substantial residual features on
scales of a few tens of pixels together with residuals from vignetting
at the bottom of the array.  These features were successfully modeled and
subtracted from the combined images using background maps generated with
the SExtractor software, version 2.2.2 \citep{Ber96}.  Since our science
goals require detection of low surface brightness emission of the galaxies
to large radii, we took particular care in making these background maps.
We masked out sufficiently large areas around each object, and then ran
SExtractor with a background mesh size of 24 pixels and filter size of
three meshes.  This combination provided the best match to the spatial
frequency of the residual features, without compromising the faint extended
emission from the galaxies.  Further linear residual features were removed
by fitting second-order polynomials (with object masking) along columns and
rows.  This step also resulted in a small reduction of the pixel-to-pixel
rms noise.

Weight maps (for use in the determination of the noise properties and in the
structural analysis) were generated in the final combination of the exposures
with the task {\em drizzle}.  These reflect the effective total integration
time at each output pixel in the drizzling procedure (i.e., accounting for
the dropsize and the relative size of the input and output pixels as well
as the bad pixel and cosmic ray masking).  We normalized the weight maps of
each galaxy to a maximum of unity.  We cropped the final images and weight
maps to a region $19\farcs 55 \times 18\farcs 05$ to exclude the edges with
relative weight $\la 0.4$.

\subsection{PSF}   \label{Sub-psf}

For the morphological analysis, accurate knowledge of the PSF in our
data is important.  We empirically determined the resulting PSF based
on stellar profiles in our final reduced images.  In total, there are
four suitable stars (isolated, sufficiently bright but unsaturated) in
three of our NIC2 pointings.  Their profiles have a Gaussian core and
show Airy wings at larger radii (around $0\farcs 23$).  To increase the
S/N, we averaged the peak-normalized stellar profiles.  From a Gaussian fit
to the resulting empirical PSF, the FWHM is $0\farcs 145$ with ellipticity
of 0.08 (or FWHM of $0\farcs 143$ and identical ellipticity of 0.08 for a
Moffat fit).  Relative to the average PSF, the deviations of the radial
profiles of the stars are at most $15\%$ (on average $\approx 5\%$) out
to a radius of $0\farcs 35$, enclosing $95\%$ of the total flux and where
the profiles are at $\approx 1\%$ of the peak value.  The curves-of-growth
of the individual stars differ by $< 10\%$ (on average $\approx 3\%$) from
that of the average PSF.  We adopted this empirical average PSF for our
analysis.

\subsection{Noise Properties and Limiting Depths}   \label{Sub-noise}

The quality of the final reduced data is very uniform among our targets.
To characterize the effective noise properties, we followed the approach
described in detail by \citet{Lab03} and \citet{FS06a}.  For uncorrelated
Gaussian noise, the effective standard deviation from the mean of the
background for an aperture of area $A$ is
simply the pixel-to-pixel rms $\overline{\sigma}$ scaled by the linear size
$N = \sqrt A$ of the aperture, $\sigma(N) = N\,\overline{\sigma}$.  However,
instrumental features, the data reduction, and the drizzling and combination
procedures have added significant systematics and correlated noise in the
final data.  We thus derived the function $\sigma(N)$ directly from the data
by measuring the fluxes of $\approx 240$ non-overlapping circular apertures.
We placed the apertures at random in the maps but so as to avoid pixels
associated with sources down to $\rm S/N \approx 3$, corresponding to a
surface brightness of $\mu(H_{\rm 160,AB}) = 23.4~{\rm mag~arcsec^{-2}}$.
Series of measurements were repeated with different aperture diameters
ranging from $0\farcs 1$ to 1\arcsec.

For each aperture size, the distribution of fluxes is symmetric around 
the background value (about 0), and is well reproduced with a Gaussian
profile.  We determined the background rms variations from the width of
the best-fit Gaussian to the observed distribution for each aperture size.
While at any fixed spatial scale, the noise properties over the images are 
consistent with being Gaussian, the variations with $N$ deviate appreciably 
from the scaling expected for Gaussian noise, being systematically larger
and increasing non-linearly.  The real noise behaviour in our NIC2 images
can be approximated by the polynomial model:
\begin{equation}
\sigma(N) =
\frac{N\,\overline{\sigma}\,\left(a + b\,N\right)}{\sqrt w},
\label{Eq-noise}
\end{equation}
where the weight term $1/\sqrt w$ accounts for the spatial variations
in noise level related to the exposure time and is taken from the weight
maps.  The coefficient $a$ represents the effects of correlated noise,
which are dominated by the resampling of the images to the final pixel
scale through the drizzling procedure.  The curvature in the $\sigma(N)$
relation quantified by the coefficient $b$ indicates a noise contribution
that becomes increasingly important on larger scales.  For our data, this
probably originates mostly from instrumental features, and residuals from
the background subtraction and flat-fielding.
Qualitatively, the same second-order polynomial behavior is seen
in ground- and space-based optical and near-IR imaging data analyzed by
\citet[their Figure~4]{Lab03}, \citet[their Figure~6]{FS06a}, and
\citet[their Figure~3]{Wuy08}.

The noise properties are very similar in all NIC2 maps.  The derived
pixel-to-pixel rms values are identical within $3\%$.  The mean and
standard deviation of the best-fit polynomial coefficients among the
six images are 1.20 and 0.15 for $a$, and 0.039 and 0.017 for $b$.
Limiting magnitudes and photometric uncertainties from our NIC2 data are
computed through Eq.~\ref{Eq-noise} for the corresponding aperture size.
The effective $3\,\sigma$ magnitude limits in a ``point-source aperture''
with diameter $d = 1.5 \times {\rm FWHM} = 0\farcs 22$ (maximizing the S/N
of photometric measurements in unweighted circular apertures of point-like
sources) are $\rm \approx 28.1~mag$ (see Table~\ref{tab-obs}).

\section{GLOBAL PROPERTIES OF THE GALAXIES}    \label{Sect-global}

\subsection{General Appearance of the NIC2 Images}
            \label{Sub-look}

Figure~\ref{fig-nicmaps} presents the final reduced NIC2 images at
a sampling of $\rm 0\farcs 05~pixel^{-1}$.  The angular resolution
from the effective PSF with $\rm FWHM = 0\farcs 145$ corresponds to a
spatial resolution of $\rm 1.2~kpc$ at the median $z = 2.2$ of the targets.
We computed the geometric center as the unweighted average of the $x$ and
$y$ coordinates of pixels with $\rm S/N > 3$.  Emission from the galaxies
is detected out to radii of $0\farcs 8 - 1\farcs 2$, or $\rm 6.5 - 10~kpc$.
The averaged surface brightness of the sources over the emitting regions
(with $\rm S/N > 3$) is in the range 
$\mu(H_{\rm 160,AB}) = 22.6 - 23.0~{\rm mag~arcsec^{-2}}$,
with typical $\rm S/N = 4 - 6$ per pixel.

The spatial distribution of the $H_{160}$ band light in the
images is characterized by a diffuse and low surface brightness
component over which rich, brighter substructure is superposed.
The most prominent features are the bright and compact regions, seen
in all galaxies; each of the non-AGN disks contains several such clumps.
BX\,663 exhibits a bright central peak, and only one other and comparatively
fainter clump to the southeast; since the rest-UV and optical emission lines
of this object indicate the presence of a Type 2 AGN, the compact central
$H_{160}$ band peak may not be dominated by AGN emission and could trace
in large part stellar continuum light.
Detailed modeling of the optical to mid-IR spectral energy distribution
of this galaxy will help to place constraints on the contribution of the
stars and AGN \citep{Hai11}.
In BX\,528, the merger components
are clearly separated in our NIC2 image, with a projected distance of
$\rm \approx 8~kpc$ between the compact northwestern and the more extended
southeastern parts (hereafter BX\,528$-$NW and BX\,528$-$SE) and significant
emission along a ``bridge'' between the two components.  The fainter source
$\approx 1\farcs 8$ to the northeast is undetected in our SINFONI data so
that its association with BX\,528 is unclear.  Likewise for the source
$\approx 2^{\prime\prime}$ south of BX\,482, which could also result from
chance alignment.  In contrast, the small source $\approx 0\farcs 6$ south
of the main body of BX\,389 corresponds to an extension in our seeing-limited
H$\alpha$ maps (with about 3 times coarser resolution) and is thus likely to
be a small satellite at the same redshift and projected distance of 5~kpc
(hereafter ``BX\,389$-$S''; see \S\S~\ref{Sect-res} and \ref{Sect-disc}).
No other source is detected in the NIC2 images within
$\approx 2^{\prime\prime}$ or projected $\rm \approx 15~kpc$ of our
primary targets; this area corresponds to the deeper part of the SINFONI
data, where we can search for line emission to faint levels of candidate
companions to determine a physical association.

Other characteristics of note include the broad curved features
reminiscent of spiral arms seen in the outer isophotes of BX\,610, on the
northwest and southeast sides.  In addition, the inner isophotes in this
galaxy show a significant and systematic increase in ellipticity and change
in position angle.  From fitting elliptical isophotes to the NIC2 image
using the task {\em ellipse} in IRAF, the apparent (uncorrected for beam
smearing) ellipticity and P.A. vary from $\sim 0.8$ and $\sim 35^{\circ}$
at radii $r \la 0\farcs 5$ (or 4~kpc) to $\sim 0.1$ and $\sim 0^{\circ}$
at $r \sim 1^{\prime\prime}$ (8~kpc).  The interpretation of the inner
isophotal feature is complicated by the presence of the clumps, which
are mostly aligned along a similar P.A., but it is suggestive of
a bar-like structure as seen in local barred spirals.

\subsection{Photometric Properties}
              \label{Sub-phot}

The new $H_{160}$ band photometry from our {\em HST\/}/NICMOS data
constitute a significant addition to that available for our targets,
especially for BX\,482 without any previous near-IR imaging and MD\,41
for which the existing $K$-band data are fairly shallow.  Published data
include ground-based optical $U_{n}\,G\,\mathcal{R}$ imaging for all
sources, and near-IR $J\,K_{\rm s}$ imaging for a subset, presented by
\citet[][see also \citealt{Ste04}]{Erb06}.  The photometry was measured in
matched isophotal apertures determined from the $\mathcal{R}$ band images,
and corrected to ``total'' magnitudes in elliptical apertures scaled based
on the first moment of the $\mathcal{R}$-band light profiles \citep{Ste03}.
{\em Spitzer\/}/IRAC mid-IR photometry has subsequently been obtained for
all objects except BX\,482 \citep{Red10}.  To construct the broad-band
spectral energy distributions (SEDs) and derive the global stellar and
extinction properties of our targets, we however excluded the IRAC data
for the following reasons.  The optical and near-IR photometry provides
the most uniform data sets among the six galaxies in terms of wavelength
coverage whereas of the five objects with IRAC data, only three were
observed in all four channels.  Moreover, contamination by AGN emission
in BX\,663 and BX\,610 becomes important at the longer wavelengths probed
by IRAC and would affect the modeling results.

To include the $H_{160}$ band in the SEDs of our targets, we measured the
total magnitude in a circular aperture of diameter 3\arcsec\ on the NIC2
images (indicated in Figure~\ref{fig-nicmaps}).  Because of the small NIC2
field of view, the accuracy of the measurements in the $\mathcal{R}$-band
isophotal aperture used for the ground-based photometry on the PSF-matched
$H_{160}$ band images is compromised by the difficulty of determining the
local background sufficiently reliably: the area empty of sources signal
after convolving to the optical seeing of $\sim 1^{\prime\prime}$ is too
small.  The median difference between the total magnitudes based on the
fixed circular aperture measurement on unconvolved NIC2 maps and that in
the isophotal apertures on PSF-matched NIC2 maps is $\rm -0.16~mag$, with
the maximum for BX\,389 ($\rm -0.51~mag$) and best agreement for BX\,528
($\rm -0.002~mag$).  The photometry is given in Table~\ref{tab-targets}
and the optical/near-IR SEDs are plotted in Figure~\ref{fig-seds}.

By their primary selection, all targets satisfy the
$U_{n}\,G\,\mathcal{R}$ color criteria of ``BX'' or ``MD'' objects
\citep{Ade04, Ste04}.  Synthetic broad-band colors measured on the spectrum
of the best-fitting stellar population model to the observed SEDs (described
in \S~\ref{Sub-sedmod} below) suggest all targets would also satisfy the
$BzK = (z - K)_{\rm AB} - (B - z)_{\rm AB} > -0.2~{\rm mag}$ criterion
of $z > 1.4$ ``star-forming $BzK$ objects'' introduced by \citet{Dad04}.
In addition, two of the four sources with both $J$ and $K$ photometry
(BX\,663 and BX\,389) have the red $(J - K_{\rm s})_{\rm AB} > 1.34$
colors of ``Distant Red Galaxies'' \citep[DRGs;][]{Fra03, Dok04}, and
one (BX\,610, $(J - K_{\rm s})_{\rm AB} = 1.28~{\rm mag}$) is within
$0.3\,\sigma$ of the DRG criterion.  For these three sources and for
BX\,528, the photometry is consistent with the presence of a significant
population of evolved stars along with the young stars producing the blue
rest-UV SED.
The best-fit models to the SEDs of these four objects imply rest-frame
$(U-B)_{\rm AB} \approx 0.75 - 0.9~{\rm mag}$ colors, close to or just
at the separation adopted by, e.g., \citet{Kri09}, between ``red'' and
``blue'' objects.  In contrast, MD\,41 and BX\,482 are significantly
bluer with rest-frame $(U-B)_{\rm AB} \approx 0.5~{\rm mag}$.

Our NIC2 sample is relatively bright in the near-IR; the five targets
with $K$ band imaging have a median $K_{\rm s,AB} = 21.8~{\rm mag}$
and span a range of $\rm \approx 1.2~mag$ in observed $K$ band
($K_{\rm s,AB} = 21.1 - 22.3~{\rm mag}$) as well as in rest-frame
absolute $V$ band magnitude (as derived from the SED modeling below,
with $M_{V,{\rm AB}}$ between $-23.1$ and $\rm -21.9~mag$).
In addition, the {\em Spitzer\/}/MIPS 24\,\micron\ measurements available
for four of the targets (BX\,389, BX\,610, BX\,528, and BX\,663) indicate
that they are among the brightest 3\% of rest-UV selected $z \sim 2$ galaxies,
in terms of rest-frame 8\,\micron\ luminosity \citep{Red10}.

\subsection{SED Modeling and Stellar Properties}
            \label{Sub-sedmod}

Stellar population synthesis modeling of BX\,528, BX\,663, BX\,389,
and BX\,610 was first presented by \citet{Erb06}.  The new $H_{160}$
photometry from NIC2, and $K$ band flux for MD\,41, provide additional
constraints and allow us to extend the SED modeling to BX\,482 and MD\,41,
which had previously only three optical data points.  For consistently
derived stellar and dust extinction properties within our sample, we
re-modeled the optical/near-IR SEDs of all six objects\,\footnote{
 We verified that the SED modeling results are essentially unchanged
 when including the available IRAC photometry for the three targets
 without AGN contribution to their mid-IR SED.
}.
We followed standard procedures \citep[e.g.][]{Sha01, Erb06, FS06a},
exploring also a range of parameters and model ingredients to assess
possible systematics from the assumptions adopted.  Our specific
procedure is described in detail by \citet{FS09} and summarized here.

We used the stellar evolutionary code from \citet{BC03}.  The age,
extinction, and luminosity scaling of the synthetic model SED were
the free parameters in the fits.  The ages were allowed to vary between
50~Myr and the age of the Universe at the redshift of the source.  The
lower limit is set to avoid implausibly young solutions and corresponds
approximately to the dynamical timescale $\tau_{\rm dyn} = r / v$ from
the inclination-corrected velocity $v$ at a radius of $r \sim 10~{\rm kpc}$
of the galaxies \citep[e.g.][]{FS06b, Gen08}.  We employed the \citet{Chab03}
IMF and the reddening curve of \citet{Cal00}, and assumed a solar metallicity.
We fixed the star formation history (SFH), adopting a constant star formation
rate (``CSF'') model, which we found appropriate for most galaxies (see below,
and also \citealt{Erb06}).  We derived the formal (random) uncertainties
on the best-fit parameters from 200 Monte-Carlo simulations, where we varied
the input photometry assuming a Gaussian probability distribution for the
measurements uncertainties (see \S~\ref{Sub-noise}).  To explore the impact
of model assumptions and ingredients, we also used the evolutionary synthesis
code of \citet{Mar05}, and ran suites of models with exponentially declining
star formation rates (SFRs) with $e$-folding timescales $\tau$ between 10~Myr
and 1~Gyr, with metallicity of $1/5$ and 2.5 times solar, and with extinction
laws appropriate for the Small Magellanic Cloud \citep{Pre84, Bou85} and the
Milky Way \citep{All76}.

Table~\ref{tab-galprop} lists the best-fit properties for our adopted
set of \citet{BC03} models with CSF, solar metallicity, and \citet{Cal00}
law along with the formal uncertainties taken as the $68\%$ confidence
intervals from the Monte-Carlo simulations around the best-fit values.
Figure~\ref{fig-seds} shows the corresponding synthetic model spectra;
best-fit \citet{Mar05} models for the same CSF, metallicity, and reddening
law are also shown for comparison.
To gauge systematic uncertainties, we looked at the variations in derived
properties for the other suites of model assumptions and synthesis code
considered.
For individual galaxies, and compared to the adopted
properties, the median of the variations are as follows:
the ages are mostly younger by a factor of $\approx 2$,
the extinction in $V$ band $A_{V}$ decrease by $\rm \approx 0.2~mag$,
the stellar masses $M_{\star}$ tend to be lower by a factor of $\approx 1.5$,
the absolute and specific SFRs tend to decrease by factors of $\approx 2$.
The overall sense of these variations reflects in part the fact that, by
construction, CSF models generally lead to older ages than declining SFHs.
While the systematic uncertainties above are significant, the derived
properties of the galaxies are robust in a relative sense.  In particular,
MD\,41 and BX\,482 remain the youngest and least massive of the sample in
any case.  

Our choice of CSF is obviously very simplistic.  For the range of SFHs
above, we find that the longest timescales ($\rm \tau = 1~Gyr$ or CSF)
are favored (in a minimum $\chi^{2}$ sense) for four out of six galaxies.
MD\,41 and BX\,482, with bluest SEDs, have best fits formally obtained
for models with $\rm \tau \la 50~Myr$ but the age, $M_{\star}$, and SFR 
are within $3\,\sigma$ or less of the values for CSF.  Recent work has
emphasized the fact that declining SFHs may well be inappropriate for
high $z$ star-forming galaxies \citep{Ren09, Mar10}.  Our exploration
of constant and declining SFHs was motivated by continuity with previous
work and consistency with the SED modeling assumptions for the comparison
samples discussed in \S~\ref{Sect-comp}.

We did not account for emission line contribution in our SED modeling.
The main contribution expected for our sample is from H$\alpha$ in the
$K$ band, and these are in the range $7\% - 19\%$ (with median of $13\%$)
based on our SINFONI $K$ band observations \citep[see also][]{Erb06}.
The median [\ion{N}{2}]\,$\rm \lambda\,6584/H\alpha$ ratio is $0.23$ (with
highest ratio 0.43 for BX\,610).  SINFONI $H$ band data are available for
five targets; the [\ion{O}{3}] doublet and H$\beta$ are all detected in two
cases and their contributions to the $H_{\rm 160}$ flux densities are smaller
than for H$\alpha$ in $K$, and $3\,\sigma$ upper limits for the others are
at most $29\%$ (see \S~\ref{Sect-sinf}).  The H$\alpha$-corrected $K$ band
flux densities are indicated in Figure~\ref{fig-seds}; the impact on the
SED modeling results is not significant in view of other uncertainties
involved.  

Our targets span over an order of magnitude in stellar mass, from
$7.7 \times 10^{9}$ to $\rm 1.0 \times 10^{11}~M_{\odot}$ with median
$\rm 5.3 \times 10^{10}~M_{\odot}$.  The stellar populations that
dominate their rest-UV to optical SEDs cover the full range allowed in our
modeling (from 50~Myr to 2.75~Gyr), and are moderately obscured (with $A_{V}$ 
between 0.6 and 1.2~mag, or equivalently $E(B-V) = 0.15 - 0.30~{\rm mag}$).
Four of the six targets have in fact ages consistent with the age of the
Universe at their redshift, indicating the presence of mature stellar
populations.
The median absolute and specific SFRs are $\rm \approx 50~M_{\odot}\,yr^{-1}$
and $\rm 0.6~Gyr^{-1}$ (with ranges of $\rm 25 - 185~M_{\odot}\,yr^{-1}$ and
$\rm 0.6 - 24~Gyr^{-1}$).
Figure~\ref{fig-sedkinprop}{\em a} compares our NIC2 sample in the 
$M_{\star} - {\rm SFR}$ plane with the full SINS H$\alpha$ sample,
and a purely $K$-selected sample taken from the Chandra Deep Field
South (CDFS) in the same $z = 1.3 - 2.6$ interval and to the same 
$K_{\rm s, Vega} = 22~{\rm mag}$ limit as the SINS sample
\citep[][and also \citealt{Wuy08}]{FS09}.  The properties of these
comparison samples were derived using the same SED fitting procedure
and model assumptions as for the NIC2 sample.  Clearly, and by selection,
our NIC2 galaxies probe the actively star-forming part of the $z \sim 2$
galaxy population \citep[see also discussions by][]{Erb06, FS09}.
Their median stellar mass and SFR are comparable to those of the
SINS H$\alpha$ sample ($\approx 3 \times 10^{10}~{\rm M_{\odot}}$
and $\rm \approx 70~M_{\odot}\,yr^{-1}$).

\subsection{H$\alpha$ and Kinematic Properties}
              \label{Sub-kin}

If our targets do not particularly stand out in terms of their stellar
properties compared to the SINS H$\alpha$ sample, they do more so in
terms of their H$\alpha$ brightness and kinematics.  This distinction
arises because, as described in \S~\ref{Sect-data}, we chose them among
the sources with highest S/N and spatially best-resolved H$\alpha$ emission
from our seeing-limited SINFONI data sets.  Five of them were explicitly
selected because of their disk-like kinematics, and BX\,528 because it
shows in contrast kinematic signatures expected for counter-rotating
mergers.  The spatial extent and brightness of our targets make them
particularly well suited for a detailed study of morphologies and
reliable assessment of structural parameters from sensitive
kpc-resolution continuum imaging.

Figure~\ref{fig-sedkinprop}{\em b} compares the NIC2 galaxies with the
SINS H$\alpha$ sample in the velocity-size plane, with the half-light
radii and (inclination-corrected) circular velocities derived from
H$\alpha$ \citep{Gen08, Cre09, FS09}.  The NIC2 targets lie at the
high end of the $v_{\rm d} - r_{1/2}{\rm (H\alpha)}$ distribution.
With $r_{1/2}{\rm (H\alpha)} \approx 4 - 5~{\rm kpc}$, they are all
larger than the median of 3.1~kpc for the SINS H$\alpha$ sample.
The $v_{\rm d}$ for the disks are roughly equal to or larger than the
median for SINS galaxies of $\rm \approx 180~km\,s^{-1}$, and up to
$\rm \sim 300~km\,s^{-1}$.  For the merger BX\,528, the velocity
gradient is lower, and leads to a lower equivalent $v_{\rm d}$ of
$\rm 145~km\,s^{-1}$ \citep[see][]{FS09}.
The ratios of circular velocity to intrinsic local velocity dispersion
of the disks are in the range $v_{\rm d} / \sigma_{0} \approx 2 - 6$,
which is significantly lower than for local spiral galaxies
\citep*[values in the range $\sim 10 - 40$; e.g.,][]{Dib06, Epi10}
and suggests comparatively larger gas turbulence and geometrical
thickness.  These low $v_{\rm d} / \sigma_{0}$ ratios appear to be a
characteristic feature of early disk galaxies at $z \sim 1 - 3$
\citep[e.g.][]{FS06b, Wri07, Gen08, Sta08, Cre09, Epi09}.

The median integrated H$\alpha$ flux of the NIC2 targets
(uncorrected for extinction) is
$F({\rm H\alpha}) = 1.9 \times 10^{-16}~{\rm erg\,s^{-1}\,cm^{-2}}$
(and range from 1.1 to $\rm 3.1 \times 10^{-16}~erg\,s^{-1}\,cm^{-2}$).
This places them among the brighter half of the SINS H$\alpha$ sample (with
median $F({\rm H\alpha}) = 1.1 \times 10^{-16}~{\rm erg\,s^{-1}\,cm^{-2}}$).
In terms of integrated H$\alpha$ line widths, their median is
$\sigma({\rm H\alpha}) = 155~{\rm km\,s^{-1}}$ (spanning the
range $\rm 120 - 245~{\rm km\,s^{-1}}$), somewhat broader than
the median for the SINS H$\alpha$ sample ($\rm 130~{\rm km\,s^{-1}}$).
This difference is due in part to the large observed velocity gradients
for the sources.

\section{ANALYSIS OF THE REST-FRAME OPTICAL MORPHOLOGIES}   \label{Sect-res}

\subsection{Parametric Analysis of the Galaxies' Structure}
           \label{Sub-galfit}

\subsubsection{Methodology}   \label{Sub-galfit_meth}

In a first step to characterize quantitatively the morphologies of our
galaxies, we followed a parametric approach.  The main goal here is to
derive the global structural parameters of the sources.  We used the
code GALFIT \citep{Pen02} to fit the two-dimensional surface brightness
distributions with a \citet{Ser68} profile:
\begin{equation}
I(r) = I(0)\,\exp\,\left[-b_{n}\,(r/R_{\rm e})^{1/n}\right],
\label{Eq-sersic}
\end{equation}
where $I$ is the intensity as a function of radius $r$, $b_{n}$
is a normalization constant, and $R_{\rm e}$ is the effective radius
enclosing half of the total light of the model.  In this parametrization,
ellipticals with de Vaucouleurs profiles have a S\'ersic index $n = 4$,
exponential disks have $n = 1$, and Gaussians have $n = 0.5$.  In the
fitting procedure, GALFIT convolves the model light distribution to the
effective resolution of the data, and determines the best fit based on
$\chi^{2}$ minimization.  We input our empirically determined PSF, and
the weight maps scaled by the pixel-to-pixel rms measured on the
reduced images (see \S~\ref{Sect-data}).
To reduce the impact of large-scale residuals from flat-fielding and
background subtraction (\S~\ref{Sub-noise}), we performed the fits within
a region $7^{\prime\prime} \times 7^{\prime\prime}$ centered on the sources.

The free parameters in the fits were the S\'ersic index ($n$), the effective
radius ($R_{\rm e}$), the axis ratio ($b/a$, semi-minor over semi-major axis
radii), the position angle (P.A.) of the major axis, and the total magnitude.
Given that surface brightness fitting of faint galaxies is sensitive to
the background determination and that S\'ersic profile parameters tend
to be degenerate with the ``sky'' value \citep[see, e.g.,][]{Pen02, Dok08},
we fixed the sky level $S$ in our fits.  Moreover, asymmetric and clumpy
surface brightness distributions can strongly bias the results, so we
also fixed the coordinates $x_{0}$ and $y_{0}$ of the center position.
In order to derive realistic uncertainties, we performed a series of
fits where these fixed parameters were drawn at random from plausible
distributions.  Specifically, for every galaxy we ran GALFIT 500 times,
each time varying the $x_{0}$ and $y_{0}$ according to a uniform distribution
within a squared box of $2 \times 2$ pixels around the geometric center as
defined in \S~\ref{Sub-look}, which corresponds closely to the dynamical
center of our targets.  For the background, the fixed input $S$ values
were drawn at random from the central $10\%$ of the distribution of
individual pixel values in the regions empty of sources (with this
range reflecting the uncertainty in the background level). The best-fit
parameters were taken as the median of the results for all 500 runs,
and the $1\,\sigma$ uncertainties as the $68\%$ confidence intervals
about the median.

We performed additional series of fits to explore systematics from
the center and sky background determinations.  In all cases, the initial
guesses for the center and sky values were generated as described above.
In the first two series, one of the sky or center was chosen as a free
parameter and the other was fixed.  In a third series, both parameters
were free.  For BX\,663, MD\,41, BX\,389, and BX\,610, the best-fit
structural parameters for these additional series are all within the
68\% confidence intervals of the results for the first series with sky
and center fixed to the randomly drawn values in each iteration.  For
BX\,528 and BX\,482, the two series with the free center position lead
to significantly different results as expected from their particularly
asymmetric morphologies, with changes in $R_{\rm e}$ and $n$ by
$\ga 2\,\sigma$.

For the interacting system BX\,528, single component fits are clearly
inappropriate.  When the center is free to vary, the fits are heavily
influenced by the brighter BX\,528$-$SE component: the best-fit center
moves towards this component, the S\'ersic index increases to match the
steeper inner profile and the $R_{\rm e}$ also increases.  For this system,
we thus also ran a series fitting two S\'ersic profiles simultaneously,
fixing the respective center position and sky to random values drawn
from appropriate distributions as explained above.
Likewise, we performed two-component fits for BX\,389 to derive the
structural parameters of the small southern companion BX\,389$-$S.
The best-fit parameters for the main part of BX\,389 are almost
identical to those obtained with the single-component fits.
For BX\,482, the large and bright clump on the southeast side (which makes
up $\approx 15\%$ of the total light after subtracting the ``background''
light from the host galaxy; see Paper~II) drives the center in this direction
when allowed as a free parameter, and leads to a somewhat higher best-fit
$n$ index.  At the same time, because the northwest side of the galaxy
is overall fainter and lacks bright small-scale features, the best-fit
$R_{\rm e}$ is smaller.  We did not consider two-component fits for
BX\,482 because the kinematics as well as morphologies are more
consistent with a single system.

\subsubsection{Results for the NIC2 Targets}   \label{Sub-galfit_res}

Table~\ref{tab-galfitres} gives the results for the main structural
parameters, $R_{\rm e}$, $n$, $b/a$, and P.A., of our targets from the
single-component fits as well as from the two-component fits for BX\,528
and BX\,389.  In Figure~\ref{fig-galfitres}, the first two columns show
the best-fit values for single-component fits along with the distributions
obtained from the 500 GALFIT runs in each parameter.  In the third column,
ellipses are drawn on the NIC2 images showing the effective radius, axis
ratio, and position angle of the best-fit models.  The fourth column shows
the fit residuals, revealing the clumpy structure and large scale asymmetries
in the light distributions.  Figure~\ref{fig-galfitres2} shows the same but
for the two-component fits to BX\,528 and BX\,389.  Figure~\ref{fig-profiles}
compares the radial light profiles of the galaxies' data and of the best-fit
model (here using the two-component model for BX\,528).
The profiles correspond to the average along ellipses centered on the
best-fit $x_{\rm 0}$ and $y_{\rm 0}$ position, and with best-fit axis
ratio and position angle of the single-component models.  This tends
to smooth out the small-scale structure and asymmetric features of the
surface brightness distributions; the detailed substructure will be
considered in the next subsection.

All of our targets qualify as morphologically late-type, disk-dominated
systems, often defined as having $n < 2 - 2.5$ \citep[e.g.,][]{Bel04, Tru06}.
The only exceptions are BX\,528$-$NW and BX\,389$-$S that have $n \approx 3$
but the large uncertainties make them within $1\,\sigma$ of the above
criterion.  A remarkable feature of the four non-AGN galaxies with
disk-like H$\alpha$ kinematics is their shallow inner light profiles.
With S\'ersic indices in the range $\approx 0.15 - 0.6$, these are
significantly less centrally concentrated than exponential disks (by
definition $n = 1$).  The residuals are overall smallest for MD\,41
and BX\,389, the two most edge-on disks based on their axis ratios
($b/a \approx 0.30$) from the fits to the $H_{160}$ band images as well as
from their H$\alpha$ morphologies and kinematics \citep{FS06b, Gen08, Cre09}.
The spiral-like features in BX\,610 noted in \S~\ref{Sub-look} are most
obvious in the outer parts beyond the effective radius.  The residuals for
the single- and double-component fits to BX\,528 indicate that neither case
provides a very good representation.  There is clearly significant light
emitted along the ``bridge'' between the SE and NW components.

BX\,663 has the highest $n$ index among our sample, reflecting the fairly
prominent central peak and shallower profile of the extended emission at
larger radii.  We explored two-component models, meant to represent the
case of a disk $+$ bulge and of a disk $+$ unresolved nuclear source.
A free $n$ component was considered for the disk, and the second component
was taken as either a de Vaucouleurs profile with fixed $n = 4$ or as the
empirically determined PSF.  Because the central peak is accounted for by
the $n = 4$ or PSF component, the underlying disk ends up with low $n$ of
0.24 (disk $+$ bulge) or 0.63 (disk $+$ unresolved source), in the range
obtained for the other five SINS targets.  The disk $R_{\rm e}$ values,
however, differ by $10\%$ or less ($< 2.5\,\sigma$) from the best-fit
single-component $R_{\rm e}$.  For the disk $+$ bulge case, the
de Vaucouleurs component has $R_{\rm e} = 2.47^{+5.03}_{-0.95}~{\rm kpc}$.
In terms of $\chi^{2}$ values and of residuals, the best-fit two-component
models are indistinguishable from the best-fit single S\'ersic model, so we
consider only the single-component model for BX\,663 throughout this paper.

We find no significant correlation between structural parameters
($R_{\rm e}$, $n$, and $b/a$) and SED derived properties ($M_{\star}$,
age, $A_{V}$, absolute and specific SFR) among our targets.
Studies of larger samples based on ground-based seeing-limited or higher
resolution NICMOS/NIC3 near-IR imaging \citep[e.g.,][]{Tru06, Tof07, Fra08}
show the existence of relationships between size, stellar mass, and star
formation activity at $z \sim 2 - 3$ (though with significant scatter).
Among our five disks, MD\,41 and BX\,389 with highest $A_{V}$ values
have the lowest axis ratios, which seems consistent with the trend of
higher extinction in more edge-on star-forming disks observed locally
\citep[e.g.,][]{Mal09, Yip10}.  However, clearly, only the strongest
correlations would be discerned with our very small sample.

\subsection{Non-parametric Analysis of the Morphologies}
            \label{Sub-morph}

\subsubsection{Methodology}   \label{Sub-morph_meth}

The models considered in the previous section are admittedly very simplistic.
Simple one- or two-component axisymmetric models can reproduce reasonably
well the overall surface brightness distribution of some of our sources but
do poorly for others.  Such a parametric approach appears to be satisfactory
to estimate global parameters such as the effective radius and shape of the
radial light profile.  On the other hand, it obviously does not capture the
prominent small-scale and irregular structure of clumpy disks and merging
systems.
The difficulty of describing star-forming galaxies at $z \sim 2 - 3$
within the framework of a Hubble sequence of more regular spiral and
elliptical morphological types has been highlighted previously by
several authors \citep[e.g.,][]{Lot06, Law07a, Pet07}.  Accordingly,
we also followed a non-parametric approach to characterize in detail
morphologies of our galaxies, measuring the Gini, $M_{20}$, and
Multiplicity coefficients, which have been introduced and applied
by \citet{Abr03}, \citet{Lot04, Lot06}, and \citet{Law07a}.

The Gini parameter, $G$, is a measure of the uniformity with which the flux
of a galaxy is distributed among its constituent pixels.  $G=0$ corresponds
to the case where the flux is distributed evenly among all pixels, while
$G=1$ indicates that a single pixel contains all of the flux.  The Gini
parameter is formally defined such that:
\begin{equation}
\label{Eq-gini}
\mathit{G} =
  \frac{1}{\bar{f}N_{pix} (N_{pix} - 1)}
  \sum^{N_{pix}}_{i=1} (2i - N_{pix} -1)f_i,
\end{equation}
where $\bar{f}$ is the mean sky-subtracted flux per pixel, $f_i$ is the
sky-subtracted flux in pixel $i$, and the pixels have been sorted in order
of increasing flux.

The $M_{20}$ parameter represents the second-order moment
of the brightest 20\% of a galaxy's flux. The total
second-order moment is defined such that:
\begin{equation}
\label{Eq-mtot}
M_{tot} = \sum_{i}^{N_{pix}} f_i [(x_i - x_c)^2 + (y_i - y_c)^2]
\end{equation}
where $f_{i}$ is the flux in each pixel, $x_i$ and $y_i$ indicate the
pixel location, and $x_c$ and $y_c$ indicate the centroid of the galaxy
light distribution, as determined from its first-order moments.  $M_{20}$
is then computed by sorting galaxy pixels by flux, and summing $M_i$ until
the pixel is reached at which $20\%$ of the total galaxy flux is contained
in pixels that have brighter or equal fluxes.  That is:
\begin{equation}
\label{Eq-m20}
M_{20} \equiv \log_{10} \left( \frac{\sum_{i} M_i}{M_{tot}} \right)
        \mbox{, while } \sum_{i} f_i < 0.2 f_{tot}.
\end{equation}
\citet{Lot04} introduced $M_{20}$ as an indicator sensitive to merger
signatures such as multiple nuclei.

Finally, a third classification parameter, the Multiplicity $(\Psi)$,
introduced by \citet{Law07a}, is somewhat analogous to the $M_{20}$
parameter, in that it is sensitive to the presence and distribution
of multiple clumps of flux.  $\Psi$ is conceptually akin to the total
gravitational potential energy of a massive body.  The ``potential energy''
of the pixel flux distribution can be described as
\begin{equation}
\label{Eq-mult_tot}
\psi_{tot} = 
  \sum^{N_{pix}}_{i = 1} \sum^{N_{pix}}_{j=1,j\neq i} \frac{f_i f_j}{r_{ij}},
\end{equation}
where $r_{ij}$ is the distance between the $i$th and $j$th pixels.
The greatest ``potential energy'' for a given set of pixel fluxes is the 
most compact distribution, with the brightest pixels at the center, and
flux decreasing outwards.  Such a configuration would be described by
\begin{equation}
\label{Eq-mult_compact}
\psi_{compact} =
  \sum^{N_{pix}}_{i = 1} \sum^{N_{pix}}_{j =1,j\neq i}
  \frac{f_i f_j}{r^{\prime}_{ij}},
\end{equation}
where $r^{\prime}_{ij}$ is the distance between the $i$th and $j$th pixels 
in the maximally compact configuration.  The multiplicity, $\Psi$, is then
defined to be
\begin{eqnarray}
\label{Eq-multiplicity}
\Psi = 100 \log_{10} \left[ \frac{\psi_{compact}}{\psi_{tot}} \right].
\end{eqnarray}
\citet{Law07a} argued that $\Psi$ provides larger dynamic range than
$M_{20}$.  Illustrative examples of the relationship between multiplicity
and overall appearance are provided by both \citet{Pet07} and \citet{Law07a}.

Some care must be taken in assigning pixels to a galaxy light distribution,
especially when considering galaxies over a range of redshifts, where the
differential effects of cosmological surface brightness dimming must be
taken into account.  We search for pixels associated with a galaxy within
a $1\farcs 5$ radius of the galaxy centroid.  Pixel selection is performed
on a lightly smoothed version of the data, which has been convolved with a
circularly symmetric Gaussian kernel with a 1 pixel standard deviation.
We select pixels to be associated with the galaxy if the smoothed flux is
more than $1.5\left(\frac{1+z_{high}}{1+z}\right)^3 \times \sigma$ above
the sky, with $\sigma$ calculated from the raw data, and where $z_{high}$
is the highest redshift for the subset of galaxies under consideration.
The redshift-dependent factor counteracts the effects of surface brightness
dimming in a given passband.  Once pixels are assigned to a galaxy light
distribution, the raw, unsmoothed data are used for morphological analysis.

\subsubsection{Results for the NIC2 Targets}   \label{Sub-morph_res}

Figure~\ref{fig-morph1} shows the resulting Gini, $M_{20}$, and $\Psi$
coefficients for the SINFONI sample.  To control for surface brightness
dimming effects, each galaxy was analyzed as if it would be observed at
the maximum redshift in the sample, i.e. $z = 2.43$.  The Gini values
range from $G = 0.25 - 0.36$, with a median of $G_{\rm med} = 0.33$.
The $\Psi$ values are in the range $2.8 - 9.8$, with a median of
$\Psi_{\rm med} = 5.4$.  Finally, the $M_{20}$ values range from
$-1.1$ to $-0.6$ with a median of $M_{\rm 20,med} = -0.7$.

In order to interpret these numbers, we compare the NIC2 morphologies
for objects in our sample with those derived from much larger samples
of rest-UV selected galaxies at similar redshifts with morphologies from
optical imaging with the {\em HST\/} Advanced Camera for Surveys (ACS).
The ACS F814W ($i_{814}$-band) rest-UV morphologies from \citet{Pet07}
provides one such comparison sample, spanning a comparable redshift range
to that of the objects in our NIC2 sample.  To make a controlled comparison,
we restrict the analysis to the subset of objects in the \citeauthor{Pet07}
sample with the same average pixel-based S/N ratio\,\footnote{
 To determine the average pixel-based S/N, we calculated the
 average ratio over all object pixels of the flux in each pixel
 and the pixel-to-pixel noise in the background.}.
This subset of objects in the \citet{Pet07} sample has an identical median
$G$ value to the objects in our sample, and median $\Psi$ and $M_{20}$ values
that are very similar ($\Psi_{\rm med} = 7.0$ and $M_{\rm 20,med} = -0.8$).
The dataset used to analyze the ACS $i_{814}$-band morphologies of rest-UV
selected galaxies by \citet{Pet07} has similar depth to the summed $BViz$
images featured in \citet{Law07a}, where rest-UV selected objects were
also analyzed.  Accordingly, these authors found roughly identical rest-UV
morphological results.  Overall, the rest-frame optical morphologies for the
objects in our sample are similar to the typical rest-frame UV morphologies
for $z \sim 2$ star-forming galaxies.  

One of the objects in our sample, MD\,41, has existing deep ACS F814W
imaging.  In Figure~\ref{fig-md41_panel}, both ACS and NIC2 images are
shown side-by-side at their respective original spatial resolution,
along with an RGB color composite after PSF matching of the ACS map to
the resolution of the NIC2 map (as described in \S~\ref{Sub-disc_md41}).
The average pixel S/N of MD\,41 is very similar in the two images,
making for a fair comparison of the wavelength-dependent morphology.
Overall, the morphologies in the two bands are strikingly similar.
The most notable difference consists of the two brightest clumps
in the ACS map at the southwestern edge of MD\,41, which have only faint
counterparts in the NIC2 image and, accordingly, exhibit bluer colors than
the rest of the galaxy.  In addition, two compact (radius $\sim 0\farcs 2$)
sources offset by $\sim 1^{\prime\prime}$ from the main body of the galaxy
appear in the ACS image, alone, indicating significantly bluer colors than
the main region of galaxy emission.  We have not yet determined if these
sources are associated with MD\,41.  In particular, a search of the SINFONI
data cube does not reveal any emission line at these locations and within
$\rm \pm 1000~km\,s^{-1}$ of the H$\alpha$ line of MD\,41, which could
support a physical association.  

The Gini values measured in the ACS and NIC2 images of MD\,41
are both $G \sim 0.3$, while the multiplicity is slightly lower for the
rest-frame optical image (6.3) than the rest-frame UV image (9.7 with the
two offset objects included and 8.1 without).  The $M_{20}$ values are
comparable as well, with $-0.7$ in the rest-frame optical and $-0.9$ in
the rest-frame UV.  While the parameters and appearance in each band are
not identical, we stress that the differences are not significant.  MD\,41
does not display a strong morphological ``$k$-correction,'' according to
which the rest-frame UV and optical morphologies would trace light from
very different stellar populations.
Similar comparative studies of the rest-UV and optical morphologies of
samples of $z \sim 1.5 - 4$ galaxies based on ACS optical and recent WFC3
$H_{160}$ band imaging are presented by \citet{Ove10} and \citet{Cam10}.

For another object in the sample, BX\,482, the SINFONI H$\alpha$ map
obtained with LGS-AO and $0\farcs 17$ PSF FWHM \citep{Gen08,FS09} is
of sufficient resolution and S/N to measure the detailed morphological
parameters using the same methodology and compare with the NIC2 $H_{160}$
band properties.  We plot the morphological parameters derived from the
H$\alpha$ map of BX\,482 in Figure~\ref{fig-morph1}, along with the
NIC2-based parameters.  The H$\alpha$ line and $H_{160}$ band morphologies
are quite similar in terms of $G$, $\Psi$, and $M_{20}$, with the prominent
high surface brightness clump appearing in both SINFONI and NIC2 images.

We searched for correlations between the morphological ($G$, $\Psi$,
and $M_{20}$) and stellar population parameters ($M_{\star}$, age, $A_{V}$,
absolute and specific SFR).  The most significant ($3\,\sigma$) trend is
found between stellar mass and $G$, such that more massive galaxies are
characterized by higher $G$ values.  Correlations between $G$ and other
gauges of galaxy evolutionary state such as age and specific SFR, are not
readily apparent within our sample.  However, our sample of six objects
is obviously too small for discerning all but the strongest trends.
Furthermore, with only actively star-forming galaxies in our sample, we
do not probe the full dynamic range of $z \sim 2$ star-forming histories.
In section~\ref{Sect-comp}, we return to the connection between morphology
and other galaxy properties, considering our SINS sample alongside $z \sim 2$
galaxies selected using very different criteria \citep{Dok08, Das08, Kri09}.

\section{COMPARISON WITH EMISSION LINE PROPERTIES FROM SINFONI}
         \label{Sect-sinf}

In this section, we combine the NIC2 images and SINFONI emission
line data of our six targets to compare the H$\alpha$ morphologies
and structural parameters with those from the rest-frame optical
emission, and determine the contribution from [OIII] and H$\beta$
line emission to broad-band $H_{160}$ flux densities.

\subsection{Comparison of Morphologies}
            \label{Sub-Hamorph_morph}

For proper comparison of the $H_{160}$ band and H$\alpha$ maps, we first
convolved the NIC2 images to the spatial resolution of the SINFONI data.
The SINFONI FOV is too small to include neighboring stars observed
simultaneously with the galaxies, and the PSFs are based on the average
of broad-band images synthesized from the SINFONI data cubes of stars
observed immediately before or after the target.  These non-simultaneous
PSF calibrations are however estimated to provide good representations of
the PSF of the targets' data, to $\approx 20\%$ in terms of the FWHM
\citep{FS09}.  We computed
the smoothing kernels with a Lucy-Richardson deconvolution algorithm.
Curve-of-growth analysis of the convolved NIC2 and original SINFONI PSFs
indicates that the fraction of enclosed flux agrees to better than 5\% in
circular apertures with diameters in the range $0\farcs 15 - 1\farcs 5$
in all cases.

After convolution, we resampled the NIC2 images to the pixel scale of the
SINFONI data where appropriate ($\rm 0\farcs 125~pixel^{-1}$ for all sets
but the AO $\rm 0\farcs 05~pixel^{-1}$ data of BX\,482) and aligned them
with the emission line maps relying primarily on the outer isophotes.
The detailed emission line and broad-band morphologies can plausibly differ
for any galaxy.  However, for our sources, and especially at a seeing-limited
resolution of $\approx 0\farcs 5$, the overall light distribution turns out
to be very similar.  We verified the registration with line-free continuum
maps extracted from the SINFONI data cubes for the brighter sources.
In contrast to the line maps, the continuum maps have lower S/N so that
mostly only the central brighter regions or the continuum peak are reliable.
The alignment based on line and continuum in those cases agrees well within
$1 - 2$ SINFONI pixels.

Figures~\ref{fig-nic_sinf1} and \ref{fig-nic_sinf2} show, for each
target, the original $H_{160}$ band image along with the H$\alpha$ line
map and velocity field with contours of the convolved $H_{160}$ band image
overplotted.  The most detailed comparison is possible for BX\,482 with
AO-assisted SINFONI data.  Both the lower surface brightness component
and the brighter sub-structure are comparable between the $H_{160}$
band and H$\alpha$ emission.  This similarity is reflected in the nearly
equal values of Gini, Multiplicity, and $M_{20}$ coefficients from the
analysis in \S~\ref{Sub-morph} (see Figure~\ref{fig-morph1}).  In the
details, there are however some noticeable differences.  In particular,
a few of the brighter peaks, or clumps, visible in the NIC2 or in the
H$\alpha$ image have no obvious counterparts in the other map.  The
properties of the clumps identified in these two images are studied
in Paper~II.  The central regions of BX\,482 also appear to be slightly
brighter in $H_{160}$ band emission than in H$\alpha$ emission.

For the other galaxies with lower resolution SINFONI data, the small-scale
substructure is smeared out but the overall $H_{160}$ band and H$\alpha$
distributions at $0\farcs 4 - 0\farcs 6$ resolution are very similar.
For BX\,389, the southern companion that is clearly separated from the main
galaxy in the original NIC2 image is still apparent at lower resolution as
an extension in the outer isophotes.  The well-resolved SE and NW merger
components of BX\,528, and the bridge of emission connecting them, blend
together and the PSF-matched NIC2 image strikingly resembles the fairly
regular and centrally concentrated H$\alpha$ line map.  While in most cases,
the general large-scale features of the morphologies are recognizable at
$\approx 0\farcs 5$ resolution, BX\,528 offers an example where the nature
of the system can only be reliably assessed with the additional information
from either the kinematics or higher spatial resolution data.

\subsection{Comparison of Structural Parameters}
            \label{Sub-Hamorph_struct}

We performed 2D S\'ersic model fits on both the H$\alpha$ maps and the
PSF-matched NIC2 images using GALFIT, following the procedure described
in \S~\ref{Sub-galfit_meth}.  For BX\,482, we ran fits on the AO-assisted
as well as the seeing-limited H$\alpha$ data.  For BX\,528 and BX\,389,
we considered only single component fits since, at the $\approx 0\farcs 5$
resolution of the SINFONI data, the merger components are unresolved in the
former case, and the southern companion is blended with the primary galaxy
in the latter.  For consistency, fits on the convolved NIC2 images were
restricted to a smaller region corresponding to the deepest area with
effective $\approx 3\farcs 8 \times 3\farcs 8$ FOV of the SINFONI data
at the $\rm 0\farcs 125~pixel^{-1}$ scale, and
$\approx 2\farcs 3 \times 2\farcs 3$ FOV of the AO data at
$\rm 0\farcs 05~pixel^{-1}$ for BX\,482.
This reduced area potentially affects the sky background estimates,
degenerate with the profile shape and half-light radius.  The smaller
FOV as well as the broader PSF both contribute to the significantly
larger uncertainties for the fits performed on the lower resolution
maps compared to those using the maps at the original NIC2 resolution.
Table~\ref{tab-galfitres_smoo} reports the best-fit structural parameters
and Figure~\ref{fig-galfitres3} compares the results from H$\alpha$ with
those from the PSF-matched NIC2 data.

Overall, there is good agreement between the measurements based on the
different data sets, with differences almost always within the (larger)
$1\,\sigma$ uncertainties of the results obtained from the lower resolution
maps.  BX\,663 is the most deviant source, with largest offsets in S\'ersic
index (smaller by $\rm 1.3 - 1.6$), half-light radius (smaller by factors
of $1.14 - 1.35$), and P.A. (differences by $\approx 15^{\circ} - 25^{\circ}$)
between the lower resolution H$\alpha$ and PSF-matched $H_{160}$ band maps
and the original resolution $H_{160}$ band image.  On the other hand, the
parameters derived from the H$\alpha$ and PSF-matched $H_{160}$ maps are
in closer agreement, with $R_{e}$ and $n$ for H$\alpha$ higher by $18\%$
and 0.3, respectively, and $b/a$ lower by $13\%$, all within the $1\,\sigma$
uncertainties.
To investigate the cause of the differences in parameters obtained from
the original $H_{160}$ band image of this galaxy and from the smoothed
and resampled version, we created a pure $n = 2$ S\'ersic model with total
magnitude, $R_{\rm e}$, $b/a$, and P.A. equal to the values of the best fit
to the original NIC2 map of BX\,663.  The mock galaxy was pasted into an
empty area of the full NIC2 map, which was then convolved to the SINFONI
PSF, and finally resampled to $\rm 0\farcs 125\,pixel^{-1}$.
The input parameters are recovered within $< 1\,\sigma$ by the fitting
procedure in the $\rm 0\farcs 05\,pixel^{-1}$ data, both at the original
NIC2 and the lower SINFONI resolution (with differences of $\leq 4\%$ in
$R_{\rm e}$ and $< 0.2$ in $n$).  Significantly lower $R_{\rm e}$ and $n$
index are derived after resampling to the larger pixel scale with, however,
the best-fit parameters for the mock galaxy in excellent agreement with those
of the real PSF-matched and resampled $H_{160}$ band image of BX\,663 (within
$< 1\,\sigma$, with $4\%$ difference in $R_{\rm e}$ and 0.12 in $n$).  These
tests indicate that the information on the steep inner profile of BX\,663
leading to $n = 2$ is lost with the coarser $\rm 0\farcs 125\,pixel^{-1}$
sampling despite the good $\approx 0\farcs 4$ resolution of its SINFONI data.

For the other five sources, a comparison of the fits to the smoothed
and original $H_{160}$ maps indicates that the half-light radii, S\'ersic
indices, and P.A.'s agree on average to $< 1\%$, 0.1, and $\rm 4^{\circ}$,
respectively, with differences among individual galaxies of at most $2\%$,
0.2, and $7^{\circ}$.  The axis ratios are higher by $\approx 30\%$ on
average (in the range $5\% - 55\%$) for the fits to the smoothed NIC2
data.  Among these five sources, the differences between fits to the
$\approx 0\farcs 5$ resolution H$\alpha$ and PSF-matched $H_{160}$ maps
are on average $3\%$ for $R_{\rm e}$, 0.08 for $n$, and $5^{\circ}$ for
P.A. (at most $17\%$, 0.12, and $11^{\circ}$).  The axis ratios from the
H$\alpha$ maps are higher than those from the smoothed $H_{160}$ data by
$20\%$ on average (in the range $3\% - 30\%$).  For BX\,482, the differences
between the parameters obtained based on the seeing-limited and AO H$\alpha$
maps are in the same ranges as for the smoothed versus unsmoothed $H_{160}$
band maps for the other sources (excluding BX\,663).
The close agreement between the parameters derived from the H$\alpha$
and $H_{160}$ band emission for BX\,482, either at $\approx 0\farcs 15$
or $\approx 0\farcs 5$ resolution, is unsurprising in view of their
similar morphologies in the high resolution SINFONI and NIC2 data
(Figures~\ref{fig-morph1} and \ref{fig-nic_sinf2}).

One important implication of the S\'ersic model fits is that the global 
H$\alpha$ and rest-frame $\rm \approx 5000$\,\AA\ light distributions do not
differ significantly from each other.  This result is further supported by
the comparison of the radial light profiles extracted along ellipses with
center, axis ratio, and P.A. from the best-fit model to the $H_{160}$ band
maps in Figure~\ref{fig-profiles2}.  In this Figure, the lower resolution
H$\alpha$ and PSF-matched $H_{160}$ maps are used for all sources, except
for BX\,482 where the profiles are extracted from the AO-assisted H$\alpha$
data and original $H_{160}$ band image.  At $\approx 0\farcs 5$ resolution,
the H$\alpha$ and $H_{160}$ band profiles follow closely each other, with
differences becoming important only beyond $\approx 1.5 - 2\,R_{\rm e}$
or typically $\rm 8 - 10~kpc$.  For the low flux levels reached at these
large radii, systematic residuals in background subtraction and flatfielding
potentially affect the profiles in a substantial way.
At radii $\la 1.5 - 2\,R_{\rm e}$, some of the (small) discrepancies may
reflect differences in the detailed morphologies as well as uncertainties
in the exact SINFONI PSF and thus the convolution kernel applied to smooth
the NIC2 data.  For the higher resolution profiles of BX\,482, the ring-like
feature that is dominated by the very bright clump on the southeast side
of the galaxy is prominent in both H$\alpha$ and rest-frame
$\rm \approx 5000$\,\AA; as seen also from Figure~\ref{fig-nic_sinf2},
the regions in the vicinity of the galaxy center are comparatively
fainter in H$\alpha$ than in $H_{160}$ band light.

The comparisons above do not indicate significant beam-smearing
effects between angular resolution of $\rm FWHM \approx 0\farcs 15$
and $\approx 0\farcs 5$ for the objects in our NIC2 sample.  The
galaxies are sufficiently large that their overall radial light
distributions remain well resolved at $\approx 0\farcs 5$ resolution.
The most noticeable difference is the systematic shift to higher axis
ratios for all sources.  The coarser $\rm 0\farcs 125\,pixel^{-1}$ sampling
leads to a significant decrease in S\'ersic index for BX\,663, the object
with steepest inner light profile ($n = 2$) at $\rm 0\farcs 05\,pixel^{-1}$.
The parametric fits do not reveal any major difference between H$\alpha$
and broad-band rest-frame $\approx 5000$\,\AA\ emission in terms of
$R_{\rm e}$, $n$, and P.A. for the five sources with $n < 1$.

\subsection{Emission Line Contributions}
           \label{Sub-linecontrib}

In view of the similarity between the $H_{160}$ band and H$\alpha$ maps
for our targets, it is important to assess the contribution of emission
lines to the broad-band emission.  At the redshift of our targets, the
NIC2 F160W bandpass includes the [\ion{O}{3}]\,$\lambda\lambda\,4959,5007$
doublet and H$\beta$.  For all sources except BX\,528, SINFONI seeing-limited
observations of these lines with the $H$ band grating were taken as part
of the SINS survey \citep{FS09}.  These data are analyzed and discussed by
\citet{Bus11}, to which we refer for details.  The [OIII] lines are detected
in all galaxies but BX\,663, while H$\beta$ is detected in MD\,41 and BX\,610.
Some of the non-detections can be attributed to the galaxies' lines being
redshifted to the wavelength of bright telluric lines (in particular
H$\beta$ for BX\,663 and BX\,482).  For all non-detections, we derived
upper limits on the fluxes.

Table~\ref{tab-lineprop} reports the fractional emission line
contribution based on the integrated $H_{160}$ band flux density
and line fluxes (or $3\,\sigma$ upper limits).  In all five sources
with SINFONI $H$-band data, H$\beta$ is the weakest line.  The largest
line contribution is inferred for BX\,389, where the [\ion{O}{3}] lines
make up $24\%$ of the total $H_{160}$ band flux density and, with the
$3\,\sigma$ upper limit on H$\beta$, the total line contribution is
estimated at $< 29\%$.  BX\,610 has the smallest contributions, amounting
to $\approx 6\%$ for all three lines combined.  The broad-band $H_{160}$
emission of our NIC2 targets is thus clearly dominated by the continuum,
with only a modest to very small fraction from the [\ion{O}{3}] and
H$\beta$ lines.

While the general appearance of the targets in [\ion{O}{3}] emission, 
and in H$\beta$ where detected, resembles that in H$\alpha$, there are in
some cases measurable differences.  These gradients in line ratios reflect
spatial variations in line excitation or gas-phase oxygen abundance, and
in dust extinction across the targets \citep{Gen08, Bus11}.
To assess possible spatial variations in the emission line contributions,
we also examined the galaxies on a pixel by pixel basis after convolving
the NIC2 maps to the PSFs of the SINFONI $H$ band data, resampling them to
$\rm 0\farcs 125~pixel^{-1}$, and aligning them following the same procedure
as for H$\alpha$.  We considered only pixels with $\rm S/N \geq 3$ in both
$H_{160}$ flux density and line flux (the summed fluxes over these pixels
represent typically $> 75\%$ of the integrated measurements of the galaxies
in 3\,\arcsec -diameter apertures).
The mean and median of the [\ion{O}{3}]\,$\lambda\,5007$ contributions for
the four sources where the line is detected are within $\approx 7\%$ of the
estimates from the integrated fluxes.
The top (75\%) quartile of the pixel line contributions range from 7\%
(BX\,610) to 16\% (BX\,389).  For H$\beta$, detected in MD\,41 and BX\,610,
the mean and median contributions in individual pixels are within $\leq 5\%$
of those determined from the total fluxes, with top quartiles of 7\% for
MD\,41 and 2\% for BX\,610.  We conclude that spatial variations in line
emission are not likely to affect significantly the observed $H_{160}$
band morphologies.

\section{COMPARISON WITH OTHER $z \sim 2$ SAMPLES}    \label{Sect-comp}

As we found in \S~\ref{Sect-res}, there are hardly any significant trends
of structural parameters or morphological properties with global stellar and
dust extinction properties among our six SINS galaxies.  However, the small
sample size leads to limited statistics and range in parameter space covered.
In order to expand our analysis and ultimately place the morphological
results for our NIC2 sample of rest-UV selected galaxies in the larger
context of the $z \sim 2$ galaxy population, we compare their properties
to those of galaxy samples selected using very different photometric
criteria.  We restrict this comparison to samples in the same redshift
range that have been characterized using the same type of observations,
i.e. {\em HST\/}/NIC2 F160W imaging.
The two comparison samples are 19 $K$-selected galaxies presented
by \citet{Dok08} and \citet{Kri09}, and 11 24\,\micron -selected
dusty, IR-luminous galaxies from \citet{Das08}.

The parent sample for the \citet{Dok08} and \citet{Kri09} analyses
is described by \citet{Kri06, Kri08a} and consists of 36 massive
galaxies with $K_{\rm Vega} < 19.7~{\rm mag}$, $1.6 \leq z \leq 3.0$,
and $\log(M_{\star}/{\rm M_{\odot}}) > 10.5$.  The stellar masses are
given by \citet{Kri08a} for a \citet{Sal55} IMF; we divided them by 1.7
for consistency with our adopted \citet{Chab03} IMF throughout the analysis
below.  All galaxies in this $K$-selected sample have spectroscopy with the
{\em Gemini} near-IR spectrograph (GNIRS), and 19 objects obtained early in
the survey, at $2 < z < 2.6$, have NIC2 imaging with the F160W filter.
Based on their GNIRS spectra, nine of these objects present no detectable
H$\alpha$ emission, with an upper limit of $W({\rm H\alpha}) = 10$\,\AA.
The remaining ten objects have detectable rest-frame optical emission lines,
indicating either star formation or AGN activity
\citep[for six and four objects, respectively;][]{Kri09}.
The range of activity in this roughly stellar mass-selected sample of 19
objects is reflected in their structural parameters.  The nine quiescent
galaxies are very compact, nearly circularly symmetric, and show little
if any substructure except for one source.  They are characterized by a
median circularized effective radius\,\footnote{
  The circularized effective radius is defined as
  $R_{\rm e, circ} \equiv R_{\rm e}\,\sqrt{b/a}$.}
of $\rm 0.9~kpc$ (ranging from 0.5
to 2.4~kpc), and a median S\'ersic index, $n$, of 2.3 (ranging from 0.5
to 4.5).  The six star-forming galaxies are signicantly more extended
and less centrally-concentrated, and appear more similar to our SINS
targets.  Their median $R_{\rm e,circ}$ is $\rm 2.8~kpc$ (ranging from
1.5 to 5.0~kpc) and median $n$ index is 0.9 (ranging from 0.0 to 3.3).
The four AGN sources resemble the quiescent galaxies on average in terms
of structure, with a median size of $R_{\rm e,circ} \approx 1.1~{\rm kpc}$
and S\'ersic index of $n \approx 2.5$.

As discussed by \citet{Kri09},
in terms of establishing evolutionary connections within this $K$-selected
NIC2 sample, half of the star-forming galaxies appear to constitute direct
structural precursors to the AGN and quiescent galaxies, based on the fact
that the majority of their stellar mass may be contained within a compact
core.  However, the remaining star-forming galaxies lack evidence for a
strong, central mass concentration and therefore any direct structural link
with the quiescent objects \citep{Kri09}.  The median stellar mass for the
quiescent galaxies is $\rm 1.6 \times 10^{11}~M_{\odot}$, implying central
stellar mass densities $\sim 100$ times higher than for $z \sim 0$ galaxies
with similar stellar mass, characterized by significantly larger effective
radii ($\rm \sim 5~kpc$).   Understanding the formation and evolution of
such compact, quiescent galaxies at high $z$ is the basis of much
ongoing observational and theoretical work
\citep[e.g.,][]{Tru07, Cim08, Wel08, Damj09, Hop09, Man10, Wuy10}.

The parent sample for the objects we analyzed from \citet{Das08}
consists of 52 IR-bright sources in the {\em Spitzer} Extragalactic
First Look Survey, with 24\,\micron\ fluxes greater than 0.9~mJy,
$\nu F_{\nu}({\rm 24\mu m})/\nu F_{\nu}({\rm 8\mu m}) > 0.5$, and
$\nu F_{\nu}({\rm 24\mu m})/\nu F_{\nu}({\rm 6440\mbox{\AA}}) > 1$.
The latter two criteria were designed to select dusty starbursts at
$z \ga 1$, or obscured AGNs.  Thirty-three such sources at $z > 1.5$
were imaged with NIC2 and the F160W filter, 11 of which are in the
$2.0 \leq z \leq 2.5$ range of interest here.  All of these sources have
rest-frame 14\,\micron\ luminosities greater than $\rm 10^{12}~L_{\odot}$
(derived from {\em Spitzer}/IRS data), implying total IR luminosities
greater than $\rm \sim 10^{13}~L_{\odot}$.  Each is classified as one
of multiple types based on its mid-IR IRS spectrum.  The sample
considered here contains the following types:
``PAH,'' indicating a spectrum dominated by PAH emission (one object);
``AGN,'' indicating a spectrum characterized primarily by the AGN
continuum (two objects);
``obscured,'' indicating a spectrum dominated by the SiO absorption
feature at $\rm 9.7~\mu m$ (four objects);
or, finally, ``mixed," in which no single starburst or AGN component
dominates the spectrum (four objects).  The median NIC2 $H_{160}$ band
magnitude for these 11 objects is $H_{\rm 160,AB} = 22.19~{\rm mag}$,
ranging from 21.51 to 24.10~mag, and their median half-light radius
(estimated using the SExtractor software package; \citealt{Ber96})
is 2.4~kpc, ranging from 1.5 to 4.3~kpc.  These sources show varied
morphologies, from compact and regular to multiple systems, to disturbed
and/or irregular; the majority, however, tends to be more compact and
concentrated than our SINS targets.  \citet{Das08} assign a qualitative
description of the morphology of each object in their NIC2 sample, to
assess the importance of merger events.  Of the 11 objects described
here, five are classified as exhibiting no signs of interaction, five
are ``Distorted,'' and one is ``Binary.''

These comparison samples observed with the same NICMOS camera and filter,
and spanning the same redshift range as our SINS NIC2 targets, enable
in principle a controlled comparison of morphological properties.
However, slightly different analysis techniques, integration times,
and pixel sampling are featured in the published work.  These differences
have consequences for the derivation of the morphological properties, which
depend on the depth of the data as well as on the pixel sampling and S/N
threshold applied for the analysis.  We therefore obtained the actual images
for these samples (van Dokkum 2008, private communication; Yan 2008, private
communication; Kriek 2010, private communication) and performed a uniform
analysis of our NIC2 sample alongside each of the comparison samples.

\subsection{Comparison of Structural Parameters}
           \label{Sub-comp_galfit}

As a first step in our comparison, we considered the global structural
parameters and, more specifically, the effective radius $R_{\rm e}$ and
S\'ersic index $n$.
While each of the SINS targets was observed for four orbits, 17 of the 19
$K$-selected targets were observed for three orbits and the two brightest
objects for two orbits.  In the 24\,\micron-selected sample, five objects
were observed for two orbits, and six for only one orbit.  Moreover, the
published images from \citet{Dok08} and \citet{Kri09} were drizzled to
$0\farcs 038$ pixels, while those of \citet{Das08} to $0\farcs 076$ pixels.
To match both the imaging depth and sampling of the two comparison samples,
we created 1-, 2-, and 3-orbit subsets of the NIC2 data of our SINS targets,
and also drizzled them to $0\farcs 038$ and $\rm 0\farcs 076~pixel^{-1}$.
We repeated our single-component S\'ersic model fitting as described in
\S~\ref{Sub-galfit_meth} for each combination of exposure time and pixel
sampling.  Because of the differences in depth and pixel sampling of the
\citet{Dok08} and \citet{Kri09} NIC2 data with respect to those of
\citet{Das08}, we compare our SINS targets separately with the
$K$-selected and 24\,\micron -selected samples.

For the $K$-selected galaxies, \citet{Dok08} and \citet{Kri09} derived
the structural parameters using the same code as we employed (GALFIT)
and a conceptually analogous approach was followed in treating the sky
background.  We thus adopted the
best-fit $R_{\rm e}$ and $n$ parameters as published by these authors,
converting the circularized effective radii to standard effective radii.
None of our conclusions is affected if we convert our radii to circularized
radii instead.  For the SINS galaxies, the difference in structural
parameters between the 3- and 4-orbit subsets is negligible.  The pixel
scale has fairly little impact on the derived parameters of extended,
well-resolved sources but more so for compact ones.  Accordingly, for all
but one of our SINS targets, the resulting best-fit $R_{\rm e}$ and $n$
obtained from the $0\farcs 038$ and $\rm 0\farcs 05~pixel^{-1}$ images
are nearly identical and within the (small) $1 - 2\,\sigma$ uncertainties,
depending on the source and parameter.  For BX\,663, the AGN host with a
compact central peak, the changes are significantly larger than for the
other five SINS sources although they are formally within $1.5\,\sigma$
of the best-fit values for each pixel scale\,\footnote{
 Specifically, the best-fit $R_{\rm e}$ and $n$ for the finer
 $\rm 0\farcs 038~pixel^{-1}$ sampling are $\rm 3.6^{+6.2}_{-0.3}~kpc$
 and $1.31^{+2.15}_{-0.38}$, respectively,
 compared to $\rm 4.5^{+7.7}_{-0.8}~kpc$ and $2.00^{+2.11}_{-0.59}$
 for the $\rm 0\farcs 05~pixel^{-1}$ one.}.

For the 24\,\micron -selected dusty IR-luminous sample, \citet{Das08}
did not perform single-component S\'ersic profile fitting for their
analysis.  We thus applied our fitting procedure using their images
to obtain the $R_{\rm e}$ and $n$ for comparison with our sample.
The resulting effective radii agree on average within $20\%$, or
$< 1\,\sigma$ based on our fitting uncertainties, with those estimated
using the SExtractor software by \citeauthor{Das08}.  The most important
difference (by $30\%$, or $3\,\sigma$) occurs for the largest, most diffuse
and asymmetric source (MIPS\,8242).  For our SINS targets, and compared
to the nominal fits for the 4-orbit $\rm 0\farcs 05~pixel^{-1}$ data,
the differences resulting from the shallower subsets and larger
$\rm 0\farcs 076~pixel^{-1}$ sampling are typically $\la 3\,\sigma$.
Again, BX\,663 shows the largest changes with $R_{\rm e}$ smaller by
$\rm \approx 0.6 - 1.2~kpc$ and $n$ lower by $\approx 0.7 - 1$,
depending on the subset.

Figure~\ref{fig-galfit_comp} plots the best-fit $n$ index versus
$R_{\rm e}$ of our SINS sample along with those of the $K$-selected
and 24\,\micron -selected comparison samples.
In terms of size and S\'ersic index, the quiescent sample of \citet{Dok08}
is clearly distinct from our SINS galaxies, with overall significantly
smaller $R_{\rm e}$ and typically higher $n$ values.  The \citet{Kri09}
objects are intermediate between the passive objects of \citeauthor{Dok08}
and our actively star-forming SINS sources with, as already discussed by
\citeauthor{Kri09}, a clear separation between the AGN sources that have
sizes and $n$ indices in a narrow range and fully overlapping with the
passive objects whereas the purely star-forming sources span wider ranges
shifted towards larger sizes and lower $n$ values.

Quantitatively, our SINS objects lie at the large size $-$ low $n$ end
with sample median $R_{\rm e,med} = 5.1~{\rm kpc}$ and $n_{\rm med} = 0.35$
(or 5.7~kpc and 0.33, respectively, when excluding the AGN source BX\,663).
In the $K$-selected comparison sample, the six star-forming sources have 
median values of $R_{\rm e,med} = 4.3~{\rm kpc}$ and $n_{\rm med} = 0.95$,
the four AGN sources have $R_{\rm e,med} = 1.4~{\rm kpc}$ and
$n_{\rm med} = 2.46$, and the nine passive objects have
$R_{\rm e,med} = 1.5~{\rm kpc}$ and $n_{\rm med} = 2.30$.
Our AGN source BX\,663 is not as compact as the \citeauthor{Kri09} AGNs
but, among our SINS sample, it reflects the trend towards lower $R_{\rm e}$
and higher $n$ of AGN relative to star-forming galaxies seen in the
$K$-selected sample.  One source (HDFS1-1849) from \citeauthor{Dok08}
appears to be an ``outlier'' with respect to the other eight passive
$K$-selected objects: its $R_{\rm e} = 4.4~{\rm kpc}$ and $n = 0.5$
is clearly more characteristic of the SINS and \citet{Kri09}
star-forming galaxies.  As we will see below, it is also more similar
to the star-forming sources in terms of the non-parametric Gini, $\Psi$,
and $M_{20}$ coefficients.  This object is anomalous within the sample of
quiescent galaxies insofar as its broad-band SED fit indicates in fact
active star formation.  The lack of emission lines that formed the basis
of the ``quiescent'' classification may only be due to this object's dusty,
edge-on nature \citep{Kri09}.  Excluding this object, the median values for
the eight remaining passive $K$-selected sources are
$R_{\rm e,med} = 1.4~{\rm kpc}$ and $n = 2.55$.

With the \citet{Das08} sample, we can explore variations in different
regimes of bolometric luminosity.  While the four SINS objects with
{\em Spitzer}/MIPS data are among the most luminous of the rest-UV
selected high $z$ population, their observed 24\,\micron\ fluxes and
total IR luminosities are still on average almost an order of magnitude
lower than those of the 24\,\micron-selected sources of \citet{Das08}.
Overall, the 24\,\micron -selected dusty IR-luminous sample is distinct from
our SINS objects, and appears to be intermediate between the rest-UV SINS
and $K$-selected star-forming galaxies and the $K$-selected passive and AGN
objects.  The 24\,\micron -selected sources have a median size and S\'ersic
index of $R_{\rm e,med} = 2.6~{\rm kpc}$ and $n_{\rm med} = 1.45$.
These values are comparable to the median rest-frame optical
structural parameters derived from NIC2 imaging for a sample of
high-$z$ submillimeter-selected galaxies by \citet{Swi10}, which
overlap in bolometric luminosity with the \citeauthor{Das08} objects.
Sampled at the same $\rm 0\farcs 076~pixel^{-1}$ scale, and 1- and
2-orbit integration times, the SINS sample has about twice larger
$R_{\rm e,med} = 5.4~{\rm kpc}$ and significantly lower $n_{\rm med} = 0.37$.
The trend for the 24\,\micron -selected sources to lie towards smaller radii
and steeper inner profiles with respect to the SINS galaxies may reflect the
prevalence of AGN activity among the \citeauthor{Das08} sample (10 out of 11
objects).

\subsection{Comparison of Morphological Coefficients}
           \label{Sub-comp_morph}

Clearly, the various samples probe different regimes in structural
parameters, and this appears to be related to both the level of star
formation and AGN activity.  To further quantify the differences between
our SINS galaxies and the comparison samples, we analyzed the images through
the non-parametric approach described in \S~\ref{Sub-morph_meth}, using the
relevant data sets for the SINS targets matching the depth and pixel sampling
of the $K$- and 24\,\micron -selected samples.  In addition, to take into
account the redshift range our targets and remove the effects of surface
brightness dimming, particularly important in this approach, we selected
pixels for our analysis as if all objects were observed at the maximum
redshift among the three samples, i.e. $z = 2.6$.

We estimated Gini, $\Psi$, and $M_{20}$ for galaxies in the
three samples, and the results are shown in Figure~\ref{fig-morph2}.
All panels contain our SINS galaxies, and the top two panels show $G$
versus $\Psi$ and $G$ versus $M_{20}$, using the $K$-selected sample for
comparison.  The bottom two panels show the same parameters, but feature
the 24\,\micron-selected sample for comparison.  For the top two panels,
the SINS targets have been reanalyzed using 3- and 2- orbit subsets of
the data at $\rm 0\farcs 038~pixel^{-1}$, in order to match the depth and
sampling of the \citet{Dok08} and \citet{Kri09} images.  Given the compact
and axisymmetric nature of the quiescent $K$-selected subsample featured
in \citeauthor{Dok08}, we applied the same non-parametric analysis to our
empirical PSF, constructed from multiple stars as described in
\S~\ref{Sub-psf} but from our $0\farcs 038$ pixel set of images.
This analysis indicates that while the quiescent, $K$-selected objects
are extremely compact, the majority of them can be distinguished from
unresolved point sources, with significantly lower Gini values.
For the bottom two panels, our SINS targets have been reanalyzed using
2- and 1-orbit subsets of the data at $\rm 0\farcs 076~pixel^{-1}$, in
order to match the depth and sampling of the \citet{Das08} images.
The non-parametric statistics were calculated as well for the empirical
PSF sampled with this pixel scale.  In all panels, the shallower 3-, 2-,
and 1-orbit SINS data subsets result in lower Gini coefficient for a
given object, while higher values are obtained for $\Psi$ and $M_{20}$.

The non-parametric morphological analysis reinforces the qualitative
visual impression from the images and the results from the S\'ersic profile
fitting of the different samples of galaxies.  The $K$-selected passive and
AGN sources occupy a disjoint region of $G - \Psi - M_{20}$ morphological
parameter space from that of the SINS targets.  They are virtually {\em all}
offset towards higher Gini, and both lower $\Psi$ and $M_{20}$ values, with
sample median values of $G_{\rm med} = 0.54$, $\Psi_{\rm med} = 0.8$, and
$M_{\rm 20,med} = -1.8$ for the passive galaxies and essentially identical
$G_{\rm med} = 0.53$, $\Psi_{\rm med} = 0.8$, and $M_{\rm 20,med} = -1.8$
for the AGN sources.
As was the case for the structural parameters, only a single galaxy from
the $K$-selected passive sample (HDFS1-1849) even approaches the lower Gini
(median of 0.28), and higher $\Psi$ (median of 7.8) and $M_{20}$ (median of
$-0.5$) values of our SINS sample\,\footnote{
 These median values represent the ensemble of 3- and
 2-orbit subsets of the SINS data shown in Figure~\ref{fig-morph2}.
}.
In contrast to the passive and AGN objects, the six $K$-selected
star-forming objects show considerably more overlap with the SINS
sample.  The median values for the $K$-selected star-forming sources are
$G_{\rm med} = 0.34$, $\Psi_{\rm med} = 3.6$, and $M_{\rm 20,med} = -1.4$,
which are intermediate between the parameters describing the SINS sample
and the $K$-selected passive and AGN objects.

One slight caveat to these results is the difference in typical S/N among
the samples, with the $K$-selected quiescent and AGN objects $\sim 2 - 3$
times higher in average surface brightness than the SINS targets, and the
$K$-selected star-forming sources $\sim 1.5$ times brighter.  As discussed
by \citet{Pet07} and \citet{Abr07},
using a fixed surface brightness threshold (effectively
what we have done), will lead to a ``truncation bias'' in the Gini
coefficients estimated for fainter systems, in that their pixel flux
distribution functions will be sampled over a smaller dynamic range than
those of more luminous objects.  To test for the significance of this
effect, we applied a higher surface brightness threshold to the $K$-selected
samples, such that the resulting pixel flux distribution functions span the
same dynamic range as those of our SINS targets.  We find $G_{\rm med} = 0.45$
for the truncated pixel flux distribution functions for $K$-selected passive
and AGN sources, still significantly higher than the median of the SINS sample,
and still with no overlap in morphological parameter space.  Applying the same
type of truncation to the $K$-selected star-forming galaxies, we find very
little effect on the sample median properties, with $G_{\rm med} = 0.33$.

Turning to the properties as a function of bolometric luminosity,
in the $G - \Psi - M_{20}$ parameter space, the 24\,\micron-selected dusty
IR-luminous sample separates on average from the SINS sample in the same
sense as the $K$-selected sample, if not as significantly as the passive and
AGN objects.  The median parameter values for the 24\,\micron-selected dusty
IR-luminous sources are $G_{\rm med} = 0.44$, $\Psi_{\rm med} = 2.1$, and
$M_{\rm 20,med} = -1.2$.  Sampled at the same pixel scale and exposure time,
the SINS targets have lower Gini (median of 0.30), and higher $\Psi$ (median
of 6.4) and $M_{20}$ (median of $-0.6$).  The typical S/N of objects in the
two samples is very similar, so no issue of truncation bias applies in this
comparison.  These quantitative morphological results confirm again the
qualitative visual impression that our SINS targets have systematically
more irregular, extended, and clumpy morphologies than those of the
24\,\micron-selected dusty IR-luminous sample of \citet{Das08}.

Luminous dusty systems with $L_{\rm IR} > 10^{12}~{\rm L_{\odot}}$
are commonly associated with major mergers in the local universe
\citep[e.g.,][]{Gen01}.  The 24\,\micron-selected sample of \citet{Das08}
falls well above this luminosity threshold, yet the processes giving rise
to extreme bolometric luminosities at high redshift may be more diverse,
based on theoretical grounds \citep[e.g.,][]{Dave10}, and supported by
observational evidence \citep[e.g.,][]{Das08, Swi10}.  \citeauthor{Das08}
find roughly 50\% of their full NIC2 sample to appear morphologically
``quiescent.''  Furthermore, consistent with the results of \citet{Kri09},
these authors show that objects in their sample with mid-IR AGN signatures
also tend to have more compact rest-frame optical morphologies than those
with starburst-dominated spectra.  Given that all but one object in the
\citeauthor{Das08} $z = 2.0 - 2.5$ sub-sample show evidence of AGN activity,
the 24\,\micron-selected dusty IR-luminous galaxies included in
Figure~\ref{fig-morph2} are among the more compact objects in the
full \citeauthor{Das08} sample.  \citeauthor{Das08} hypothesize that
24\,\micron-selected IR-luminous objects, with warmer mid-IR/far-IR colors
than the submillimeter-selected IR-luminous high $z$ population \citep{Cha03},
may represent the merged, dynamically-relaxed counterparts of the (major-)
merging submillimeter galaxies.  On the other hand, the rest-frame optical
structural parameters and sizes of submillimeter galaxies as presented by
\citet{Swi10} are very similar to those of the 24\,\micron-selected
IR-luminous objects.  A more careful morphological comparison of the
two bolometrically-luminous samples is clearly required to construct
an evolutionary sequence.

\subsection{The Connection Between Structure and Stellar Populations
            in $z \sim 2$ Galaxies}
            \label{Sub-comp_morph_sedmod}

In analogy with trends seen in the local universe \citep{Kau03, She03},
the structural properties of $z \sim 2$ galaxies appear to be correlated
with their stellar populations.  Based on imaging with both NICMOS/NIC3
from space \citep{Zir07, Tof07}, and with VLT/ISAAC \citep{Fra08, Tof09}
and UKIRT/WFCAM \citep{Wil10} from the ground, it has been shown that
quiescent, high redshift galaxies have systematically smaller sizes and
higher stellar mass surface densities than actively star-forming systems.
Based on the sample of 19 $K$-selected galaxies discussed in the previous
subsections, \citet{Kri09} used their higher resolution NICMOS/NIC2 imaging
to demonstrate that specific SFR and rest-frame $U - B$ color are strongly
correlated with galaxy structural parameters.  We here examine the structural
properties in connection with the stellar populations for our SINS sample,
complemented with the \citet{Dok08} and \citet{Kri09} $K$-selected samples
for which stellar properties from SED modeling are available.

As proxies for galaxy stellar populations, we consider the stellar
mass and specific SFR.  Within our SINS NIC2 sample, the absence of
correlation between morphology and stellar population properties includes
parameters derived from S\'ersic models ($R_{\rm e}$ and $n$) as well as
the non-parametric morphological coefficients ($G$, $\Psi$, and $M_{20}$).
This result echoes previous ones from \citet{Law07a} and \citet{Pet07}
demonstrating a lack of correlation between stellar populations and
rest-UV morphologies for UV-selected galaxies at $z \sim 2$.  However,
as described by \citet{Pet07} the UV-selected galaxies do not span the
full range of stellar population properties that exist at $z \sim 2$
\citep[see also, e.g.,][]{Dok06, Dam11}.  The $K$-selected sample with
NIC2 imaging from \citet{Dok08} and \citet{Kri09} ideally augments our
SINS NIC2 sample in this respect, considerably extending the range probed
in specific SFR by over two orders of magnitude, in the same redshift range
and at comparable stellar masses as our SINS targets.

Figures~\ref{fig-galfitsedmod} and \ref{fig-morphsedmod} combine the
morphological properties of our SINS and $K$-selected samples, now plotted
as a function of the stellar mass and specific SFR derived from SED modeling.
The SINS and $K$-selected samples overlap in stellar mass over the range from
$\approx 3 \times 10^{10}$ to $\rm \approx 10^{11}~M_{\odot}$.
Over this range, there is significant diversity in structural parameters,
with the SINS targets exhibiting the largest $R_{\rm e}$, smallest $n$,
largest $\Psi$ and $M_{20}$, and smallest $G$ values.  At the other extreme,
we find the AGN and quiescent $K$-selected galaxies, with the star-forming
$K$-selected galaxies showing intermediate properties.  More specifically,
over this interval in stellar mass, the galaxies cover a wide range in
$R_{\rm e}$, consistent with the spread of about an order of magnitude in
rest-frame optical sizes for $z \sim 2$ galaxies of similar masses found in
previous studies \citep[e.g.,][]{Tru06, Zir07, Tof07, Tof09, Fra08}.

On the other hand, the $K$-selected and SINS samples separate more
cleanly in specific SFR.  A clear sequence of increasing specific SFR is
represented, respectively, by the quiescent and AGN $K$-selected galaxies,
the star-forming $K$-selected galaxies, and the SINS sample.  In turn,
significant connections with specific SFR are found for every morphological
coefficient considered here, with systematically higher $R_{\rm e}$, $\Psi$,
and $M_{20}$, and systematically lower $n$ and $G$, as the specific SFR
increases.  These strong trends with star formation activity are consistent
with the previous work from \citet{Tof07, Tof09, Zir07, Fra08}, but are
based on higher resolution NIC2 imaging.
Furthermore, they extend towards higher specific SFR values the NIC2-based
correlations between morphology and stellar populations from \citet{Dok08}
and \citet{Kri09}.  The results presented here reinforce the conclusion
by \citet{Fra08} that galaxy structural parameters are better correlated
with specific SFR and color than with stellar mass.
These results highlight the fact that stellar mass alone is not a good
predictor of structural parameters, nor of detailed morphologies as we
find from our non-parametric morphological analysis.

In that respect, it is also interesting to consider morphologies as a
function of the stellar mass surface density of the galaxies.  Indeed,
in the local universe, the specific SFR (as well as rest-frame optical
colors) is observed to be much more closely related to stellar mass surface
density than to stellar mass \citep[e.g.,][for the SDSS sample]{Kau03, Kau06},
and these relationships are found to persist out to at least $z \sim 2$
\citep{Fra08, Mai09}.
This trend is again reflected among the SINS and $K$-selected samples
discussed here.  Figure~\ref{fig-Mdens_sSFR} plots the specific SFR as
a function of the projected stellar mass surface density $\Sigma(M_{\star})$
of the objects.
We calculated the surface density assuming that the observed 1.6\,\micron\
light (rest-frame $\approx 5000$\,\AA) from the NIC2 images traces well
the stellar mass (as we will demonstrate for one of our SINS targets in
\S~\ref{Sub-disc_md41}), and thus that half of the total stellar mass is
enclosed at $r < R_{\rm e}$.  We find a clear trend of decreasing specific
SFR with increasing mass surface density, which is fully consistent with
the median relationship at $z \sim 2$ derived by \citet{Fra08}.  Our six
actively star-forming SINS targets lie at the lower end of projected mass
surface density with median of
$\log(\Sigma(M_{\star})\,[{\rm M_{\odot}\,kpc^{-2}}]) = 9.0$
(or 8.8 when excluding the AGN source BX\,663) and the nine quiescent
galaxies of \citet{Dok08} lie at the higher end with median of
$\log(\Sigma(M_{\star})\,[{\rm M_{\odot}\,kpc^{-2}}]) = 10.5$
(or 10.6 when omitting the anomalous HDFS1-1849).  The ten emission
line galaxies of \citet{Kri09} span the intermediate range, with sample
median of $\log(\Sigma(M_{\star})\,[{\rm M_{\odot}\,kpc^{-2}}]) = 9.5$.
Among these objects, the subsamples of six star-forming and four AGN
sources have, respectively, lower and higher mass surface densities
(median values of
$\log(\Sigma(M_{\star})\,[{\rm M_{\odot}\,kpc^{-2}}]) = 9.2$ and 10.3).
Since the structural and morphological properties of the SINS and
$K$-selected samples considered here are strongly connected with the
specific SFR, it follows from the specific SFR versus $\Sigma(M_{\star})$
trend that they are also closely related to the stellar mass surface density.
We note here that the high-$z$ submillimeter-selected galaxies of
\citet{Swi10} have properties implying on average both higher specific
SFR {\it and} $\Sigma(M_{\star})$ than the SINS galaxies, and do not
follow the median relationship of \citet{Fra08}, pointing towards
differences in their assembly histories.

\section{DISCUSSION}    \label{Sect-disc}

\subsection{Rest-frame UV Structure, Colors, and Mass Distribution of MD\,41}
           \label{Sub-disc_md41}

In this section, we focus on MD\,41, the target among our NIC2 sample
with additional high resolution optical imaging with {\em HST\/}/ACS.
We first combine the ACS and NIC2 images to compare the structural
parameters derived from the rest-frame UV and optical surface brightness
distributions.  We then use the observed optical-to-near-IR color map to
construct the spatially-resolved stellar mass map of the galaxy.

\subsubsection{Rest-frame UV Structural Parameters}
               \label{Sub-disc_md41_struct}

In \S~\ref{Sub-morph}, we presented the ACS F814W image that exists for one
of our NIC2 targets, MD\,41.  The fairly similar appearance of this galaxy
in the ACS $i_{814}$ and NIC2 $H_{160}$ bands (Figure~\ref{fig-md41_panel})
is supported by our quantitative non-parametric analysis, which yields very
comparable $G$, $\Psi$, and $M_{20}$ values for both images.
Here we extend the comparison to the structural parameters between the
$H_{160}$ and $i_{814}$ band images.  This comparison also complements the
one in \S~\ref{Sub-Hamorph_struct}, where we found that the size, S\'ersic
index, and axis ratio are very similar between the H$\alpha$ and rest-frame
$\approx 5000$\,\AA\ continuum emission for all our targets at a common
(PSF-matched) $\approx 0\farcs 5$ resolution.  At the redshift of MD\,41
($z = 2.17$), the ACS $i_{814}$ band probes rest-frame $\approx 2600$\,\AA\
light, predominantly tracing young massive stars similarly to H$\alpha$.
Therefore, the ACS$+$NIC2 imaging allows us to examine the relative surface
brightness distributions of different stellar populations at significantly
higher resolution than possible with the SINFONI seeing-limited H$\alpha$
data available for MD\,41, although with the caveat that extinction effects
are more important at rest-frame $\approx 2600$\,\AA\ than at the
$\sim 5000 - 6500$\,\AA\ mapped by the $H_{160}$ band and H$\alpha$
observations.

We repeated our S\'ersic model fitting procedure (\S~\ref{Sub-galfit_meth})
on the ACS $i_{814}$ band image of MD\,41.  For consistent comparison with
the results from the NIC2 $H_{160}$ band image, we first convolved the ACS
data to the spatial resolution of the NIC2 data.  We constructed an empirical
PSF for the ACS data using ten bright but unsaturated, isolated stars
identified across the full map.  From a Gaussian fit, the PSF FWHM is
$0\farcs 099$ with ellipticity of 0.02 (or $\rm FWHM = 0\farcs 080$ and
same ellipticity for a Moffat profile fit).  The smoothing kernel was
computed through a Lucy-Richardson deconvolution algorithm, and the
curves-of-growth of the NIC2 and smoothed ACS PSFs agree to better than
10\% at radii $0\farcs 1 < r < 1\farcs 5$.  No spatial resampling of
the data is needed since both ACS and NIC2 data sets were drizzled to
$\rm 0\farcs 05$ pixels.
For accurate registration between the NIC2 and ACS maps, we relied on the
star and a number of very compact sources visible in both $H_{160}$ and
$i_{814}$ band emission within the smaller NIC2 FOV.  The right panel of
Figure~\ref{fig-md41_panel} shows an RGB color composite with the $H_{160}$
band image in red, and the PSF-matched $i_{814}$ band image in green$+$blue.

The best-fit parameters based on the smoothed ACS map are:
$R_{\rm e}(i_{814}) = 7.44^{+0.59}_{-0.71}~{\rm kpc}$,
$n\,(i_{814}) = 0.27^{+0.45}_{-0.09}$,
$b/a\,(i_{814}) = 0.30^{+0.12}_{-0.02}$, and
${\rm P.A.}(i_{814}) = 38.8^{+1.3}_{-1.6}~{\rm deg}$.
Compared to the parameters derived from the NIC2 image
(Table~\ref{tab-galfitres}), the effective radius is thus $\approx 30\%$
larger, the S\'ersic index is $\rm \approx 0.25$ lower, and the axis
ratio and P.A. are virtually identical.  The larger size and lower light
concentration index derived from the $i_{814}$-band image could be due to
higher extinction in the central regions and/or more extended distribution
of the young stellar population dominating the rest-UV emission.  While the
differences in derived size and $n$ index are significant with respect to
the uncertainties, the fits to the $i_{814}$ and $H_{160}$ band images are
in broad agreement.  Both indicate that MD\,41 is a large and fairly edge-on
system (implied disk inclination angle of $\approx 72^{\circ}$), with a
shallow inner light profile.  The structural analysis supports the
conclusions from the non-parametric morphological analysis that there is
no strong $k$-correction between the rest-frame UV and optical morphologies
for MD\,41.  This result is consistent with the general findings for rest-UV
selected samples at $z \sim 2$, and in contrast to those for samples selected
at longer wavelengths with typically redder rest-frame optical colors
\citep[]
 [see also \citealt{Ove10} for rest-UV selected samples at $z \sim 3 - 4$
 and rest-optical selected samples at $z \sim 2$]{Dic03, Pet07, Law07a, Tof07}.

\subsubsection{Rest-frame Colors, Mass-to-Light Ratio, and Mass Distribution}
               \label{Sub-disc_md41_colmass}

We can further exploit the ACS and NIC2 images to investigate spatial
variations in observed colors and mass-to-light ratio, and derive the
stellar mass distribution within MD\,41.  In Figure~\ref{fig-md41_colmass},
the left panel shows the observed $i_{814} - H_{160}$ color map, over the
region for which the pixel flux densities have $\rm S/N > 3$ in each of the
ACS and NIC2 map.  The observed colors are fairly uniform across most of the
disk of MD\,41, with mean and median of $(i_{814} - H_{160})_{\rm AB} = 0.8$
and 0.9~mag, respectively, and standard deviation of 0.37~mag.  This compares
reasonably well with the integrated color of 0.63 measured in a circular
aperture of 3\arcsec\ diameter\,\footnote{
 The two blue sources offset from the main part of MD\,41 to the northwest
 and southeast that are detected in the ACS image but not in the NIC2 image
 are included in this large aperture, and so contribute to this estimated
 integrated color.
}.
Only two small regions exhibit significantly different colors:
the southwestern edge of MD\,41 at the location of the two
brightest clumps in $i_{814}$ emission with bluer colors
($(i_{814} - H_{160})_{\rm AB} \approx 0.2$ on average),
and an area around the center of the galaxy characterized by redder
colors ($(i_{814} - H_{160})_{\rm AB} \approx 1.2~{\rm mag}$ on average).

To interpret the color map of MD\,41, we turn to synthetic colors and
$M/L$ ratios calculated for the $z = 2.2$ of MD\,41 from \citet{BC03}
models for a range of ages, extinction, star formation histories (SFHs),
and two stellar metallicities (solar and $1/5$ solar).
The details are given in Appendix~\ref{App-MLcol_models}.
Stellar age, extinction, and SFH are highly degenerate in observed
$i_{814} - H_{160}$ colors (or rest-frame 2600\,\AA\ $-$ 5000\,\AA\
colors at the redshift of MD\,41).  However, there is a fairly well
defined relationship between the stellar mass to rest-frame optical
light ratio (with luminosity uncorrected for extinction) and the
observed colors, with very little dependence on our choice of
metallicity.  For our purposes, we used model predictions through
the SDSS $g$-band filter, a good approximation to the rest-frame
$\approx 5000$\,\AA\ luminosity probed by the NIC2 F160W filter
(see Appendix~\ref{App-MLcol_models}).

We thus derived a mean relationship between the observed (uncorrected
for extinction) $M_{\star}/L^{\rm rest}_{g}$ ratio and observer's frame
$i_{814} - H_{160}$ colors for $z = 2.2$.  We used this relationship
to compute the $M_{\star}/L^{\rm rest}_{g}$ ratio map of MD\,41, shown
in the middle panel of Figure~\ref{fig-md41_colmass}.   Unsurprisingly,
the $M_{\star}/L^{\rm rest}_{g}$ ratio map looks similar to the 
$i_{814} - H_{160}$ map, given the roughly linear relation
(in logarithmic units for the $M/L$).
The mean and median values of individual pixels are
$\log(M_{\star}/L^{\rm rest}_{g}~{\rm [M_{\odot}\,L^{-1}_{g,\odot}]})
  \approx -0.9$, with standard deviation of $\rm \approx 0.2~dex$,
The observed integrated color gives nearly the same value, $-0.83$,
while the model with best-fit parameters from the optical/near-IR SED
modeling for MD\,41 (Table~\ref{tab-galprop}) has a somewhat lower
value of $-1.08$.
An average value of $\approx -1.3$ is measured in the regions of bluest
colors at the southwestern edge of MD\,41, and of $\approx -0.7$ in the
redder regions around the center.  The fairly small scatter among individual
pixels implies roughly constant observed $M_{\star}/L_{g}^{\rm rest}$ ratio
across most of the galaxy.

Multiplying the observed $M_{\star}/L^{\rm rest}_{g}$ map by the $H_{160}$
band image as a proxy for the extrinsic rest-frame $g$-band luminosity
then leads to the spatial distribution of stellar mass across MD\,41.
The map of projected stellar mass surface density $\Sigma(M_{\star})$
is shown in the right panel of Figure~\ref{fig-md41_colmass}.
It is remarkably similar to the $H_{160}$ band image itself, indicating
that the observed 1.6\,\micron\ emission does trace closely the stellar
mass distribution in MD\,41.  The typical (mean, median) values of
individual pixels is
$\log(\Sigma(M_{\star}\,[{\rm M_{\odot}\,kpc^{-2}}])) \approx 8.0$,
with standard deviation of 0.3~dex.  At the largest radii probed,
and especially on the southwestern edge, the surface density drops to 
$\log(\Sigma(M_{\star})) \approx 7.5$,
and it increases towards the center by about an order of magnitude to 
$\log(\Sigma(M_{\star})) \approx 8.4$.
The deprojected surface density from the total stellar mass of
MD\,41, assuming half of it is enclosed within a radius of $R_{\rm e}$,
is $\log(\Sigma(M_{\star}, r < R_{\rm e})) = 7.6$; for this fairly inclined
system, this increases significantly when considering the projected surface
density to $\log(\Sigma(M_{\star}, r < R_{\rm e})) = 8.1$, virtually the
same as inferred on average at individual locations across the galaxy.
Compared to the properties derived for massive $1.5 < z < 2.5$ galaxies
by \citet[][see also Figure~\ref{fig-Mdens_sSFR}]{Fra08}, MD\,41 lies
at the low surface mass density end, consistent with its high specific
SFR of $\rm 24~Gyr^{-1}$ and blue estimated integrated rest-frame
$(U - B)_{\rm AB} \approx 0.5$ color.

The total stellar mass obtained by summing up the inferred stellar mass of
individual pixels with $\rm S/N > 3$ in the $i_{814}$ and $H_{160}$ band
images is $\rm 7.3 \times 10^{9}~M_{\odot}$, or $95\%$ of the value derived
from the modeling of the integrated optical/near-IR SED (\S~\ref{Sub-sedmod}).
The sum of the $H_{160}$ band flux densities over the same pixels corresponds
to $65\%$ of the total as measured in a 3\,\arcsec -diameter aperture
(\S~\ref{Sub-phot}).  Multiplying the $\rm S/N > 3$ pixels' luminosity
densities by the $M_{\star}/L_{g}^{\rm rest}$ from the best-fit SED model
of MD41 results in a stellar mass of only $\rm 5.0 \times 10^{9}~M_{\odot}$.
This difference with respect to the spatially-resolved estimate reflects
the difference between the pixel-averaged $M_{\star}/L_{g}^{\rm rest}$ and
that from the integrated SED noted above.  It suggests the total SED-derived
stellar mass may be underestimated by $\approx 30\%$ because dustier and/or
more evolved, redder stellar populations are under-represented in the
effectively luminosity-weighted integrated rest-UV and optical photometry
of MD\,41.  A similar effect between stellar masses computed by integrating
spatially resolved mass maps and by modeling unresolved photometry was found
for nearby galaxies \citep*{Zib09}.
This effect may be more important for MD\,41 (possibly also for BX\,482)
because the SED fits rely more heavily on the rest-UV photometry than
for the other galaxies with better rest-optical coverage (three near-IR
bands compared to two for MD\,41 and one for BX\,482). 

Our analysis of ACS $i_{814}$ and NIC2 $H_{160}$ band imaging of MD\,41
shows modest variations in $M_{\star}/L_{g}^{\rm rest}$ ratio, within a
factor of $\sim 1.5$ over most of the galaxy and on spatially-resolved
scales as small as $\rm \approx 1.2~kpc$.  In addition, the results
indicate that the 1.6\,\micron\ (rest-frame $\approx 5000$\,\AA) is
a fairly good tracer of the stellar mass distribution of MD\,41.
The spatially-resolved color-based mass map suggests nevertheless that
luminosity-weighting biases may have led to a $\approx 30\%$ underestimate
of the total stellar mass based on modeling of the integrated SED of MD\,41.
It will undoubtedly be interesting to investigate this effect more
systematically with larger high $z$ galaxy samples from combined
high-resolution optical and near-IR imaging in future studies.

\subsection{The Sizes of our SINS NIC2 Sample Galaxies}
           \label{Sub-sizes}

Size is a fundamental property of galaxies.  The redshift evolution of
disk sizes and scaling relations involving size (e.g., size-velocity,
size-mass) set sensitive constraints on the coupling between the
baryonic and dark matter, on the mechanisms through which disks
acquire their angular momentum, and on the physical processes
that drive galaxy growth
\citep*[e.g.,][]{Whi78, Fal80, Blu84, Dal97, Mo98, Nav00}.
Size determinations, however, can be affected by various factors,
and reconciling observed sizes with theoretical predictions remains
a challenge \citep*[e.g.,][]{Dut07, Dut09b, Dut11, Som08, Sal09, Bur10}.
In the following, we consider whether the sizes derived for our NIC2
sample are affected by their particular morphologies and interpret them
in the broader context of the size distribution of the general $z \sim 2$
galaxy population.  Our sample has high quality data that allow, for the
first time, a reliable and consistent comparison of rest-frame optical
continuum and H$\alpha$ sizes, and we discuss implications for disk
formation at high redshift.

\subsubsection{Rest-frame Optical Continuum and H$\alpha$ Sizes}
               \label{Sub-sizes_robust}

It is clear from our NIC2 images that galaxies at $z \sim 2$ can have
complex, clumpy, and irregular morphologies even in the rest-frame
optical \citep[see also, e.g.,][]{Pap05, Tof07, Zir07, Das08, Kri09}.
Depending on the importance of the substructure, the overall radial
light profiles may or may not be well reproduced by a simple smooth model
(see Figures~\ref{fig-galfitres}, \ref{fig-galfitres2}, \ref{fig-profiles}),
and this substructure may also contribute to the observed scatter in light
concentration index.  This can complicate the measurement and interpretation
of sizes in a broader context.
For our NIC2 targets, we estimate in Paper~II that the clumps identified
in rest-frame optical continuum light contribute individually typically
only $\sim 2\%$ of the total emission of the host galaxy; their total
contributions range between $\sim 10\%$ and $\sim 25\%$ ($15\%$ on average),
depending on the object.  Therefore, for these six galaxies, the impact of
the substructure associated with the clumps, spread across the galaxies,
probably does not affect the size measurements in an important way.

As we found in \S~\ref{Sub-galfit_res}, the kinematically identified
disks have global radial light profiles that deviate appreciably from pure
exponential disks, with best-fit S\'ersic indices $n \approx 0.15 - 0.6$,
or $n \approx 2$ for the AGN source BX\,663.  The impact of fitting
exponential disks to our NIC2 targets is, however, relatively modest.
Excluding the major merger BX\,528, single-component S\'ersic model
fits to the $0\farcs 145$ resolution $H_{160}$ band images with fixed
$n = 1$ yield the following changes:
the effective radii mostly increase, with mean and median
$R_{\rm e}(n = 1) / R_{\rm e}(n~{\rm free}) \approx 1.15$,
and range of $0.8 - 1.45$.  All $n < 1$ sources have
$R_{\rm e}(n = 1) > R_{\rm e}(n~{\rm free})$ (with mean/median
factor $\approx 1.2$ and range of $1.1 - 1.45$), whereas BX\,663
with $n = 2$ has $R_{\rm e}(n = 1) = 0.8 \times R_{\rm e}(n~{\rm free})$.
For fits to the H$\alpha$ maps at $\approx 0\farcs 5$ resolution, where
all five disks have a best-fit $n < 1$ (Table~\ref{tab-galfitres_smoo}),
the $R_{\rm e}(n = 1)$ are larger compared to $R_{\rm e}(n~{\rm free})$
by a factor of 1.1 on average and median, ranging from 1.0 to 1.14.

As discussed in \S~\ref{Sub-comp_morph_sedmod}, the rest-frame optical
sizes derived for our massive and actively star-forming targets, though
large, are fully consistent with the trend of increasing effective radius
at higher specific SFR and lower stellar mass surface density observed
among the general massive galaxy population at high redshift.  While the
objects studied in this paper include some of the larger galaxies from
the SINS survey (Figure~\ref{fig-sedkinprop}), they lie well within the
distribution of sizes derived for $1.5 < z < 2.5$ galaxies at very similar
rest-frame optical wavelengths (close to $g$-band) and in the relevant
stellar mass range of $\log(M_{\star}\,[{\rm M_{\odot}}]) = 9.9 - 11.0$
by \citet{Fra08}, which span $R_{\rm e,circ} \sim 0.5 - 8~{\rm kpc}$.
Our targets have $R_{\rm e,circ} = 2.6 - 4.3~{\rm kpc}$, which places
them within the upper half of the full distribution.
The substantial scatter by roughly an order of magnitude in rest-frame
optical (circularized) effective radius at fixed stellar mass is seen at
all redshifts up to $z \sim 2.5$ and is attributed, at least in part, to
the wide range of star formation rates and histories among galaxies
\citep[e.g.,][and see Figure~\ref{fig-galfitsedmod}]
      {She03, Kau03, Tru06, Fra08, Tof09}.
This scatter as well as the relationships between size, star formation
activity, stellar mass, and stellar mass surface density mean that sample
properties and selection biases need to be considered for consistent
comparisons with other galaxy samples as well as theoretical predictions.

The rest-frame $\approx 5000$\,\AA\ continuum probed by our NIC2 data is
more sensitive to lower-mass, longer-lived stars making up the bulk of
the stellar mass while the H$\alpha$ emission is dominated by the young,
short-lived stars.  One might expect to be able to discern with these
data possible differences between the distributions of stellar mass and of
star formation, if any.  For all our SINS NIC2 galaxies, however, we find
very close agreement between the parameters derived from the $H_{160}$ band
and H$\alpha$ maps, with on average the same effective radius and S\'ersic
index to within $3\%$ and 0.08, respectively (\S~\ref{Sub-Hamorph_struct}
and Figure~\ref{fig-galfit_comp}).
This similarity in size and global profile shape is probably not due
to beam-smearing effects, which we found to be not very significant for
our targets based on the $H_{160}$ band images at the original NIC2
resolution and convolved to the typical $\approx 0\farcs 5$ resolution
of the SINFONI data as well as on the seeing-limited and AO-assisted
H$\alpha$ maps of BX\,482.

These results, albeit for a small sample, are interesting to consider in
the context of models of disk formation, where a signature of ``inside-out''
scenarios for the growth of stellar disks is that the half-mass radius
of the recently formed stars, ``$R_{\rm e}({\rm SF})$,'' is larger than
the half-mass radius from the total stellar mass, ``$R_{\rm e}(M_{\star})$.''
In particular, in the semi-analytical models of
\citet[][see also \citealt{Dut09a}]{Dut11}, the $R_{\rm e}({\rm SF})$ is
on average a factor of $\approx 2$ larger than $R_{\rm e}(M_{\star})$
for $z \sim 2$ disks.  Using cosmological hydrodynamical simulations,
\citet{Sal09} find a similar difference by a factor of $\approx 1.8$ at
$z \sim 2$.  To account for radial variations in mass-to-light ratio across
their model disks, \citet{Dut11} also compute half-light radii in several
rest-frame optical and near-IR bands, and find that the differences between
half-light radii and half-mass radius of recent star formation decreases
towards bluer bands.  For the $V$ band, the bandpass that most closely
matches the rest-frame emission probed by the $H_{160}$ observations
of our targets, the models of \citeauthor{Dut11} imply on average
$R_{\rm e}({\rm SF}) / R_{\rm e}(V) \approx 1.26$.
In these models, ``recent star formation'' refers to stars formed in
the last $\rm \approx 30 - 100~Myr$, and we assume here that it is
traced reliably by H$\alpha$.  Excluding the major merger BX\,528,
the rest-frame optical radii derived from the NIC2 images matched to the
PSF of the SINFONI H$\alpha$ data in Table~\ref{tab-galfitres_smoo} give
on average $R_{\rm e}({\rm H\alpha}) / R_{\rm e}(H_{160}) = 1.06$
(and 1.02 for the median), ranging from $\approx 0.95$ (MD\,41, BX\,482)
to $\approx 1.15$ (BX\,663, BX\,610).  These results for fits where $n$
was left as a free parameter are appropriate for comparison with the
\citet{Dut11} models, in which disks are not formally exponential.
For completeness, the same comparison based on fits with $n$ fixed
to 1 gives an average $R_{\rm e}({\rm H\alpha}) / R_{\rm e}(H_{160})$
ratio of 1.10 (median of 1.08), in the range $\approx 1.0 - 1.2$.
Formally, the uncertainties of individual objects would allow ratios
of $\approx 1.25$ but the trend of ratios $\approx 1.05 - 1.1$ (with
standard deviation of 0.1) for the ensemble of our five $z \sim 2$
disks indicates little, if any, systematic difference between the
H$\alpha$ and rest-frame optical sizes.

We conclude from the considerations above that the size measurements
of our SINS NIC2 targets appear robust against possible biases due to
substructure, or to specific assumptions on the light profile as tested
with fixed $n = 1$ fits to our disks.  We estimated
the impact of substructure or of adopting $n = 1$ models to all be within
typically $\sim 15\%$.  Based on the NIC2 and SINFONI data of the five disks
among our sample, we do not detect a clear effect whereby the distribution
of the actively star-forming regions would be characterized by a larger
effective radius relative to that of stars tracing more closely the bulk of
the stellar mass.  It would be premature to generalize this finding from a
small number of galaxies and draw conclusions on the ``inside-out'' scenario
of disk formation, but our high quality NIC2 and SINFONI data provides a first
set of direct empirical constraints at $z \sim 2$ for theoretical models.
The sizes of all our six galaxies, with $R_{\rm e} \approx 4 - 6~{\rm kpc}$
(or $R_{\rm e,circ} \approx 2.5 - 4.5~{\rm kpc}$) are not surprisingly large
when placed in the broader context of the massive $z \sim 2$ galaxy population,
taking into account their properties in conjunction with the existing trends
of larger sizes at high specific SFR and low stellar mass surface density.

\subsubsection{Possible Caveats}
               \label{Sub-sizes_caveats}

There are several possible caveats to the analysis presented in the
previous subsection.  In particular, our simple S\'ersic model fits
do not account for multi-component structure.  As a specific example,
if there is a significant bulge, the $R_{\rm e}$ from single-component
model fits will tend to underestimate the $R_{\rm e}$ for the disk itself.
The $H_{160}$ band radial light profiles do not indicate the presence of a
significant bulge component in our disks (Figure~\ref{fig-profiles}), and
this is further supported by the low S\'ersic indices and relatively small
Gini coefficients (Figures~\ref{fig-galfit_comp} and \ref{fig-morph2}).
The exception may be BX\,663 but it is unclear whether, and by how much,
the AGN in this object may contribute to the central emission peak.
The two-component fits for BX\,663 explored in \S~\ref{Sub-galfit_res}
imply a very minor fraction of $6\%$ of the total $H_{160}$ band light
for a putative nuclear point source but a more important $35\%$ for a
putative bulge.  In both cases, however, the $R_{\rm e}$ of the disk
component is not significantly different from that of the best-fit
single-component model (with $n = 2$).  Also, as we determine in Paper~II,
the clump closest to the center in each of the other four disks, which
could be interpreted as a young bulge, makes a marginal contribution of
$< 5\%$ to the total rest-frame $\approx 5000$\,\AA\ light, and even only
$\approx 13\%$ for the more prominent central peak of BX\,663\,\footnote{
 From detailed modeling of the H$\alpha$ kinematics
 of SINS disks with high quality SINFONI data, including four of our
 targets, \citet{Gen08} found a trend of increasing central dynamical
 mass concentration with stellar evolutionary stage, which could be
 interpreted as evidence for growing bulges/inner disks.  The inferred
 ratio of enclosed dynamical mass within 3~kpc and 10~kpc ranges
 from $< 0.15$ for MD\,41, to 0.2 for BX\,482, and 0.4 for BX\,389
 and BX\,610.  Disentangling the relative contributions of a bulge
 and the underlying disk from these data is uncertain but these
 estimates would provide conservative upper limits to bulge-to-disk
 mass ratios.}.
It thus appears unlikely that the sizes inferred for the disks in
our SINS NIC2 sample would be significantly affected by the presence
of a bulge, neglected in our S\'ersic model fits.

Another caveat is that the effects of dust obscuration are ignored. 
Extinction may significantly modify the surface brightness distributions,
and hence the best-fit sizes.  In particular, higher extinction towards
the central parts of the sources will lead to shallower observed inner
profiles.  For BX\,610, our seeing-limited SINFONI $H$ and $K$ band
data are of sufficient S/N to constrain the radial variations of the
$\rm H\beta/H\alpha$ flux ratio \citep{Bus11}.  The Balmer decrement
suggests indeed an increase in visual extinction by $\rm \approx 1~mag$
in the inner $\rm \approx 5~kpc$, so that the intrinsic inner light
profile could be steeper and the effective radius could be smaller.
Moreover, differential extinction towards the \ion{H}{2} regions and the
stars could affect the relative rest-frame optical and H$\alpha$ effective
radii.  The existing data for our NIC2 targets do not allow us, however,
to constrain on $\sim$ kpc scales the effects of spatial variations in
dust extinction and of differential obscuration between the \ion{H}{2}
regions and the bulk of the stellar populations.

Finally, the depth of the $H_{160}$ band and H$\alpha$ maps may also
influence the best-fit S\'ersic model parameters, including the $n$ index
and $R_{\rm e}$.  Our SINFONI and NIC2 data sets are fairly deep in terms
of what can be achieved with these instruments (see \S\S~\ref{Sub-sinfdata}
and \ref{Sub-noise}), and the radial light profiles are traced to
$\ga 2 \times R_{\rm e}$ (Figures~\ref{fig-profiles} and \ref{fig-profiles2}).
Nonetheless, the $3\,\sigma$ surface brightness limits correspond to fairly
high H$\alpha$ and rest-frame $g$ band luminosities of
$L_{\rm H\alpha} \approx 4.3 \times 10^{39}~{\rm erg\,s^{-1}\,kpc^{-2}}$
and $L_{g}^{\rm rest} \approx 6.7 \times 10^{8}~{\rm L_{g,\odot}\,kpc^{-2}}$
at the median $z = 2.2$ of our NIC2 targets, and it is conceivable that
fainter extended emission in the outer parts is undetected in our data.

\subsection{Relation Between Kinematics and Morphologies}
           \label{Sub-disc_kinmorph}

Arguably, one of the key aspects in the study of high redshift galaxies
is a characterization of the mass assembly process, and, in particular,
pinpointing evidence for the occurrence of major mergers and interactions
or, in their absence, for the establishment of disk-like systems, which
are a prediction of models of galaxy formation.  There are several
inherent difficulties in studies based on morphologies, including
bandshifting, intrinsically complex structure, and uncertainties
in the timescales over which key merger signatures are observable
\citep[e.g.,][]{Tof07, Elm07, Elm09, Lot08, Lot10a, Lot10b}.
Therefore, kinematics appear to be a prerequisite for a robust
distinction although there are possible complications as well
\citep*[for instance, particular merger configurations mimicking
   disk-like kinematics; e.g.,][]{Law06, Sha08, Rob08, FS09, Epi10}.
Morphologies and kinematics together make for a powerful combination,
mitigating the respective potential limitations as well as providing
a more complete picture of individual galaxies
\citep[see also][]{Nei08}.

The kinematic classification of our six SINS targets, all disks
except BX\,528, was first based on SINFONI data sets taken with
the $\rm 0\farcs 125~pixel^{-1}$ scale, at $\approx 0\farcs 5$ or
$\rm \approx 4.1~kpc$ resolution ($0\farcs 39$ and 3.2~kpc for the NGS-AO
data of BX\,663).  Initially based on qualitative assessment of both the
H$\alpha$ velocity field and velocity dispersion map \citep{FS06b}, the
classification into disk- and merger-like systems was quantitatively
confirmed through the application of kinemetry \citep{Sha08}, with the
exception of BX\,663 due to S/N limitations outside of the inner regions.
To date, BX\,482 is the only target among our NIC2 sample followed-up with
AO and the $\rm 0\farcs 05~pixel^{-1}$ scale, with effective resolution of
$0\farcs 17$.  Its kinematic classification as a disk is further confirmed
with the higher resolution data, although more detail in the H$\alpha$
emission and kinematic maps is resolved.  Fourteen other galaxies from
the full SINS survey were classified by kinemetry, of which four have
follow-up AO observations at $\sim 0\farcs 15 - 0\farcs 20$ resolution
\citep{Sha08, Gen08, FS09, Gen10, Bus11}.
In all cases, as for BX\,482, the kinematic classification is the same
whether it is based on the seeing-limited or the AO data sets.

At first glance, the NIC2 $H_{160}$ images of our six targets
at $0\farcs 145$ resolution ($\rm \approx 1.2~kpc$) reveal rich
sub-structure at all surface brightness levels.
Qualitatively, multiple bright compact components
can be interpreted as distinct interacting/merging units, asymmetries
in the more diffuse emission or distorted isophotes on large scales
can be attributed to tidal features \citep[e.g.][]{Lot04, Con05}.
On the other hand, these characteristics may also result from internal
dynamical processes within gas-rich disks, such as bar- and spiral-like
features from non-axisymmetric perturbations or luminous kpc-sized clumps
resulting from disk instabilities and tidal-like features induced by the
mutual interactions between massive clumps \citep[e.g.,][]{Bour07, Bour08}.
It may therefore not be immediately obvious whether a system is a disk or
a merger from simple visual inspection.  
Quantitatively, as found in \S~\ref{Sub-galfit_res}, all the kinematically
classified disks among our NIC2 sample have S\'ersic indices $n \leq 2$
(and even $< 1$ when excluding the AGN source BX\,663), consistent with
structurally late-type, disky systems.

Among our NIC2 sample, BX\,610 appears to be a particularly striking
example where the rest-frame optical morphology reflects tell-tale
signatures of internal dynamical processes.  Indeed, its appearance,
as outlined in \S~\ref{Sub-look}, exhibits evidence for a bar and
features reminiscent of spiral arms, and there is no evidence for any
nearby companion in our NIC2 and SINFONI data, nor of recent merging.
This galaxy is one of the most robust examples of rotating disk kinematics
among the SINS survey \citep{FS06b, Gen08, Sha08, Cre09}, with the H$\alpha$
velocity field exhibiting the classical ``spider diagram'' of isovelocity
contours, a nearly axisymmetric and centrally peaked velocity dispersion map,
and a close correspondence of the kinematic center and major axis with the
geometrical center and morphological major axis (from both the rest-frame
optical and H$\alpha$ emission).  BX\,610 shows several prominent clumps and
direct observations of CO molecular gas indicate a very high gas to baryonic
mass fraction of $\sim 55\%$ \citep{Tac10}.  All these properties suggest
that BX\,610 may be one such apparently isolated galaxy where clumps formed
from fragmentation in a Toomre-unstable gas-rich disk and where efficient
secular processes drive its early dynamical evolution
\citep[][see also \citealt{Car07, Bour07, Dek09, Age09, Cev10}]{Gen08}.
While BX\,610 is a particularly remarkable case, the other disks among
our NIC2 targets as well as other high $z$ samples studied at high
resolution \citep[e.g.,][and Paper II]{Bour08, Elm05, Elm09, Gen10} show
prominent clumps and properties consistent with the above theoretical
scenario.

While the majority of objects in our SINS NIC2 sample are classified as
disks, it is noteworthy to highlight examples of how the unique combination
of H$\alpha$ kinematics and rest-frame optical morphologies sheds light on
galaxy interactions.
For BX\,528, the H$\alpha$ kinematics clearly indicate its merger nature
even at the $\approx 0\farcs 6$ seeing-limited resolution of the SINFONI
data \citep{FS06b, Sha08}.  The reversal in velocity gradient across the
system could be produced by a counter-rotating binary merger.  In the NIC2
image, the system is resolved into two main ``SE'' and ``NW'' components
separated by about 1\arcsec\ (8~kpc) in projection, and connected by a
fainter ``bridge'' of emission.  Single-component S\'ersic model fits
to the $H_{160}$ map, either at the original NIC2 resolution or convolved
to the lower resolution of the SINFONI observations, imply a shallow inner
light profile with $n \sim 0.2$, evidently caused by the lower surface
brightness between the SE and NW components.  As seen in \S~\ref{Sub-look},
at $0\farcs 6$ resolution, both the H$\alpha$ and smoothed $H_{160}$ band
morphologies appear rather regular and the SE and NW components are not
discerned, making it impossible to distinguish from disk-like systems.
Moreover, BX\,528 does not stand out compared to the disks in our sample
in the $G - \Psi - M_{20}$ parameter space based on the high resolution
NIC2 image (Figure~\ref{fig-morph1}).
Thus, for BX\,528, both the H$\alpha$ kinematics and the high resolution
rest-frame optical imaging are necessary to fully characterize the nature
of this system.

Even if not as significantly as BX\,528, BX\,389 also presents
evidence for galaxy interaction (Figure~\ref{fig-nic_sinf2}).
The H$\alpha$ kinematics at $0\farcs 5$ resolution show fairly regular
motions consistent with rotation in a highly inclined disk.  The various 
line maps from the $K$ and $H$ band seeing-limited SINFONI data all exhibit
an extension to the south of the center, and the rest-frame optical line
ratios ([\ion{N}{2}]/H$\alpha$, [\ion{O}{3}]/H$\beta$) are consistent with
either a low-metallicity companion or shock excitation in a starburst-driven
galactic outflow \citep{Leh09, Bus11}.  This extension does not appear
kinematically distinct from the main part of the source, with line-of-sight
velocities similar to those along the major axis of the main body of the
galaxy, on the same side and at the same projected radii.
The NIC2 $H_{160}$ band imaging reveals that this is a separate component,
at a projected distance of 5~kpc below the galactic plane of BX\,389,
supporting the nearby satellite interpretation.  For BX\,389, therefore,
the high resolution near-IR imaging resolves the companion from the main
part of the source and the kinematics confirm the physical association
between the two components.

Assuming that the observed 1.6\,\micron\ light traces closely the stellar
mass in our NIC2 targets (see \S~\ref{Sub-disc_md41_colmass}), we can set
constraints on the mass ratios of the different components in BX\,528 and
BX\,389.  Various estimates based on the two-component S\'ersic model fits
(\S~\ref{Sub-galfit}), on simple aperture photometry, and on the fractional
light contributions of the associated ``clumps'' (presented in Paper II)
imply a mass ratio in the range 1.2:1 to 1.9:1 for BX528$-$SE relative
to BX528$-$NW, and 6:1 to 20:1 for the main part of BX\,389 relative to
the southern companion.  From the stellar mass ratio, BX\,528 is a nearly
equal-mass major merger, in agreement with the signatures and classification
from the H$\alpha$ kinematics.  In contrast, the southern companion of
BX\,389 represents a minor merger, unlikely to have strong impact on the
large-scale kinematics of the primary component and consistent with the
lack of substantial disturbances in the kinematics expected for major
mergers.

We conclude from the above that the H$\alpha$ kinematics and rest-frame
optical continuum morphologies of our NIC2 sample draw a consistent
picture for the nature of each target and are very complementary.
The kinematics, even at $\rm \sim 3 - 5~kpc$ resolution, are sufficient
(and in some cases necessary, as for BX\,528) to reliably distinguish
between disks and major mergers.  We caution that there will be specific
cases where a system's configuration will lead to ambiguous or erroneous
classification, also depending on the data quality \citep{Sha08, Law09},
but this does not appear to be the case for the objects studied in this
paper.
Morphologies (as well as kinematics) at higher $\rm \sim 1~kpc$ resolution
then reveal further details and are important to identify minor mergers,
and more subtle or smaller-scale signatures of internal dynamical
processes \citep[e.g.][]{Bour08, Gen08, Gen10}.

\section{SUMMARY}   \label{Sect-conclu}

We have presented {\em HST\/}/NIC2 F160W imaging for a unique sample
of six $z \sim 2$ star-forming galaxies with VLT/SINFONI integral
field spectroscopic measurements.  Our analysis has been based on
the combined NIC2 and SINFONI seeing-limited and AO-assisted datasets,
along with multi-wavelength ground-based photometry and, in one case,
{\em HST\/}/ACS imaging. 
Using NIC2 imaging alone we have carried out galaxy structural analyses,
using both parametric and non-parametric approaches, and placed the
structural properties of our SINS NIC2 sample in the larger context of
the correlations between morphology and star-formation activity present
at $z \sim 2$.  However, it is the full combination of datasets presented
here that has enabled investigations of unprecedented detail into the 
dynamical nature of our targets, and the importance (or lack thereof)
of mergers of different mass ratios; the relative galaxy sizes traced
by recent star formation and stellar mass; and spatial variations in
the stellar $M/L$ and stellar mass surface density.

The main scientific conclusions of this paper can be summarized as follows.

1. The rest-frame optical continuum morphology of our targets yields
classifications consistent with those based on ionized gas kinematics.
Those systems with disk-like kinematics (the majority of our sample)
are characterized by disk-like structures in rest-frame optical light.
Furthermore, for the one target (BX\,528) with kinematics indicative
of binary counter-rotating elements, the NIC2 image is clearly resolved
into two, discrete components separated by $\approx 1^{\prime\prime}$,
or 8~kpc.  While BX\,528 may provide evidence for a major merger, the
morphology and kinematics of BX\,389 provide evidence for a less dramatic 
interaction.  The SINFONI kinematics for BX\,389 indicate regular disk
rotation, but both SINFONI H$\alpha$ maps and NIC2 rest-frame optical
imaging reveal a low-luminosity companion at roughly the same line-of-sight
velocity, and a projected distance of 5~kpc --- possibly a minor merger in 
action.

2. We use both parametric (S\'ersic profile) and non-parametric
($G,\Psi,M_{20}$) methods for characterizing the NIC2 morphologies of
the SINS NIC2 sample.  The majority of these objects are characterized
by shallow ($n < 1$) inner light profiles, with median effective radius 
of $R_{\rm e} \sim 5~{\rm kpc}$.  Significant residuals are found with 
respect to the model fits for some objects due to the presence of multiple
discrete clumps in the surface-brightness distributions, giving rise to
an overall irregular appearance.  The $G$, $\Psi$, and $M_{20}$ values
for the sample also reflect the clumpy appearance, with significantly
higher $M_{20}$ and $\Psi$ values than would be derived for axisymmetric,
monolithic, and compact galaxies at similar redshift.  The relatively low
$G$ values indicate the lack of any single, prominent bulge component.

3. Given the small sample size, and the active levels of ongoing star
formation in all of our targets, it is difficult to extract strong trends
between morphology and other galaxy properties within the SINS NIC2 sample.
However, a comparison with other $z \sim 2$ samples selected using a
rest-frame optical (observed $K$-band) or mid-IR (observed 24\,\micron)
limit reveals that our galaxies follow the broader trends of increasing
size, $M_{20}$, and $\Psi$, and decreasing $n$, $G$, and stellar mass 
surface density, with increasing specific star formation rate. At the
same time, the larger ensemble of galaxies reflects a considerable
diversity in structure at fixed stellar mass.  While these trends have
been presented in other work using seeing-limited or lower-resolution
NIC3 images, the sample with adequate resolution for detailed profile
fits (i.e., NIC2) is still very small, and our objects represent an
extension of the deep $z \sim 2$ NIC2 dataset to higher specific star
formation rate.

4. The data presented here allow for a comparison between rest-frame
optical morphologies and those derived from tracers of active star
formation such as H$\alpha$ emission and rest-frame UV light.
Some theoretical models \citep{Sal09, Dut11} predict that the size of the
galaxy obtained from tracers dominated by young stellar populations,
such as H$\alpha$ emission, should be systematically larger than that
traced by rest-frame optical light.  For the SINS NIC2 sample (excluding
the one putative major merger, BX\,528), the H$\alpha$ and NIC2 $H_{160}$
band sizes reveal no significant differences.  For the one object with
AO-assisted SINFONI data obtained at comparable resolution to our NIC2
imaging (BX\,482), the non-parametric coefficients for H$\alpha$ and
rest-frame optical continuum emission are very similar.
Furthermore, the average NIC2 $G$, $\Psi$, and $M_{20}$ coefficients
for our UV-selected SINS NIC2 targets are very similar to those based
on rest-UV ACS morphologies for a larger sample of UV-selected galaxies
at $z \sim 2$ \citep{Pet07}.  MD\,41 is the only galaxy in our sample
with both NIC2 F160W and ACS F814W imaging.  The rest-UV morphology
for this object is broadly similar to that in the rest-frame optical,
although, in detail, its derived $R_e$ and $n$ are slightly larger
and lower, respectively.  Overall, however, the parametric and
non-parametric coeffients for this object are in agreement.
For UV-selected $z \sim 2$ galaxies with active star formation, the
rest-frame optical appearance thus seems very similar to that traced
by rest-frame UV light and ionized gas.  We have also established
that the NIC2 $H_{160}$ band images are truly tracing {\em continuum}
emission, and are not dominated by nebular line emission.

The power of combining $\sim$kpc-scale resolution rest-frame optical and
UV imaging together with integral field spectroscopy at seeing-limited
and AO-assisted resolution has been demonstrated here for a small sample
of galaxies.
With such a dataset, it is possible to characterize simultaneously the
past and present assembly of mass, and the history of star formation.
A statistical sample of such measurements will provide crucial inputs
into models of the formation and evolution of massive galaxies, and 
will be possible with current and planned facilities in space and on
the ground.

\acknowledgments

We are grateful to M. Franx, L. Simard, A. Renzini, S. Wuyts, S. Genel,
A. Sternberg, P. van Dokkum, and the entire SINS team for many stimulating
discussions and insightful comments on various aspects of this work.  
We wish to thank in particular P. van Dokkum, L. Yan, and M. Kriek for
kindly providing their reduced NIC2 images for the analysis, and N. Reddy
for useful discussions on the IR properties of the sample.
We thank S. Toft for advice on reducing NICMOS data and J. Greene
for help with using GALFIT.
We also thank the referee for a very constructive report and useful
suggestions.
N.M.F.S. acknowledges support by the Minerva program of the MPG.
N.B. is supported by the Marie Curie grant PIOF-GA-2009-236012 from the
European Commission.
G.C. acknowledges support by the ASI-INAF grant I/009/10/0.

\clearpage

\appendix

\section{DERIVATION OF THE RELATIONSHIP BETWEEN MASS-TO-LIGHT RATIO
         AND COLORS FOR MD\,41}
         \label{App-MLcol_models}

In \S~\ref{Sub-disc_md41}, we combined the NIC2 $H_{160}$ band data with
ACS $i_{814}$ band imaging available for one of our targets, MD\,41, to
investigate spatial variations in color, mass-to-light ratio, and stellar
mass on resolved scales of 1.2~kpc.  It is well known that observed colors
and $M/L$ ratio of stellar populations are correlated, and this behavior
can be used to infer stellar masses
\citep[see, e.g.,]
  [for applications to derive the cosmic stellar mass density
   or spatially-resolved mass maps of nearby galaxies]{Rud06, Zib09}.
In this appendix, we present the derivation of the relationship we used
to interpret quantitatively the $i_{814} - H_{160}$ color map of MD\,41.

We employed \citet{BC03} models to generate the observed $i_{814} - H_{160}$
colors of stellar populations appropriate for $z = 2.2$, corresponding to
the redshift of MD\,41.  The colors were synthesized from the redshifted
model spectra using the total transmission curves of the NIC2 F160W and ACS
F814W bandpasses.  We considered a variety of star formation histories (SFHs)
ranging from single stellar populations (``SSP'') with all stars formed
instantaneously and no subsequent star formation, exponentially declining
SFRs with $e$-folding timescales from 10~Myr to 1~Gyr, and constant star
formation (CSF).  The models were computed for a grid of ages from 1~Myr
to 5~Gyr, and of extinctions with $A_{V}$ from 0 to 3~mag.  We adopted the
\citet{Chab03} IMF and the \citet{Cal00} reddening law, and considered two
different metallicities with $Z = {\rm Z_{\odot}}$ and $\rm 0.2\,Z_{\odot}$.

For every parameter combination (SFH, age, $A_{V}$, $Z$), the models
provide the ratio of stellar mass to rest-frame optical luminosity.
We consider here the ``effective'' $M/L$ ratio, i.e., where the
luminosity refers to the light that reaches the observer after being
attenuated by dust.  We further consider the luminosity at rest-frame
wavelengths close to the range probed by our NIC2 F160W data of MD\,41.
Theses choices are relevant because we ultimately combine the derived $M/L$
map with the observed $H_{160}$ band image (least affected by extinction and
by light from young stars) to derive the stellar mass map.  For our purposes,
we approximated the rest-frame optical luminosity seen through the NIC2
F160W filter (mean wavelength of 1.61\,\micron, range with peak-normalized
transmission $> 50\%$ of $\rm 1.40 - 1.80~\mu m$, corresponding to rest-frame
5080\,\AA\ and $4415 - 5675$\,\AA, respectively) with the model predictions
through the SDSS $g$-band filter (mean wavelength of 4695\,\AA\ and range
of $4020 - 5330$\,\AA).  The $g$-band is $\approx 380$\,\AA\ bluer than
the rest-frame range probed by our NIC2 data, but it is closer in mean
wavelength and has larger overlap than the $V$ band (with mean wavelength
of 5510\,\AA\ and range of $5010 - 5865$\,\AA).  Over the entire range of
SFHs and ages considered, the differences between $M_{\star}/L_{g}^{\rm rest}$
and $M_{\star}/L_{V}^{\rm rest}$ ratios are of at most a factor of two, and
similarly over the range of extinction assumed.  Metallicity has a very minor
impact in comparison.  These differences in $M/L$ ratios constitute however
extremes as they are for the full range of model parameters.  More 
realistically, the fairly narrow range in observed $i_{814} - H_{160}$
colors across MD\,41 suggest only modest variations in age and extinction.
With the assumption of ages $\rm \sim 10^{7} - 10^{8}~yr$
and $A_{V} \sim 1~{\rm mag}$, more consistent with the results from
the modeling of the integrated optical to near-IR SED of MD\,41 in
\S~\ref{Sub-sedmod}, the
$M_{\star}/L_{g}^{\rm rest}$ ratio is $\approx 20\%$ lower than the
$M_{\star}/L_{V}^{\rm rest}$ ratio, with little dependence on the
assumed SFH.
Since our NIC2 F160W data probe a rest-frame wavelength range intermediate
between the $g$ and $V$ bands, the numbers above represent upper limits
on the offset between $M_{\star}/L_{g}^{\rm rest}$ we use and the actual
stellar mass to rest-frame optical luminosity ratio from our data.

Figure~\ref{fig-ihcol_lmlg} plots the observed $i_{814} - H_{160}$ colors
versus $M_{\star}/L_{g}^{\rm rest}$ ratios derived from the models for solar
metallicity.  Curves with different colors correspond to different SFHs.
Stellar age increases along the curves from blue to red $i_{814} - H_{160}$
colors, and from low to high $M/L$ ratios.  Different line styles are used
to plot models with different extinction (shown for $A_{V} = 0$, 1, 2, and
3~mag in the Figure).  The main effect of increasing extinction is to shift
the model curves along a direction that is roughly parallel to the age
tracks.  The model curves occupy a fairly well defined locus in the observed
$i_{814} - H_{160}$ versus $M_{\star}/L_{g}^{\rm rest}$ parameter space, 
reflecting the strong degeneracy between stellar age, extinction, and
SFH in these properties.  This degeneracy can be exploited to derive
the mean effective $M/L$ ratio for a given observed color, which,
multiplied by the observed luminosity, yields a stellar mass estimate
without requiring any knowledge about the actual SFH, age, or extinction.

We computed the ``mean'' relationship by taking the average of the
minimum and maximum values in $\log(M_{\star}/L_{g}^{\rm rest})$
over all model curves within equally-spaced bins of width 0.1~mag in
$i_{814} - H_{160}$ colors.  By proceeding in this way, we reduced
the impact of the discreteness of our grids of SFHs and extinction.
Over the ranges in colors where models for all SFHs and all ages
for a fixed extinction value are represented, the average of the
minimum and maximum $\log(M_{\star}/L_{g}^{\rm rest})$ is very close
to the weighted mean over all models, where the weights are obtained from
the time span of each model within a given color bin; the differences
are at most 0.08~dex in $\log(M_{\star}/L_{g}^{\rm rest})$.  Varying
the $A_{V}$ value has little impact since extinction effects are mostly
along the relationship.  The mean relationship for solar metallicity is
plotted in Figure~\ref{fig-ihcol_lmlg} (white-filled circles and thick
solid black line), with error bars corresponding to the standard deviation
of all models within each color bin.  The relationship derived for $1/5$
solar metallicity is also shown for comparison (grey-filled circles and
thin solid black line), but is very similar to the one for solar
metallicity.
For our analysis of MD\,41, we adopted the $i_{814} - H_{160}$
versus $\log(M_{\star}/L_{g}^{\rm rest})$ relationship derived
for $Z = {\rm Z_{\odot}}$ models.  The standard deviation of the
models suggest that the relationship is accurate to 0.15~dex in
$\log(M_{\star}/L_{g}^{\rm rest})$ over the range of observed
colors across MD\,41 (indicated by the shaded histogram in
Figure~\ref{fig-ihcol_lmlg}).

\clearpage


\setcounter{figure}{0}
\setcounter{section}{0}

\begin{deluxetable}{llccccccc}
\tabletypesize{\small}
\tablecolumns{9}
\setlength{\tabcolsep}{0.5cm}
\tablewidth{0pt}
\tablecaption{Galaxies observed
              \label{tab-targets}}
\tablehead{
   \colhead{Object} & 
   \colhead{$z_{\rm H\alpha}$\,\tablenotemark{a}} & 
   \colhead{Kinematic Type\,\tablenotemark{b}} &
   \colhead{$U_{n}$} & 
   \colhead{$G$} & 
   \colhead{$\mathcal{R}$} & 
   \colhead{$J$} &
   \colhead{$H_{\rm 160}$} &
   \colhead{$K_{\rm s}$} \\
   \colhead{} & 
   \colhead{} & 
   \colhead{} & 
   \colhead{(mag)} & 
   \colhead{(mag)} & 
   \colhead{(mag)} & 
   \colhead{(mag)} & 
   \colhead{(mag)} &
   \colhead{(mag)} 
}
\startdata
$\rm Q1623-BX528$                     & 2.2683   &  Major merger &
     $24.52 \pm 0.18$ & $23.81 \pm 0.14$ & $23.56 \pm 0.13$ & 
     $22.44 \pm 0.17$ & $22.33 \pm 0.06$ & $21.61 \pm 0.20$ \\
$\rm Q1623-BX663$\,\tablenotemark{c}  & 2.4332   &  Disk         &
     $25.40 \pm 0.24$ & $24.38 \pm 0.17$ & $24.14 \pm 0.15$ & 
     $23.41 \pm 0.30$ & $22.79 \pm 0.10$ & $21.78 \pm 0.21$ \\
$\rm SSA22a-MD41$                     & 2.1704   &  Disk         &
     $24.81 \pm 0.11$ & $23.50 \pm 0.06$ & $23.31 \pm 0.05$ & 
     \ldots           & $22.64 \pm 0.05$  & $22.30 \pm 0.36$ \\
$\rm Q2343-BX389$                     & 2.1733   &  Disk         &
     $26.39 \pm 0.39$ & $25.13 \pm 0.19$ & $24.85 \pm 0.16$ & 
     $23.82 \pm 0.24$ & $23.11 \pm 0.10$ & $22.04 \pm 0.25$ \\
$\rm Q2343-BX610$\,\tablenotemark{c}  & 2.2103   & Disk          &
     $24.67 \pm 0.21$ & $23.92 \pm 0.12$ & $23.58 \pm 0.11$ & 
     $22.35 \pm 0.13$ & $22.09 \pm 0.06$ & $21.07 \pm 0.13$ \\
$\rm Q2346-BX482$                     & 2.2571   & Disk          &
     $24.44 \pm 0.12$ & $23.54 \pm 0.10$ & $23.32 \pm 0.09$ & 
     \ldots           & $22.34 \pm 0.07$ & \ldots           \\
\enddata
\tablecomments
{
Magnitudes are given in the AB system.
Optical, $J$ and $K_{\rm s}$ magnitudes are taken from \citealt{Erb06}
(with the exception of $\rm SSA22a-MD41$ for which the $K_{\rm s}$
magnitude was measured on NTT/SOFI imaging available from the ESO
archive).
The $H_{160}$ band magnitudes were measured from the new NICMOS/NIC2
data presented in this paper.
}
\tablenotetext{a}
{
Systemic vacuum redshift derived from the integrated H$\alpha$ line
emission in the SINFONI data \citep{FS09}.
}
\tablenotetext{b}
{
Classification according to the H$\alpha$ kinematics from SINFONI
\citep{Sha08, FS06b, FS09}.
}
\tablenotetext{c}
{
Spectral signatures of an AGN are detected in the optical and near-IR
spectrum of $\rm Q1623-BX663$ and the large-scale H$\alpha$ kinematics
suggest the host galaxy is a low-inclination rotating disk
\citep{Erb06, FS06b, FS09}.
The {\em Spitzer\/}/MIPS fluxes of $\rm Q1623-BX663$ and $\rm Q2343-BX610$
are indicative of obscured AGN activity \citep{Red10}.
}
\end{deluxetable}

\begin{deluxetable}{llrc}
\tabletypesize{\small}
\tablecolumns{4}
\setlength{\tabcolsep}{0.5cm}
\tablewidth{400pt}
\tablecaption{Log of NICMOS/NIC2 observations
              \label{tab-obs}}
\tablehead{
 \colhead{Object} & 
 \colhead{Date observed} & 
 \colhead{P.A.\,\tablenotemark{a}} & 
 \colhead{$3\,\sigma$ Limit in $d = 0.22^{\prime\prime}$\,\tablenotemark{b}} \\
   \colhead{} & 
   \colhead{} & 
   \colhead{(degrees)} & 
   \colhead{(mag)}
}
\startdata
$\rm Q1623-BX528$  &
   2007 Apr 27    &  $-5.43$ &  28.07 \\
$\rm Q1623-BX663$  &
   2007 Apr 28-29 & $-20.43$ &  28.17 \\
$\rm SSA22a-MD41$  &
   2007 May 07-08 &  $30.13$ &  27.98 \\
$\rm Q2343-BX389$  &
    2007 Jul 10   &  $16.91$ &  28.08 \\
$\rm Q2343-BX610$  &
    2007 Sep 24   & $-45.43$ &  28.14 \\
$\rm Q2346-BX482$  &
    2007 Jul 18   &   $4.57$ &  28.22 \\
\enddata
\tablecomments{
The total on-source integration time is 10,240\,s for all targets.
}
\tablenotetext{a}
{
Position angle of $y$-axis (in degrees East of North).
}
\tablenotetext{b}
{
Effective $3\,\sigma$ background noise limiting magnitudes (AB system)
in a ``point-source'' circular aperture with diameter $d = 0\farcs 22$,
or $1.5 \times {\rm FWHM}$ of the NIC2 PSF as measured from stellar
profiles in the data.  The limiting depths were determined from the
analysis of the noise properties of the reduced data presented in
\S~\ref{Sub-noise}.
}
\end{deluxetable}

\begin{deluxetable}{lcccccc}
\tabletypesize{\small}
\tablecolumns{7}
\tablewidth{400pt}
\tablecaption{Stellar and dust properties of the galaxies
              \label{tab-galprop}}
\tablehead{
   \colhead{Object} & 
   \colhead{Age} &
   \colhead{$A_{V}$} &
   \colhead{$M_{\star}$} &
   \colhead{$M_{V,\,{\rm AB}}$\,\tablenotemark{a}} &
   \colhead{SFR} &
   \colhead{sSFR\,\tablenotemark{b}} \\
   \colhead{} &
   \colhead{(Myr)} &
   \colhead{(mag)} &
   \colhead{($\rm 10^{10}~M_{\odot}$)} &
   \colhead{(mag)} &
   \colhead{($\rm M_{\odot}\,yr^{-1}$)} &
   \colhead{($\rm Gyr^{-1}$)}
}
\startdata
$\rm Q1623-BX528$   & $2750^{+96}_{-2110}$       &
         $0.6 \pm 0.2$             & $6.95^{+0.17}_{-3.61}$  &
         $-22.90^{+0.04}_{-0.12}$  & $42^{+29}_{-16}$          &
         $0.6^{+1.7}_{-0.1}$     \\
$\rm Q1623-BX663$   & $2500^{+147}_{-800}$      &
         $0.8 \pm 0.2$             & $6.40^{+0.22}_{-2.28}$  &
         $-22.68^{+0.05}_{-0.20}$  & $42^{+13}_{-12}$          &
         $0.7^{+0.3}_{-0.2}$     \\
$\rm SSA22a-MD41$   & $50^{+31}_{-0}$           &
         $1.2 \pm 0.2$             & $0.77^{+0.10}_{-0.03}$  &
         $-22.37^{+0.04}_{-0.09}$  & $185^{+3}_{-70}$          &
         $24^{+1}_{-9}$            \\
$\rm Q2343-BX389$   & $2750^{+224}_{-1945}$      &
         $1.0 \pm 0.2$             & $4.12^{+0.77}_{-2.16}$  &
         $-21.94^{+0.09}_{-0.13}$  & $25^{+17}_{-2}$           &
         $0.6^{+1.2}_{-0.1}$       \\
$\rm Q2343-BX610$   & $2750^{+173}_{-650}$      &
         $0.8 \pm 0.2$             & $10.0^{+2.7}_{-0.6}$    &
         $-23.10^{+0.12}_{-0.04}$  & $60^{+26}_{-1}$           &
         $0.6 \pm 0.2$             \\
$\rm Q2346-BX482$   & $321^{+485}_{-141}$       &
         $0.8 \pm 0.2$             & $1.84^{+0.79}_{-0.46}$  &
         $-22.65^{+0.07}_{-0.08}$  & $80^{+42}_{-32}$          &
         $4.3^{+3.0}_{-2.5}$       \\
\enddata
\tablecomments
{
Results from SED modeling of the optical and near-IR photometry, with
best-fitting star formation history having a constant SFR for all targets.
The formal (random) fitting uncertainties are given, derived from the 68\%
confidence intervals based on 200 Monte Carlo simulations for the default
set of \citealt{BC03} models with solar metallicity, a \citealt{Chab03} IMF,
and the \citealt{Cal00} reddening law;
systematic uncertainties (from SED modeling assumptions) are
estimated to be typically a factor of $\sim 1.5$ for the stellar masses,
$\rm \pm 0.3~mag$ for the extinctions, and factors of $\sim 2 - 3$ for
the ages as well as for the absolute and specific star formation rates.
}
\tablenotetext{a}
{
Rest-frame absolute $V$-band magnitude, uncorrected for extinction.
}
\tablenotetext{b}
{
Specific star formation rate,
i.e. the ratio of star formation rate over stellar mass.
}
\end{deluxetable}

\clearpage

\begin{deluxetable}{lcccc}
\tabletypesize{\small}
\tablecolumns{5}
\setlength{\tabcolsep}{0.5cm}
\tablewidth{450pt}
\tablecaption{Best-fit Structural Parameters from the
              NIC2 $H_{160}$ Band Images
              \label{tab-galfitres}}
\tablehead{
   \colhead{Object} &
   \colhead{$R_{\rm e}$} &
   \colhead{$n$} &
   \colhead{$b/a$} &
   \colhead{P.A.} \\
   \colhead{} &
   \colhead{(kpc)} &
   \colhead{} &
   \colhead{} &
   \colhead{(degrees)}
}
\startdata
\cutinhead{Single-component models} \\
$\rm Q1623-BX528$  & $4.86^{+0.13}_{-0.10}$ & $0.16 \pm 0.01$        &
                     $0.31 \pm 0.01$        & $-27.8^{+1.2}_{-1.1}$  \\
$\rm Q1623-BX663$  & $4.54^{+7.71}_{-0.79}$ & $2.00^{+2.11}_{-0.59}$ &
                     $0.73^{+0.04}_{-0.02}$ & $-21.8^{+4.3}_{-2.9}$  \\
$\rm SSA22a-MD41$  & $5.69^{+0.20}_{-0.13}$ & $0.54^{+0.07}_{-0.06}$ &
                     $0.31 \pm 0.01$        & $36.5 \pm 0.2$         \\
$\rm Q2343-BX389$  & $5.93^{+0.17}_{-0.12}$ & $0.36^{+0.07}_{-0.05}$ &
                     $0.30 \pm 0.01$        & $-48.6^{+0.3}_{-0.5}$  \\
$\rm Q2343-BX610$  & $4.44 \pm 0.08$        & $0.57 \pm 0.04$        &
                     $0.56 \pm 0.01$        & $16.9^{+0.5}_{-0.9}$  \\
$\rm Q2346-BX482$  & $6.22^{+0.13}_{-0.12}$ & $0.14 \pm 0.01$        &
                     $0.48 \pm 0.01$        & $-60.8^{+1.9}_{-1.5}$ \\
\cutinhead{Two-component models} \\
$\rm Q1623-BX528$ SE  & $3.18^{+0.18}_{-0.25}$ & $1.20^{+0.15}_{-0.19}$ &
                        $0.35^{-0.08}_{-0.03}$ & $-40.5^{+3.4}_{-1.7}$  \\
$\rm Q1623-BX528$ NW  & $3.57^{+1.98}_{-0.57}$ & $2.91^{+1.70}_{-1.34}$ &
                        $0.51^{+0.06}_{-0.05}$ & $-10.9^{+9.5}_{-5.2}$  \\
$\rm Q2343-BX389$ Main & $5.89^{+0.08}_{-0.09}$ & $0.32^{+0.04}_{-0.04}$ &
                         $0.27 \pm 0.01$    & $-49.0^{+0.5}_{-0.4}$  \\
$\rm Q2343-BX389$ S    & $2.71^{+2.18}_{-0.69}$ & $3.03^{+2.64}_{-1.75}$ &
                         $0.20^{+0.35}_{-0.16}$ & $-41.5^{+2.2}_{-5.4}$  \\
\enddata
\tablecomments{
Structural parameters derived from S\'ersic profile fits to
the two-dimensional rest-frame $\rm \approx 5000$\,\AA\ surface
brightness distribution from the NIC2 $H_{\rm 160}$ band images.
Single-component models were fitted to all targets.
For BX\,528, two-component fits were also performed to account
for the southeast (``SE'') and northwest (``NW'') components.
Similarly for BX\,389, where the two components represent the
main part of the galaxy and the small companion to the south,
``Main'' and ``S,'' respectively.
The results are obtained from a series of 500 GALFIT runs
as described in \S~\ref{Sub-galfit}.
The best-fit effective radius $R_{\rm e}$, the S\'ersic index $n$,
the ratio of minor to major axis $b/a$, and the position angle P.A.
(in degrees East of North) correspond to the median of the distribution
of results, and the uncertainties represent the $68\%$ confidence
intervals around the best fit.
}
\end{deluxetable}


\begin{deluxetable}{lllcccc}
\tabletypesize{\small}
\tablecolumns{7}
\setlength{\tabcolsep}{0.5cm}
\tablewidth{480pt}
\tablecaption{Best-fit Structural Parameters from the SINFONI H$\alpha$ Maps
              and PSF-Matched NIC2 $H_{160}$ Band Images
              \label{tab-galfitres_smoo}}
\tablehead{
   \colhead{Object} &
   \colhead{Map} &
   \colhead{FWHM resolution\tablenotemark{a}} &
   \colhead{$R_{\rm e}$\,\tablenotemark{b}} &
   \colhead{$n$\,\tablenotemark{b}} &
   \colhead{$b/a$\,\tablenotemark{b}} &
   \colhead{P.A.\,\tablenotemark{b}} \\
   \colhead{} &
   \colhead{} &
   \colhead{} &
   \colhead{(kpc)} &
   \colhead{} &
   \colhead{} &
   \colhead{(degrees)}
}
\startdata
$\rm Q1623-BX528$  &
   H$\alpha$ & $0\farcs 63$ & $5.06^{+0.68}_{-1.56}$ & $0.29^{+0.69}_{-0.19}$ &
                            $0.53^{+0.26}_{-0.10}$ & $-42.1^{+30.4}_{-6.5}$ \\
 & $H_{160}$ & $0\farcs 63$ & $4.91^{+1.64}_{-1.73}$ & $0.25^{+0.66}_{-0.23}$ &
                            $0.44^{+0.32}_{-0.15}$ & $-31.3^{+14.7}_{-9.0}$ \\
\hline \\
$\rm Q1623-BX663$  &
   H$\alpha$ & $0\farcs 39$ & $3.97^{+1.15}_{-0.99}$ & $0.68^{+0.73}_{-0.41}$ &
                            $0.72^{+0.17}_{-0.12}$ & $4.9^{+24.3}_{-27.4}$ \\
 & $H_{160}$ & $0\farcs 39$ & $3.37^{+0.58}_{-0.68}$ & $0.36^{+0.32}_{-0.16}$ &
                            $0.82^{+0.14}_{-0.16}$ & $-9.3^{+46.6}_{-34.0}$ \\
\hline \\
$\rm SSA22a-MD41$  &
   H$\alpha$ & $0\farcs 44$ & $5.30^{+0.99}_{-1.02}$ & $0.28^{+0.60}_{-0.19}$ &
                            $0.50^{+0.33}_{-0.09}$ & $33.2^{+9.3}_{-17.6}$ \\
 & $H_{160}$ & $0\farcs 44$ & $5.64^{+0.85}_{-1.54}$ & $0.35^{+0.55}_{-0.24}$ &
                            $0.40^{+0.41}_{-0.11}$ & $36.7^{+9.0}_{-8.3}$  \\
\hline \\
$\rm Q2343-BX389$  &
   H$\alpha$ & $0\farcs 54$ & $6.10^{+1.62}_{-1.19}$ & $0.31^{+0.86}_{-0.23}$ &
                            $0.48^{+0.35}_{-0.11}$ & $-50.9^{+14.8}_{-10.7}$ \\
 & $H_{160}$ & $0\farcs 54$ & $5.99^{+0.49}_{-0.90}$ & $0.25^{+0.61}_{-0.13}$ &
                            $0.47^{+0.33}_{-0.09}$ & $-44.6^{+10.7}_{-5.8}$  \\
\hline \\
$\rm Q2343-BX610$  &
   H$\alpha$ & $0\farcs 39$ & $5.12^{+2.01}_{-1.13}$ & $0.56^{+0.71}_{-0.35}$ &
                            $0.75^{+0.17}_{-0.10}$ & $11.9^{+20.1}_{-14.9}$  \\
 & $H_{160}$ & $0\farcs 39$ & $4.37^{+0.62}_{-0.49}$ & $0.45^{+0.26}_{-0.20}$ &
                            $0.66^{+0.22}_{-0.09}$ & $10.1^{+12.5}_{-23.0}$  \\
\hline \\
$\rm Q2346-BX482$  &
   H$\alpha$ & $0\farcs 17$ & $6.07^{+7.09}_{-1.46}$ & $0.35^{+3.47}_{-0.28}$ &
                            $0.59^{+0.21}_{-0.12}$ & $-50.2^{+32.4}_{-6.7}$  \\
 & $H_{160}$ & $0\farcs 15$ & $6.22^{+0.13}_{-0.12}$ & $0.14 \pm 0.01$        &
                            $0.48 \pm 0.01$        & $-60.9^{+1.9}_{-1.5}$  \\
 & H$\alpha$ & $0\farcs 50$ & $6.01^{+2.74}_{-1.46}$ & $0.27^{+0.70}_{-0.19}$ &
                            $0.64^{+0.20}_{-0.11}$ & $-58.0^{+23.4}_{-10.2}$  \\
 & $H_{160}$ & $0\farcs 50$ & $6.22^{+2.11}_{-0.91}$ & $0.15^{+0.50}_{-0.11}$ &
                            $0.50^{+0.17}_{-0.06}$ & $-55.4^{+13.7}_{-8.0}$  \\
\enddata
\tablecomments{
Structural parameters derived from single-component S\'ersic profile
fits to the two-dimensional H$\alpha$ maps from SINFONI and to the
$H_{160}$ band images from NIC2 convolved to the effective spatial
resolution and resampled to the pixel scale of the SINFONI data.
The results are obtained from a series of 500 GALFIT runs as described
in \S~\ref{Sub-galfit}, and the best fit value and uncertainties correspond
to the median and $68\%$ confidence intervals of the distribution of results
in each parameter.
}
\tablenotetext{a}
{
PSF FWHM of the SINFONI H$\alpha$ maps, and of the convolved
NIC2 $H_{160}$ band images.
}
\tablenotetext{b}
{
Main structural parameters derived from the GALFIT fits:
the effective radius radius $R_{\rm e}$, the S\'ersic index $n$,
the ratio of minor to major axis $b/a$, and the position angle P.A.
(in degrees East of North).
}
\end{deluxetable}

\clearpage

\begin{deluxetable}{lcccc}
\tabletypesize{\small}
\tablecolumns{5}
\tablewidth{0pt}
\tablecaption{Emission line contributions to the $H_{160}$ band flux densities
              \label{tab-lineprop}}
\tablehead{
   \colhead{Object} &
   \colhead{$f_{\rm BB}({\rm [OIII]\,\lambda\,5007})$} &
   \colhead{$f_{\rm BB}({\rm [OIII]\,\lambda\,4959})$} &
   \colhead{$f_{\rm BB}({\rm H\beta})$} &
   \colhead{$f_{\rm BB}({\rm tot})$} \\
   \colhead{} &
   \colhead{} &
   \colhead{} &
   \colhead{} &
   \colhead{}
}
\startdata
$\rm Q1623-BX528$  &\
 \ldots  &  \ldots  &  \ldots  &  \ldots  \\
$\rm Q1623-BX663$  &
 $< 0.05$           & $< 0.02$          & $< 0.12$          & $< 0.20$        \\
$\rm SSA22a-MD41$  &
 $0.11 \pm 0.01$    & $0.043 \pm 0.005$ & $0.023 \pm 0.004$ & $0.17 \pm 0.01$ \\
$\rm Q2343-BX389$  &
 $0.19 \pm 0.02$    & $0.050 \pm 0.008$ & $< 0.054$         & $< 0.29$        \\
$\rm Q2343-BX610$  &
 $0.034 \pm 0.004$  & $0.013 \pm 0.002$ & $0.018 \pm 0.002$ & $0.064 \pm 0.005$ \\
$\rm Q2346-BX482$  &
 $0.062 \pm 0.006$  & $0.019 \pm 0.003$ & $< 0.037$         & $< 0.12$        \\
\enddata
\tablecomments
{
Individual and total fractional contributions of the
[\ion{O}{3}]\,$\lambda\lambda\,4959,5007$ and H$\beta$ emission lines to
the integrated broad-band flux density measured from the NIC2 $H_{160}$
band data.  The uncertainties correspond to $1\,\sigma$ uncertainties
from the line fluxes and broad-band magnitudes.  For undetected lines,
the $3\,\sigma$ upper limit on the line flux has been used.
}
\end{deluxetable}

\clearpage


\setcounter{figure}{0}

\begin{figure}[p]
\figurenum{1}
\epsscale{1.20}
\plotone{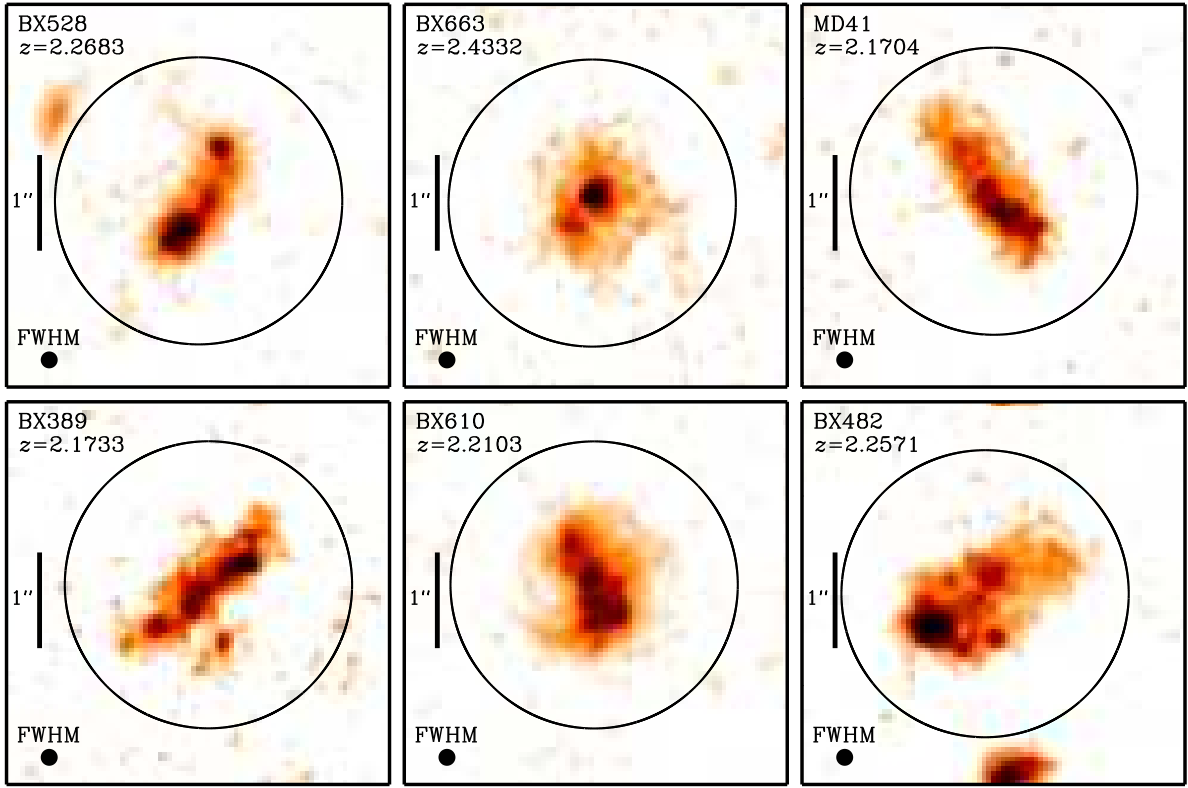}
\vspace{0.0cm}
\caption{
\small
{\em HST\/} NICMOS/NIC2 $H_{\rm 160}$ maps of the galaxies, probing the
broad-band emission around rest-frame 5000\,\AA.  The target name
and H$\alpha$ redshift are labeled in the top corner of each panel.
The color coding scales linearly with flux density from white to black
for the minimum to maximum levels displayed (varying for each galaxy).
The FWHM of the effective PSF is indicated by the filled circle at the
bottom left of each corner; the angular resolution is $0\farcs 145$,
or $\rm \approx 1.2~kpc$ at the median $z = 2.2$ of the sources.
The angular scale of the images is shown with the 1\arcsec -long
vertical bar next to each galaxy.  The black solid circle overlaid on
each map shows the aperture used to extract the integrated $H_{\rm 160}$
band flux density (diameter of 3\arcsec).
In all maps, North is up and East is to the left.
\label{fig-nicmaps}
}
\end{figure}

\clearpage

\begin{figure}[p]
\figurenum{2}
\epsscale{1.00}
\plotone{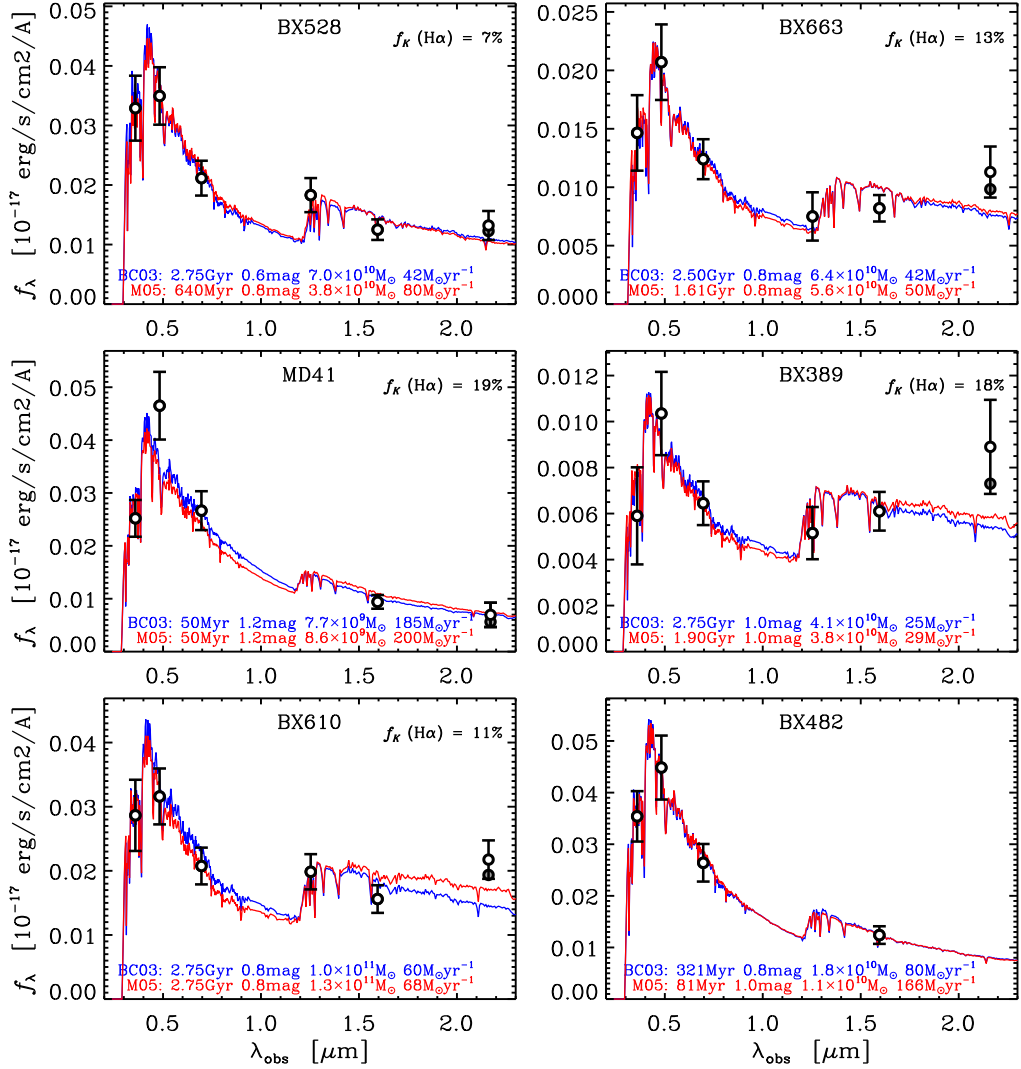}
\vspace{0.0cm}
\caption{
\small
Broad-band optical to near-IR SEDs of the galaxies, including the
$H_{\rm 160}$ flux density measured from the NIC2 data presented in
this paper.  The large white-filled circles and error bars represent the
observed photometry and $1\,\sigma$ uncertainties (Table~\ref{tab-targets}).
The grey-filled circle shows the $K$ band flux density corrected for the
contribution by the H$\alpha$ line emission, estimated from the available
SINFONI data and given explicitly in each panel.
The solid lines show the synthetic spectrum of the best-fitting stellar
population model, assuming a constant star formation rate, solar metallicity,
the \citet{Cal00} reddening law, and the \citet{Chab03} or \citet{Kro01} IMF,
obtained with the \citet{BC03} and \citet{Mar05} models (blue and red lines,
respectively).  The best-fit age, $A_{V}$, stellar mass, and SFR for each
galaxy and synthesis code are listed in each panel.
\label{fig-seds}
}
\end{figure}

\clearpage

\begin{figure}[p]
\figurenum{3}
\epsscale{1.20}
\plotone{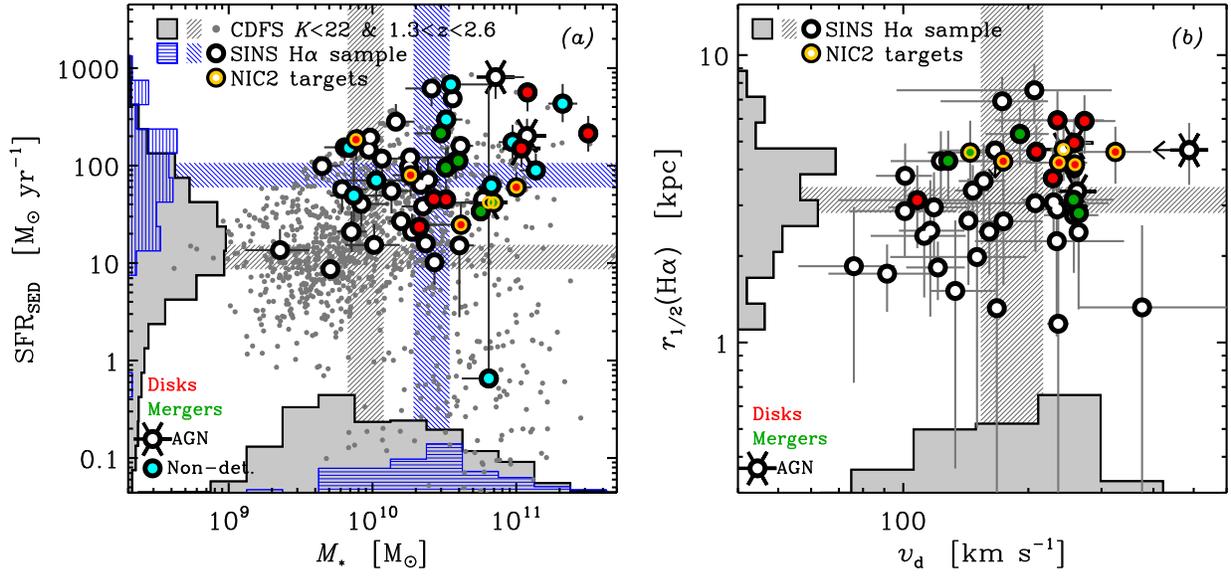}
\vspace{-0.5cm}
\caption{
\small
Stellar and dynamical properties of the galaxies observed with NIC2.
The data points for the NIC2 targets are indicated with yellow,
and the data for the comparison samples are taken from \citet{FS09}.
{\em (a)\/} Stellar mass and star formation rate derived from SED modeling,
compared to those of the SINS H$\alpha$ sample at $1.3 < z < 2.6$ and of
$K$-selected galaxies from the CDFS \citep{Wuy08} in the same redshift
interval and at $K_{\rm s} < 22.0~{\rm mag}$ (the magnitude of the
faintest of the SINS H$\alpha$ sample galaxies in the $K$ band).
The SINS data points are shown with large filled dots, their projected
distribution onto each axis with blue-hatched histograms, and their
median magnitudes as blue-hatched horizontal bars.
The CDFS data are plotted with small grey dots, grey filled histograms,
and grey-hatched bars.  The histograms are arbitrarily normalized.
The galaxies classified as disk-like and merger-like by kinemetry
\citep{Sha08} are plotted as red- and green-filled circles.  Sources
that were known to host an AGN based on optical (rest-UV) or previous
long-slit near-IR (rest-frame optical) spectroscopy are indicated with
a 6-pointed skeletal star.  SINS targets that were not detected in
H$\alpha$ line emission are marked as cyan-filled circles.
{\em (b)\/} Half-light radius and circular velocity $v_{\rm d}$ derived
from the SINFONI H$\alpha$ observations, compared to those of the
entire SINS H$\alpha$ sample.  The color scheme for data points
is as for the left panel, and the projected distributions and median
values for the SINS H$\alpha$ sample are shown with the grey histograms
and grey-hatched bars.
The NIC2 sample spans roughly an order of magnitude in stellar mass
and SFR, and lies at the high end of the distribution of H$\alpha$
sizes and circular velocities inferred for the SINS H$\alpha$ sample.
\label{fig-sedkinprop}
}
\end{figure}

\clearpage

\begin{figure}[p]
\figurenum{4}
\epsscale{1.10}
\plotone{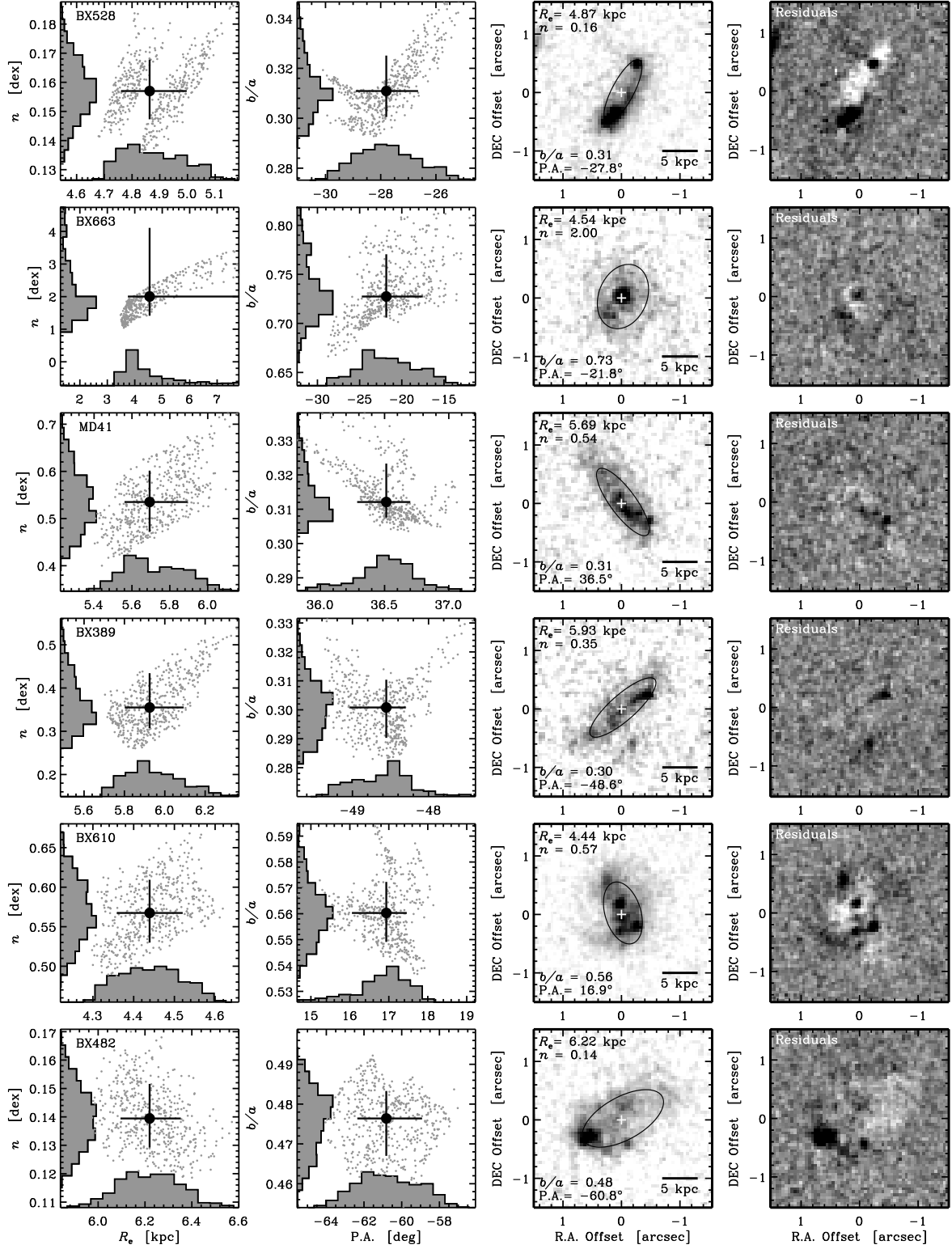}
\vspace{-1.25cm}
\caption{
\small
Results of parametric fits to the two-dimensional $H_{160}$ band surface
brightness distribution of our NIC2 sample.  The structural parameters
(effective radius $R_{\rm e}$, S\'ersic index $n$, axis ratio $b/a$, and
position angle P.A.) were derived from a series of 500 single-component
S\'ersic model fits, as described in \S~\ref{Sub-galfit_meth}. 
Each row shows the results for a given galaxy, with panels as follows.
{\em Far left\/}: $n$ versus $R_{\rm e}$, where small grey dots
show the results of the 500 iterations, histograms correspond to the
projected distributions onto each axis, and the large black dot and
error bars indicate the median and 68\% confidence intervals adopted
as best-fit parameters and $1\,\sigma$ uncertainties.
{\em Middle left\/}: Same as the far left panel but for $b/a$ versus P.A.
{\em Middle right\/}: NIC2 images with overplotted ellipse of center,
semi-major axis, axis ratio, and orientation equal to the adopted
best-fit position, $R_{\rm e}$, $b/a$, and P.A.
{\em Far right\/}: Residual map obtained by subtracting the NIC2 image
with the S\'ersic model with best-fit parameters.
All images are plotted with a linear intensity grey scale (with negative
to positive residuals ranging from white to black).
\label{fig-galfitres}
}
\end{figure}

\clearpage

\begin{figure}[p]
\figurenum{5}
\epsscale{1.10}
\plotone{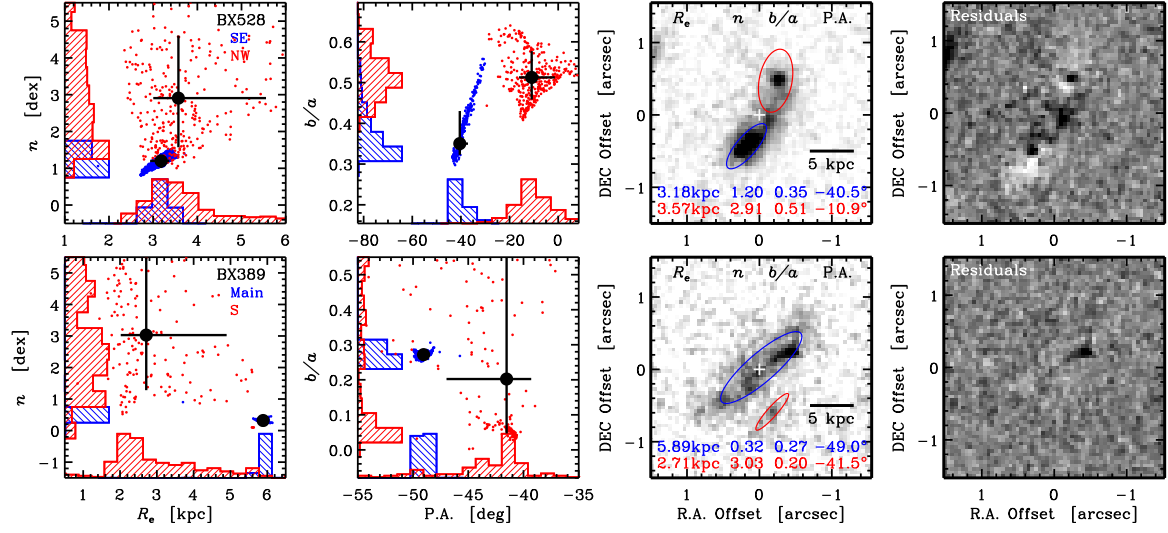}
\vspace{-0.75cm}
\caption{
\small
Same as Figure~\ref{fig-galfitres} for the two-component S\'ersic fits
for BX\,528 and BX\,389.  In the three first panels to the left and for
BX\,528, blue and red colors correspond to the results for the southeast
and northwest components, respectively.  For BX\,389, blue is used for
the main part of the galaxy, and red for the small southern companion.
\label{fig-galfitres2}
}
\end{figure}

\clearpage

\begin{figure}[p]
\figurenum{6}
\epsscale{1.10}
\plotone{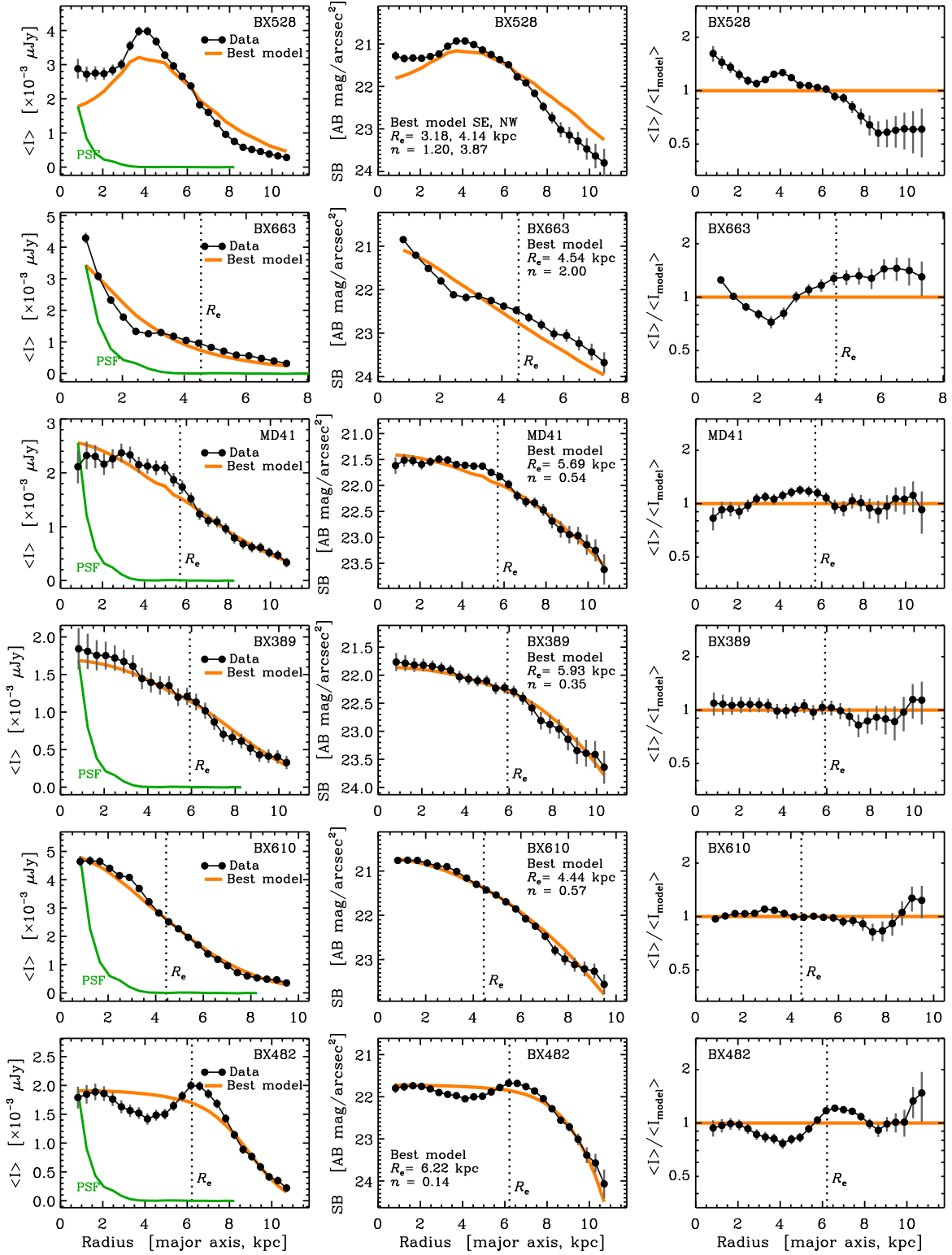}
\vspace{-1.0cm}
\caption{
\small
Radial profiles in $H_{160}$ band light of our NIC2 sample.
The profiles are azimutally-averaged along concentric ellipses with center,
semi-major axis, axis ratio, and orientation equal to the best-fit position,
$R_{\rm e}$, $b/a$, and P.A. values from the S\'ersic model fits to the 
galaxies (see \S~\ref{Sub-galfit}).
Each row shows the results of a given galaxy, with the panels as follows.
{\em (Left)\/} Mean intensity in units of $\rm \mu Jy$ as a function of
semi-major axis radius of the elliptical annuli; the black curve and dots
show the profile extracted from galaxy's image while the orange curve shows
the corresponding profile for the best-fit S\'ersic model.
The radial intensity profile of the PSF (along concentric circular annuli)
is plotted in green for reference.
{\em (Middle)\/} Same as for the left panel for the mean surface brightness
in units of $\rm mag~arcsec^{-2}$, with the best-fit structural parameters
labeled explicitly.
{\em (Right)\/} Ratio of the mean intensity profiles of the data and
of the best-fit model as a function of semi-major axis radius of the
elliptical annuli.
For all galaxies except BX\,528, single-component S\'ersic models are
considered, and the derived effective radius is indicated by a vertical
dashed line.
For BX\,528, the model curves plotted here are extracted from the image
for the best-fit two-component model but along elliptical annuli defined
by the best-fit single-component model.
The different intensity units between the left and middle panels are chosen
to highlight in turn the central brighter and outer fainter regions of the
sources.
\label{fig-profiles}
}
\end{figure}

\clearpage

\begin{figure}[t]
\figurenum{7}
\epsscale{0.80}
\plotone{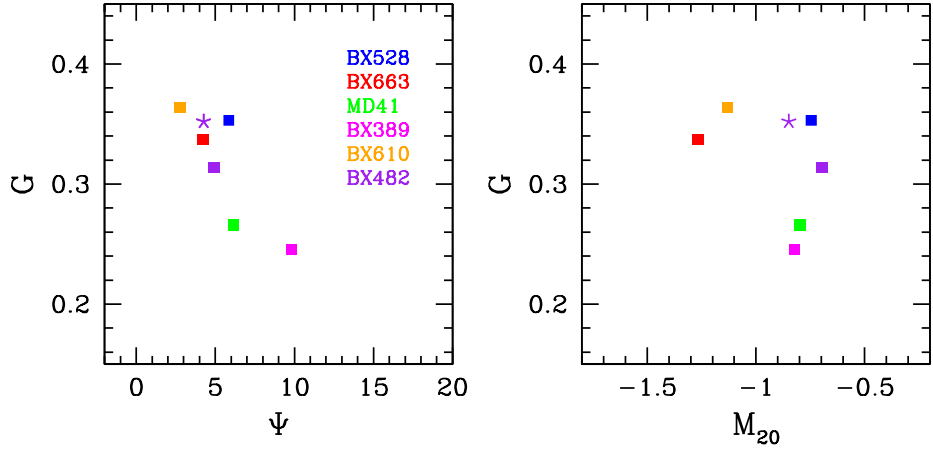}
\vspace{0.0cm}
\caption{
\small
Morphological parameters for SINS galaxies derived from the NIC2
$H_{160}$ band images.  Each galaxy was analyzed as if it was placed
at the maximum redshift of the sample, i.e., $z = 2.43$.
{\em (Left)\/} Gini versus $\Psi$ (Multiplicity).
{\em (Right)\/} Gini versus $M_{20}$.
The AO-assisted SINFONI H$\alpha$ map of BX\,482 was also analyzed in
the same manner, and the results are plotted (star symbol) along with
the NIC2 results (squares) in both panels.
\label{fig-morph1}
}
\end{figure}


\begin{figure}[b]
\figurenum{8}
\epsscale{1.10}
\plotone{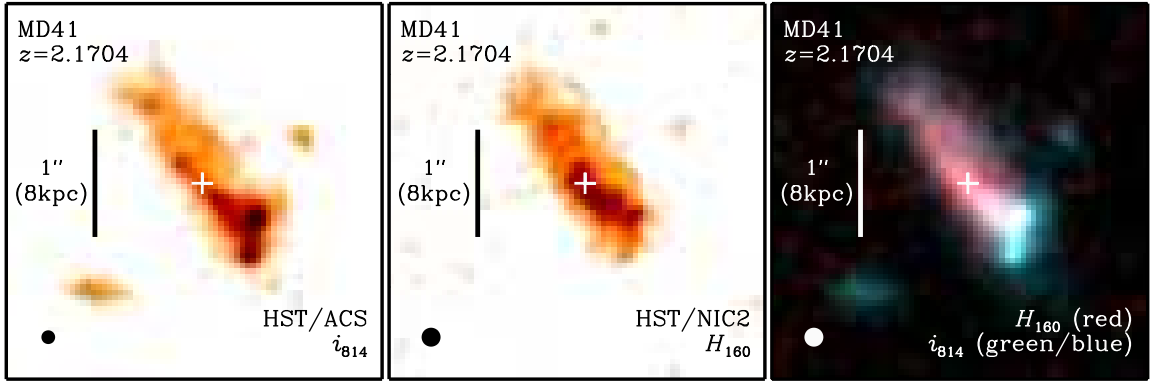}
\vspace{0.0cm}
\caption{
\small
Comparison of rest-frame UV and optical morphologies for MD\,41.
North is up and east to the left in both images, and a bar of 1\arcsec\
in length indicates the scale in the images.
{\em (Left)\/} ACS $i_{814}$ band map at the original angular resolution
of the data, with the FWHM of the PSF indicated by the filled circle at
the bottom left corner.
{\em (Middle)\/} Same as the left panel for the NIC2 $H_{160}$ band map.
{\em (Right)\/} Color-composite after accurate PSF matching and registration
of the ACS and NIC2 maps; in this RGB image, the red channel is assigned to
the $H_{160}$ band, and the green and blue channels to the $i_{814}$ band.
The colors for all images plotted correspond to a linear flux density scale.
The white cross shows the location of the geometrical center of MD\,41 as
determined from the NIC2 map.
Overall, the images are strikingly similar, although on small scales
there are noticeable differences.  In particular, the two brightest
clumps in $i_{814}$ band on the southwest edge of the galaxy are much
fainter in $H_{160}$ band and, consequently, have bluer colors than
the bulk of the source.  In addition, two compact regions appear in
the $i_{814}$ band image alone, which are offset from the main body
of the galaxy.  The colors of these compact offset sources are thus
significantly bluer than that of the main part of the galaxy.
These sources also have no counterpart in line emission within
$\rm \pm 1000~km\,s^{-1}$ of the H$\alpha$ line of MD\,41 in our
SINFONI data.  Therefore, we cannot determine whether they are 
physically associated with MD41.
\label{fig-md41_panel}
}
\end{figure}

\clearpage

\begin{figure}[p]
\figurenum{9}
\epsscale{1.20}
\plotone{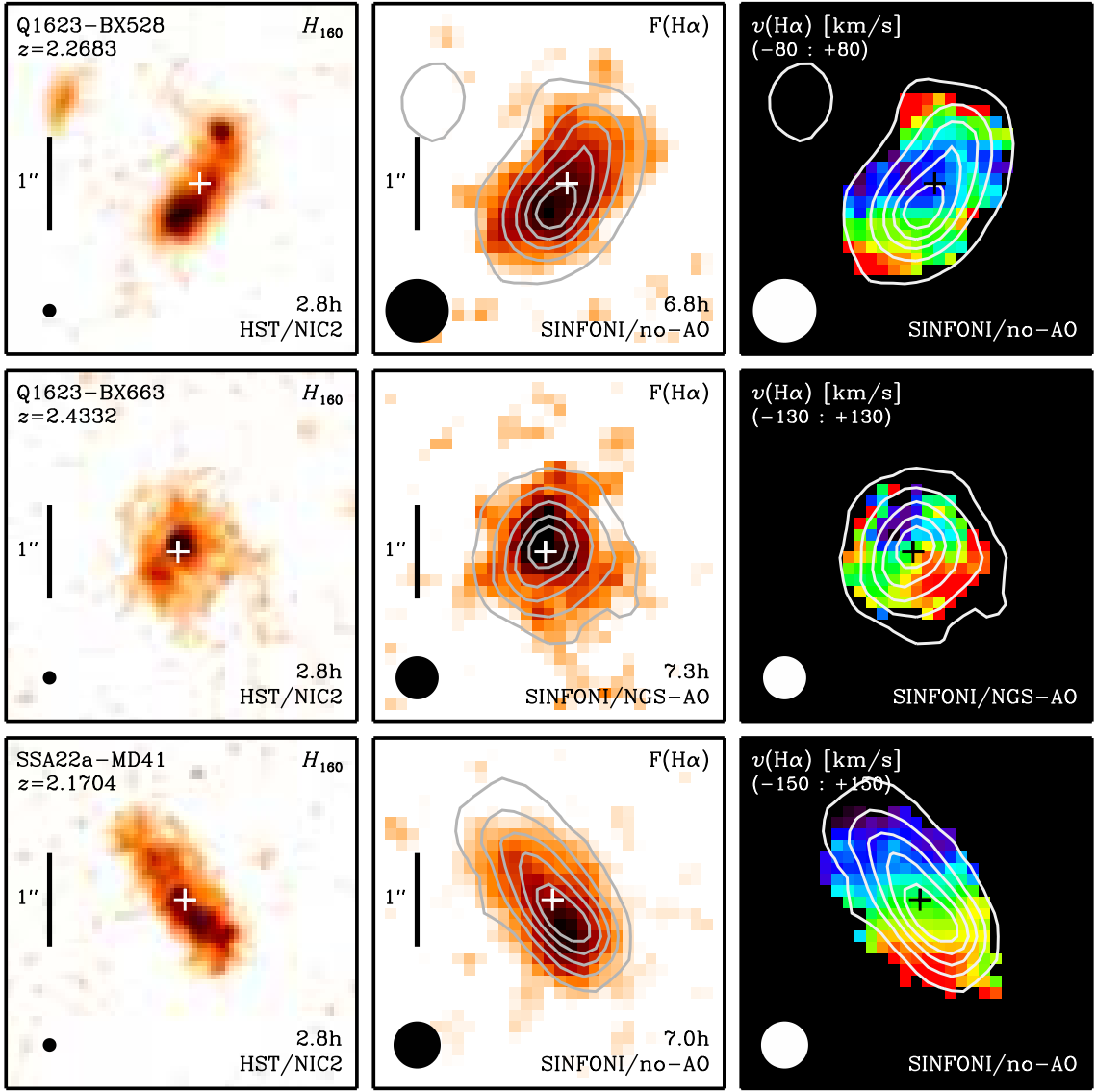}
\vspace{-4.5cm}
\caption{
\small
Comparison of $H_{\rm 160}$ band images and H$\alpha$ line maps
and velocity fields from SINFONI for three of our NIC2 targets:
BX\,528, BX\,663, and MD\,41.
Each row corresponds to a given galaxy, with the panels as follows.
{\em (Left)\/} $H_{\rm 160}$ band map at the original NIC2 angular resolution.
{\em (Middle)\/} H$\alpha$ line map with contours of the PSF-matched NIC2
$H_{\rm 160}$ band image overplotted in light grey.
{\em (Right)\/} H$\alpha$ velocity field with contours of the PSF-matched
NIC2 $H_{\rm 160}$ band image overplotted in white; the velocities are
color-coded from blue to red for blueshifted to redshifted line emission
relative to the systemic velocity and over the range labeled at the top left.
In each panel, the FWHM of the PSF of the corresponding image is shown by 
the filled circle in the bottom left corner.  The vertical bar indicates
an angular scale of 1\arcsec\ ($\rm \approx 8~kpc$ at the redshift of the
sources).  The white crosses mark the geometrical center of the sources.
The total on-source integration times are also given for the NIC2 and
SINFONI data, as well as whether the SINFONI data were obtained without
AO or with AO using a natural or the laser guide star (``NGS-AO'' and
``LGS-AO,'' respectively).
\label{fig-nic_sinf1}
}
\end{figure}

\clearpage

\begin{figure}[p]
\figurenum{10}
\epsscale{1.20}
\plotone{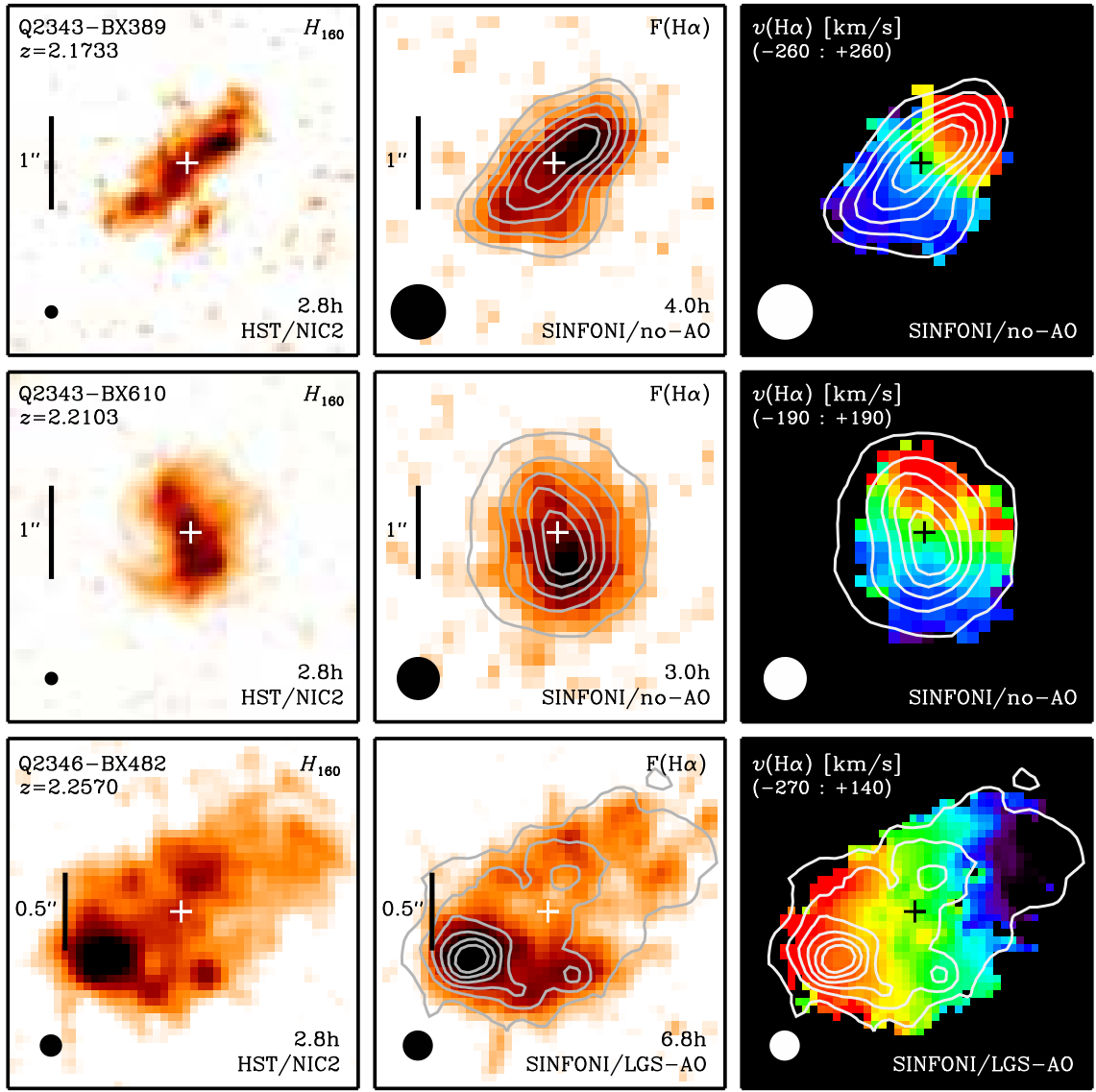}
\vspace{-4.5cm}
\caption{
\small
Same as Figure~\ref{fig-nic_sinf1} for the three other NIC2 targets:
BX\,389, BX\,610, and BX\,482.
\label{fig-nic_sinf2}
}
\end{figure}

\clearpage

\begin{figure}[p]
\figurenum{11}
\epsscale{1.20}
\plotone{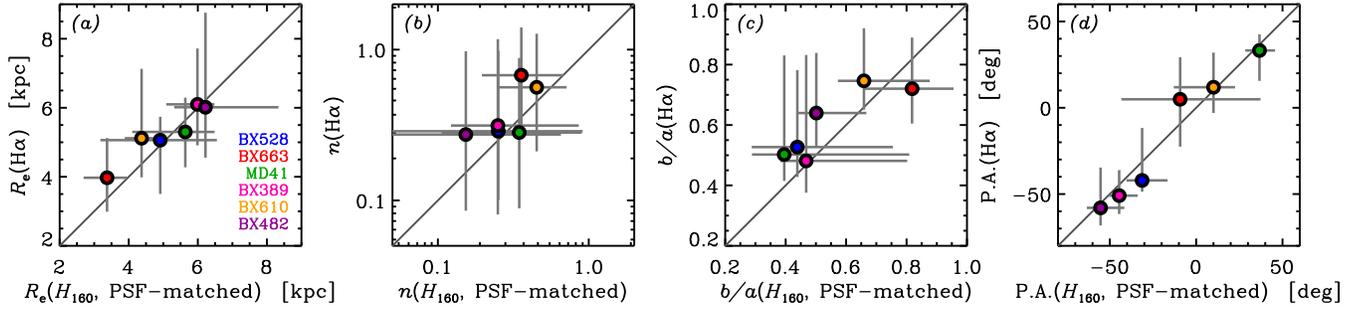}
\vspace{-0.5cm}
\caption{
\small
Comparison of the structural parameters from single-component S\'ersic
profile fits to the two-dimensional $H_{160}$ band and H$\alpha$ surface
brightness distributions.  For consistency, the fits were performed on
the $H_{160}$ band images matched to the PSF and resampled to the
$\rm 0\farcs 125~pixel^{-1}$ scale of the lower $\approx 0\farcs 5$
resolution H$\alpha$ maps.  Data for the different targets are
color-coded as indicated in the leftmost panel.
{\em (a)\/} Effective radius.
{\em (b)\/} S\'ersic index.
{\em (c)\/} Axis ratio.
{\em (d)\/} Position angle (in degrees East of North).
The solid line indicates the one-to-one relation.
The parameters derived from the H$\alpha$ and smoothed $H_{160}$ band
maps at $\approx 0\farcs 5$ resolution are in good agreement and typically
within their $1\,\sigma$ uncertainties.  They also agree well with
the values obtained from fits to the original higher resolution NIC2
images (see Table~\ref{tab-galfitres_smoo}).
\label{fig-galfitres3}
}
\end{figure}

\clearpage

\begin{figure}[p]
\figurenum{12}
\epsscale{1.05}
\plotone{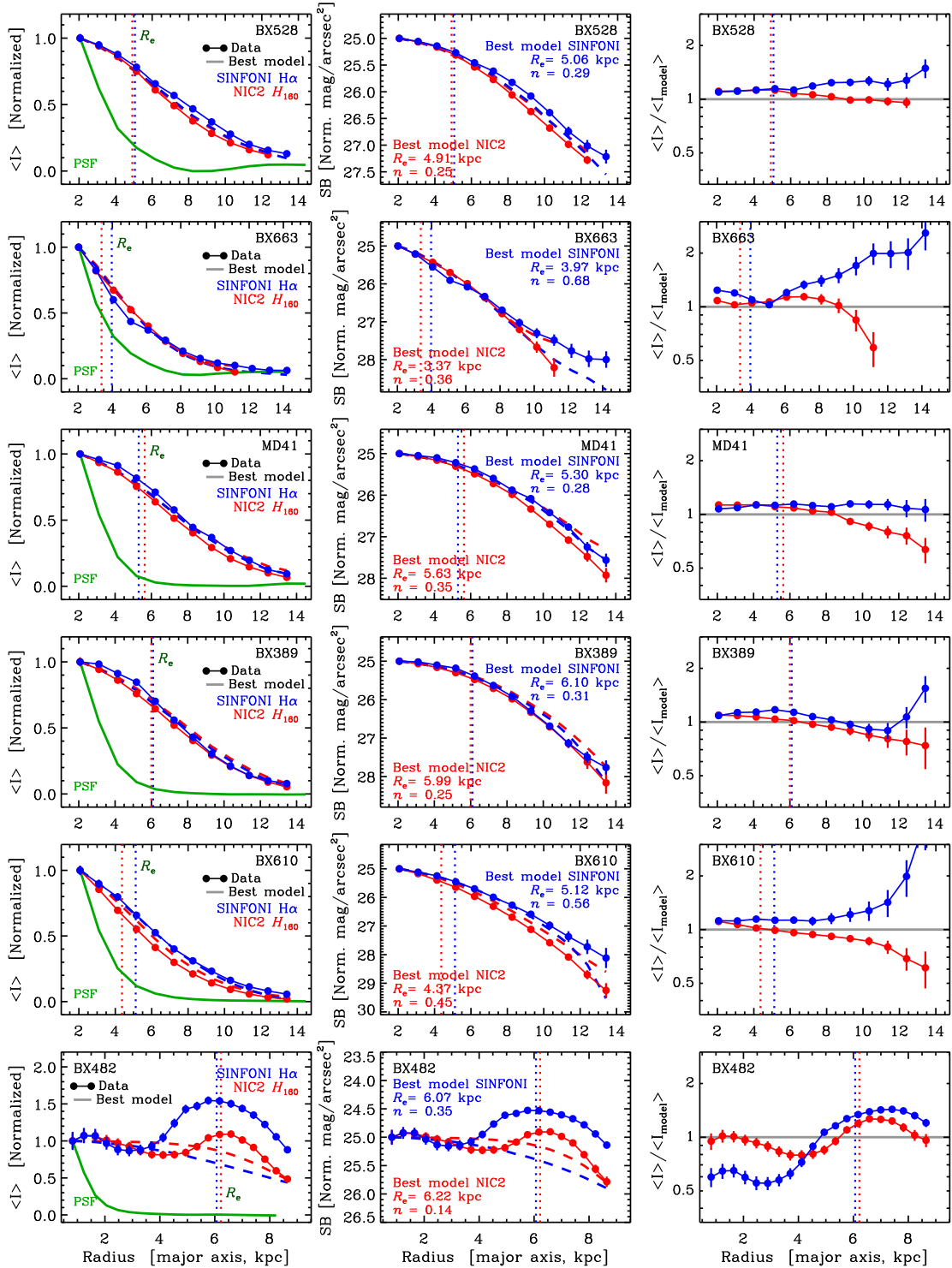}
\vspace{-1.0cm}
\caption{
\small
Radial light profiles of the H$\alpha$ and $H_{160}$ band emission of
the targets.
The profiles and best-fit single-component S\'ersic model overplotted
are based on the $\approx 0\farcs 5$ resolution SINFONI H$\alpha$ maps
(using the $\rm 0\farcs 125~pixel^{-1}$ scale, taken under seeing-limited
conditions or with NGS-AO for BX\,663) and PSF-matched NIC2 $H_{160}$
images except for BX\,482, where the AO-assisted SINFONI and original
resolution NIC2 data both at $\rm 0\farcs 05~pixel^{-1}$ are used
(see \S~\ref{Sect-sinf}).
From left to right, each row shows for a given source the mean intensity
in linear flux units, the surface brightness in magnitude units, and the
ratio of observed intensity profile to that of the best-fit S\'ersic model
(\S~\ref{Sub-Hamorph_struct}).
The data are plotted with solid lines and dots, and the models are shown
with dashed lines; blue and red colors are used for H$\alpha$ and $H_{160}$
band, respectively.
The profiles are computed along ellipses with center, axis ratio, and P.A.
corresponding to the S\'ersic model that best fits the NIC2 images; the
maximum radius is determined by the effective field of view of the SINFONI
data and so as to avoid the noisiest parts at the egdes.
The intensity and surface brightness profiles are all normalized to unity
and 25~mag, respectively, at the central position.  The ratio of observed
and model intensity profiles are as computed before normalization, and the
horizontal grey line in the right-hand panels indicates a ratio of unity.
The vertical dotted lines in all panels show the best-fit effective radii
from the S\'ersic fits, and the SINFONI PSF radial intensity profiles are
plotted in the left-hand panels in green for reference.
The close correspondence between H$\alpha$ and rest-frame
$\approx 5000$\,\AA\ radial light distributions is striking,
reflecting the similarity in the maps shown in
Figures~\ref{fig-nic_sinf1} and \ref{fig-nic_sinf2}.
\label{fig-profiles2}
}
\end{figure}

\clearpage

\begin{figure}[p]
\figurenum{13}
\epsscale{1.10}
\plotone{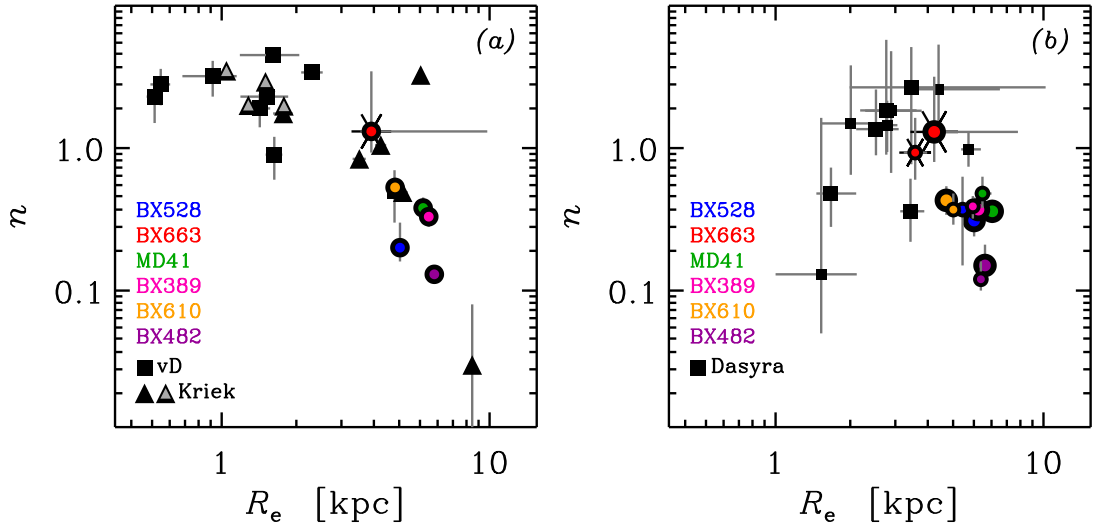}
\vspace{-0.0cm}
\caption{
\small
S\'ersic index $n$ versus effective radius $R_{\rm e}$ for our six SINS
galaxies and comparison samples at similar redshifts but with different
selection criteria.  All galaxies were observed with the same NICMOS NIC2
camera through the same F160W filter and the S\'ersic model fits were
performed in the same way for all samples, allowing consistent comparisons.
{\rm (a)\/} $n$ versus $R_{\rm e}$ for the SINS and the $K$-selected
samples (labeled ``vD'' for objects from \citealt{Dok08} and ``Kriek''
for those from \citealt{Kri09}).  The images used for all sources were
drizzled to a $0\farcs 038$ pixel scale.  Results for the SINS galaxies
based on the full 4-orbit data and 3-orbit subsets (as 17 of 19 of the
$K$-selected objects) are indistinguishable, so the parameters derived
from the 4-orbit data are plotted.
The SINS targets are plotted with filled circles following the color
scheme as given by the labels.  Squares and triangles correspond to the
quiescent and star-forming $K$-selected samples of \citet{Dok08} and 
\citet{Kri09}, respectively.
The AGN sources are distinguished with a starred circle for BX\,663
in our SINS sample, and with grey-filled triangles for those of the
\citeauthor{Kri09} sample.
{\rm (b)\/} Same as previous panel for the SINS and 24\,\micron -selected
dusty IR-luminous sample of \citeauthor{Das08} (2008; labeled ``Dasyra'').
All images used here were drizzled to $0\farcs 076$ pixels.  Large and
small symbols are used for results based on 2- and 1-orbit integrations,
respectively.  Ten of 11 sources from \citet{Das08} show signatures of
AGN activity in their mid-IR spectrum, either dominant or along with PAH
features indicative of star formation; the only star-formation dominated
object is the second largest with second highest $n$ (4.06~kpc and 2.59).
\label{fig-galfit_comp}
}
\end{figure}

\clearpage

\begin{figure}[p]
\figurenum{14}
\epsscale{1.00}
\plotone{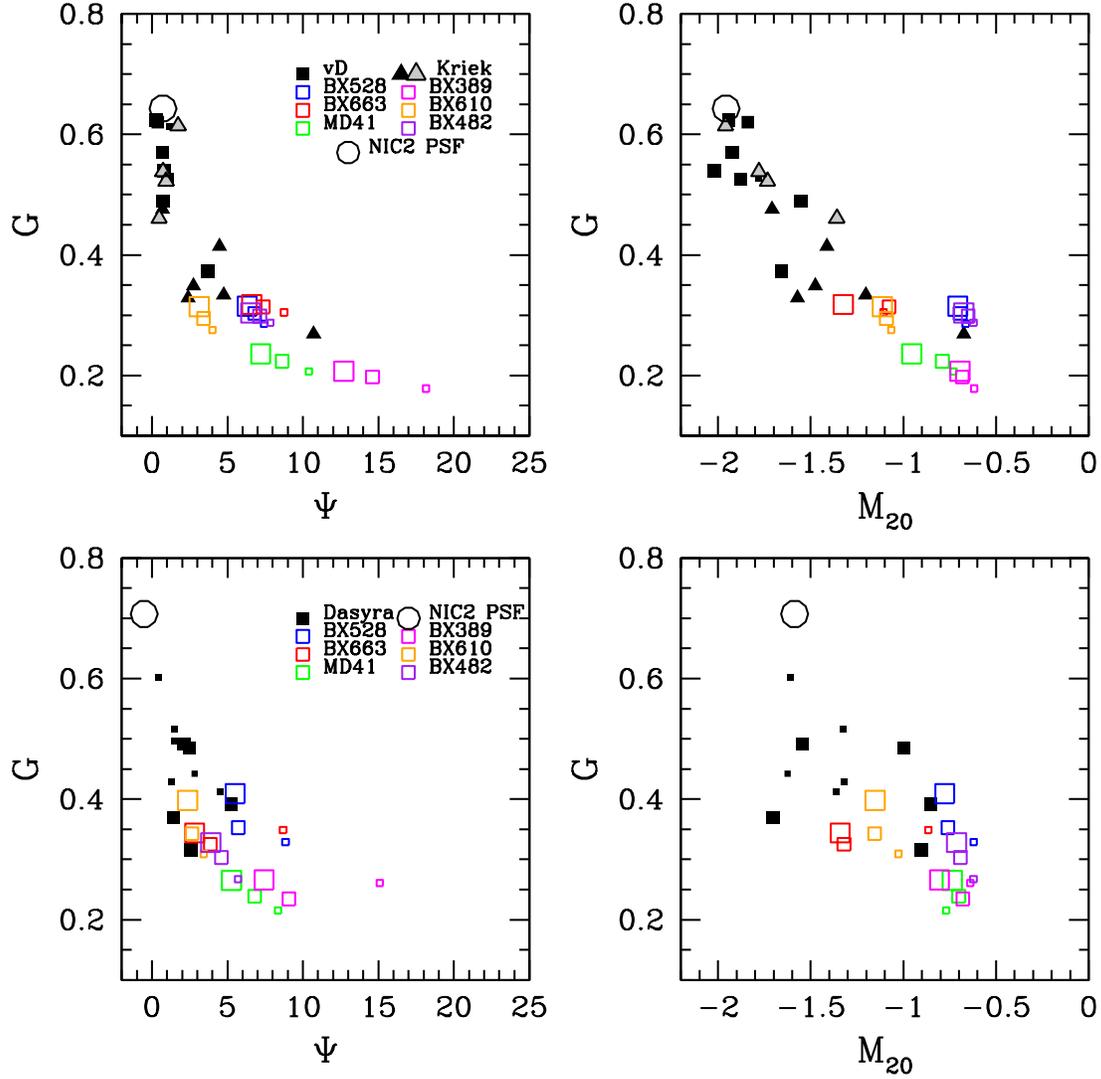}
\vspace{-0.0cm}
\caption{
\small
Morphological parameters derived from NIC2 $H_{160}$ band imaging for
our SINS targets and comparison samples at similar redshifts but with
different selection criteria.  Each galaxy was analyzed as if it was
placed at the maximum redshift of the combined samples, i.e. $z = 2.60$.
{\em (Top, Left)\/} Gini versus $\Psi$ for the SINS and $K$-selected
samples (labeled ``vD'' for objects from \citealt{Dok08} and ``Kriek''
for objects from \citealt{Kri09}).
As in Figure~\ref{fig-galfit_comp}, $K$-selected objects are indicated
with solid squares (vD) and triangles (Kriek).  For vD objects,
medium-sized symbols are used for 3-orbit exposures and small symbols
for 2-orbit exposures, while, for Kriek objects, all points represent
3-orbit exposures.
SINS objects are indicated with open squares, with the largest symbols
for 4-orbit exposures, medium-sized symbols for 3-orbit subsets of the
data, and small symbols for 2-orbit subsets.
Our NIC2 images were drizzled to the $0\farcs 038$ pixel sampling of the
\citeauthor{Dok08} and \citeauthor{Kri09} data for consistent comparison.
The non-parametric statistics for the empirical PSF constructed from
stars in our $\rm 0\farcs 038~pixel^{-1}$ NIC2 images are indicated
with an open circle.
{\em (Top, Right)\/} Gini versus $M_{20}$ for the SINS, vD, and Kriek
samples.  Symbols are as in the previous panel.
{\em (Bottom, Left)\/} Gini versus $\Psi$ for the SINS and
\citeauthor{Das08} (2008; labeled ``Dasyra'') samples.
\citeauthor{Das08} objects are indicated with solid black squares, with
medium-sized symbols for 2-orbit exposures and smaller symbols for 1-orbit
exposures.  Again, SINS objects are indicated with open squares, with the
largest symbols for 4-orbit exposures, medium-sized symbols for 2-orbit
subsets of the data, and smallest symbols for 1-orbit subsets.
Our NIC2 images were here drizzled to the $0\farcs 076$ pixel scale
of the \citeauthor{Das08} data, and the non-parametric statistics for
the empirical PSF from these images are also indicated with an open
circle.
{\em (Bottom, Right)\/} Gini versus $M_{20}$ for SINS and \citet{Das08}
samples.  Symbols are as in the previous panel.
\label{fig-morph2}
}
\end{figure}

\clearpage

\begin{figure}[p]
\figurenum{15}
\epsscale{1.00}
\plotone{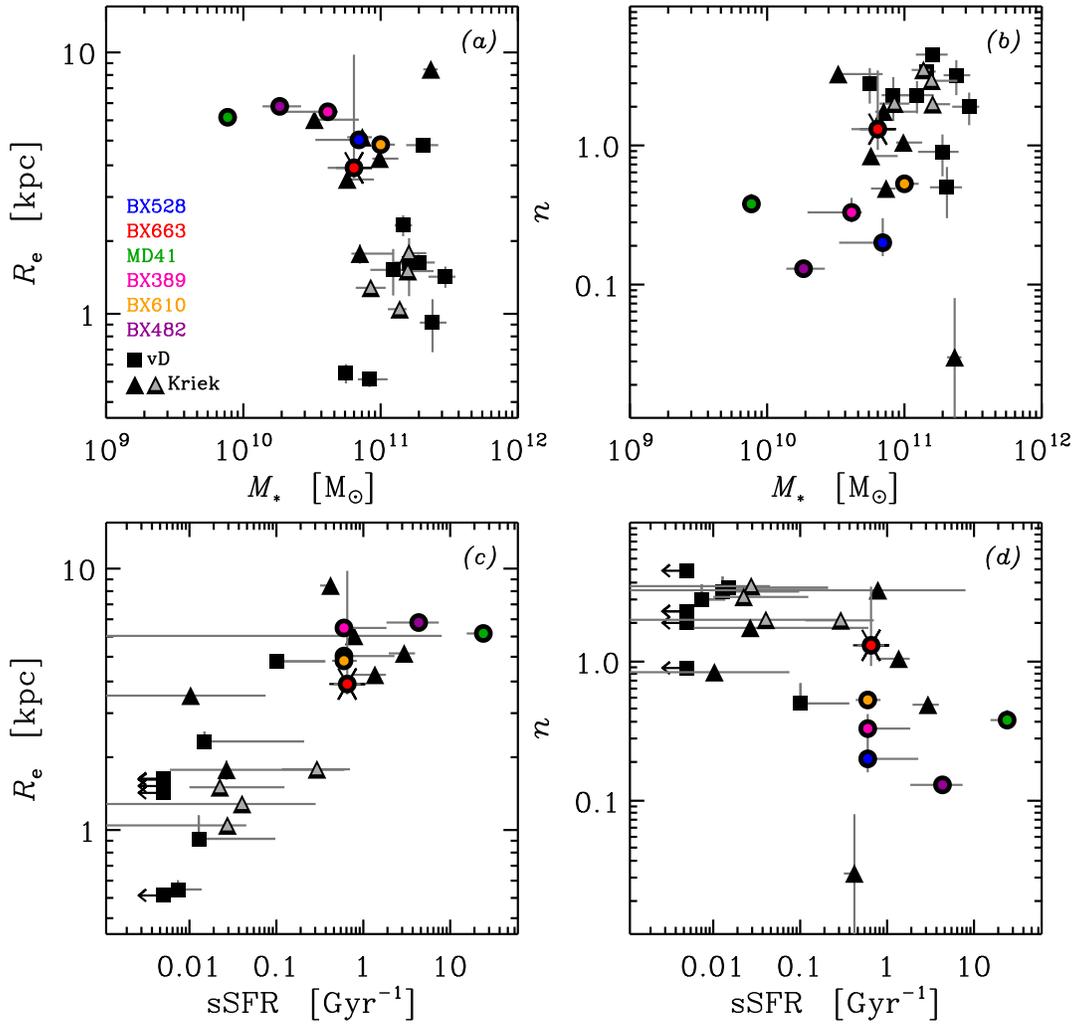}
\vspace{0.0cm}
\caption{
\small
Structural parameters of our six SINS galaxies and of the
$K$-selected comparison sample from \citet[][``vD'']{Dok08} and
\citet[][``Kriek'']{Kri09}, plotted as a function of global stellar
properties.
{\em (a)\/} Effective radius $R_{\rm e}$ versus stellar mass $M_{\star}$.
{\em (b)\/} S\'ersic index $n$ versus $M_{\star}$.
{\em (c)\/} $R_{\rm e}$ versus specific star formation rate sSFR.
{\em (d)\/} $n$ versus sSFR.
Symbols and colors are the same as in Figure~\ref{fig-galfit_comp},
and labeled in panel {\em (a)\/}.
All stellar masses are for a \citet{Chab03} IMF.
In the range of $M_{\star}$ where our SINS objects and the $K$-selected
samples overlap, the data reflect the wide range in size (and in $n$)
that is known to exist for galaxies of similar stellar masses and out
to $z \sim 3$.  Size and star formation activity are clearly correlated,
again as seen for the massive galaxy population locally and at high redshift.
\label{fig-galfitsedmod}
}
\end{figure}

\clearpage

\begin{figure}[p]
\figurenum{16}
\epsscale{1.20}
\plotone{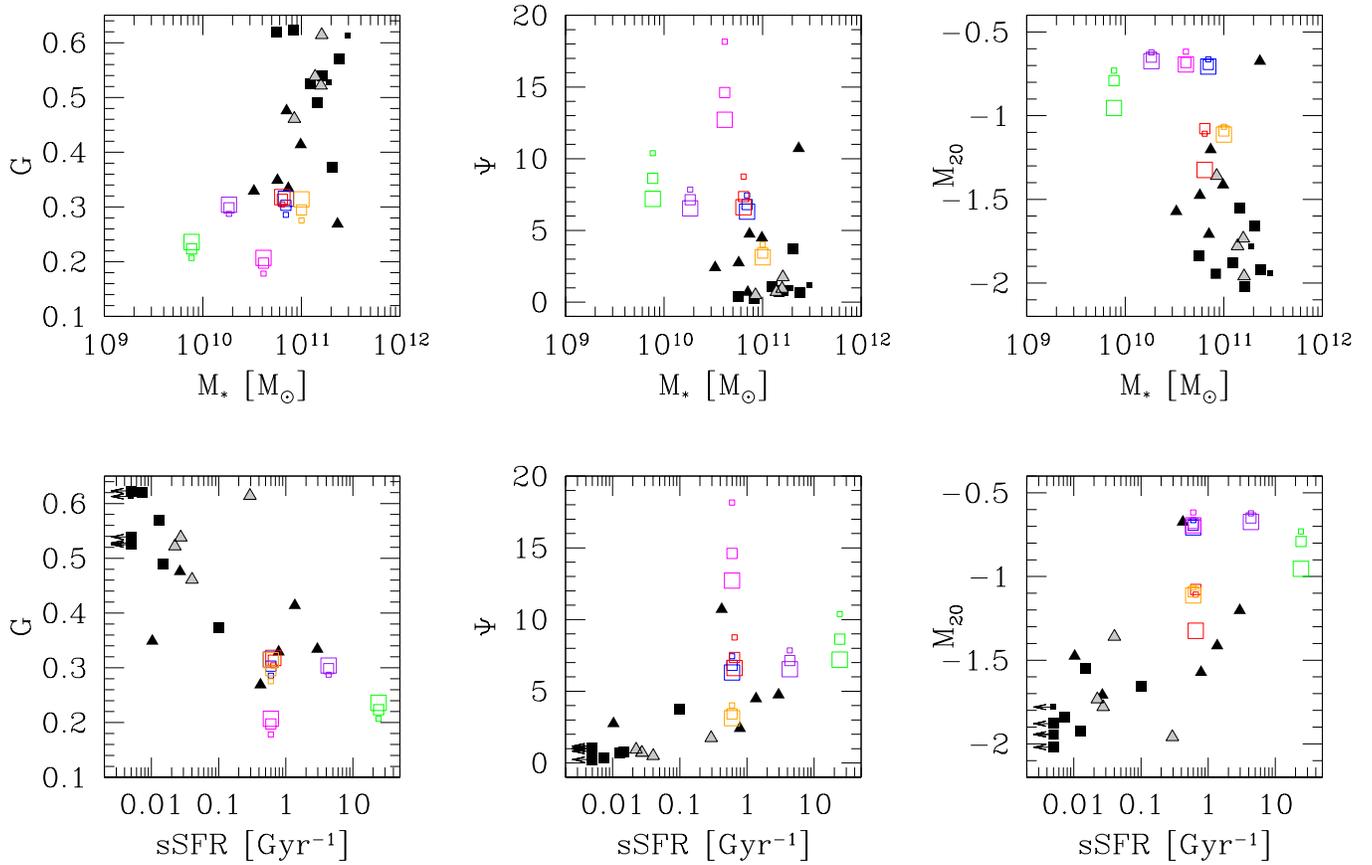}
\vspace{0.5cm}
\caption{
\small
Same as Figure~\ref{fig-galfitsedmod} for the non-parametric morphological
coefficients.
The top row shows morphological coefficients plotted as a function of
$M_{\star}$, while the bottom row shows them as a function of specific SFR.
As for the structural parameters ($R_{\rm e}$ and $n$), a large diversity
of non-parametric morphological coefficients are found in the region
of overlapping $M_{\star}$, while significant correlations are found
between morphology and  specific SFR.
\label{fig-morphsedmod}
}
\end{figure}

\clearpage

\begin{figure}[!htp]
\figurenum{17}
\epsscale{0.55}
\plotone{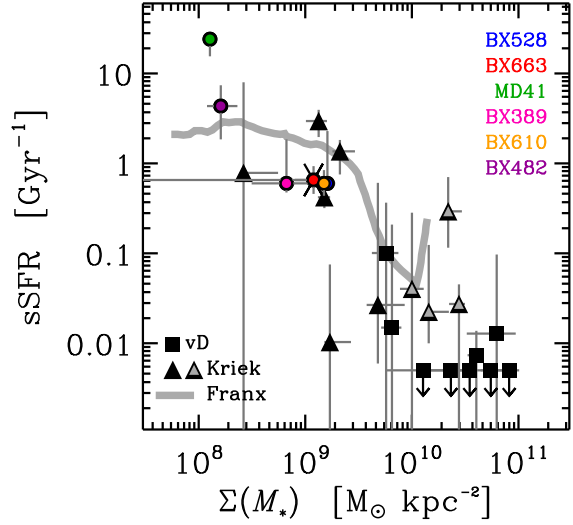}
\vspace{-0.25cm}
\caption{
\small
Same as Figure~\ref{fig-galfitsedmod} for the specific star
formation rate as a function of the stellar mass surface density.
Half of the total stellar mass is assumed to be enclosed within a
radius corresponding to the circularized effective radius
$R_{\rm e\,circ} = R_{\rm e}\,\sqrt{b/a}$, giving estimates
of the projected surface density.
There is a clear trend outlined by our SINS targets and the $K$-selected
comparison samples of \citet{Dok08} and \citet{Kri09}, which span overall
three orders of magnitude in stellar mass surface density and even more
in specific SFR (a similar trend is seen when using the $R_{\rm e}$
instead for estimates of the deprojected surface density).
For comparison with the more general massive galaxy population at $z \sim 2$,
the median relationship derived by \citet{Fra08} from a large $K$-selected
sample in GOODS-S and based on ground-based near-IR imaging to derive
rest-frame optical sizes is overplotted as a grey thick line.
Our SINS and the $K$-selected comparison samples consistently follow
this relationship.
\label{fig-Mdens_sSFR}
}
\end{figure}

\vspace{2.0cm}

\begin{figure}[!hbp]
\figurenum{18}
\epsscale{1.10}
\plotone{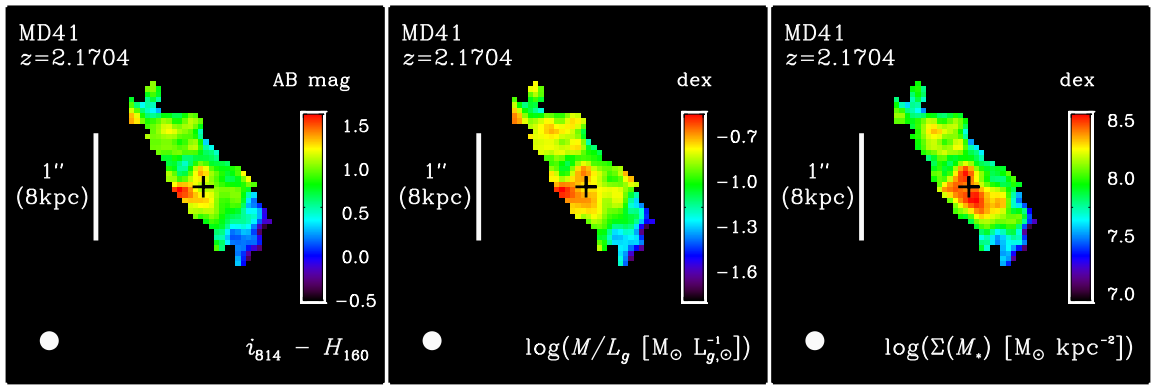}
\vspace{-0.25cm}
\caption{
\small
Color, observed mass-to-light ratio, and stellar mass maps of MD\,41.
North is up and east to the left in all panels, and a bar of 1\arcsec\
in length (approximately 8~kpc at the redshift of the source) indicates
the scale in the images.
{\em (Left)\/} Observed $i_{814} - H_{160}$ colors from the ACS/F814W 
and NIC2/F160W images, matched to the same PSF with FWHM indicated by
the filled circle at the bottom left corner.
{\em (Middle)\/} Observed stellar mass to rest-frame $g$-band light ratio,
computed from the $i_{814} - H_{160}$ colors using the relationship
derived in Appendix~A.
{\em (Right)\/} Stellar mass surface density, derived from the
$M_{\star}/L_{g}^{\rm rest}$ ratio of the previous panel and the
NIC2 $H_{160}$ band map.
Pixels with a $\rm S/N > 3$ in each of the ACS/F814W and NIC2/F160W
images are shown, while others are masked out.
\label{fig-md41_colmass}
}
\end{figure}

\clearpage

\begin{figure}[p]
\figurenum{19}
\epsscale{0.60}
\plotone{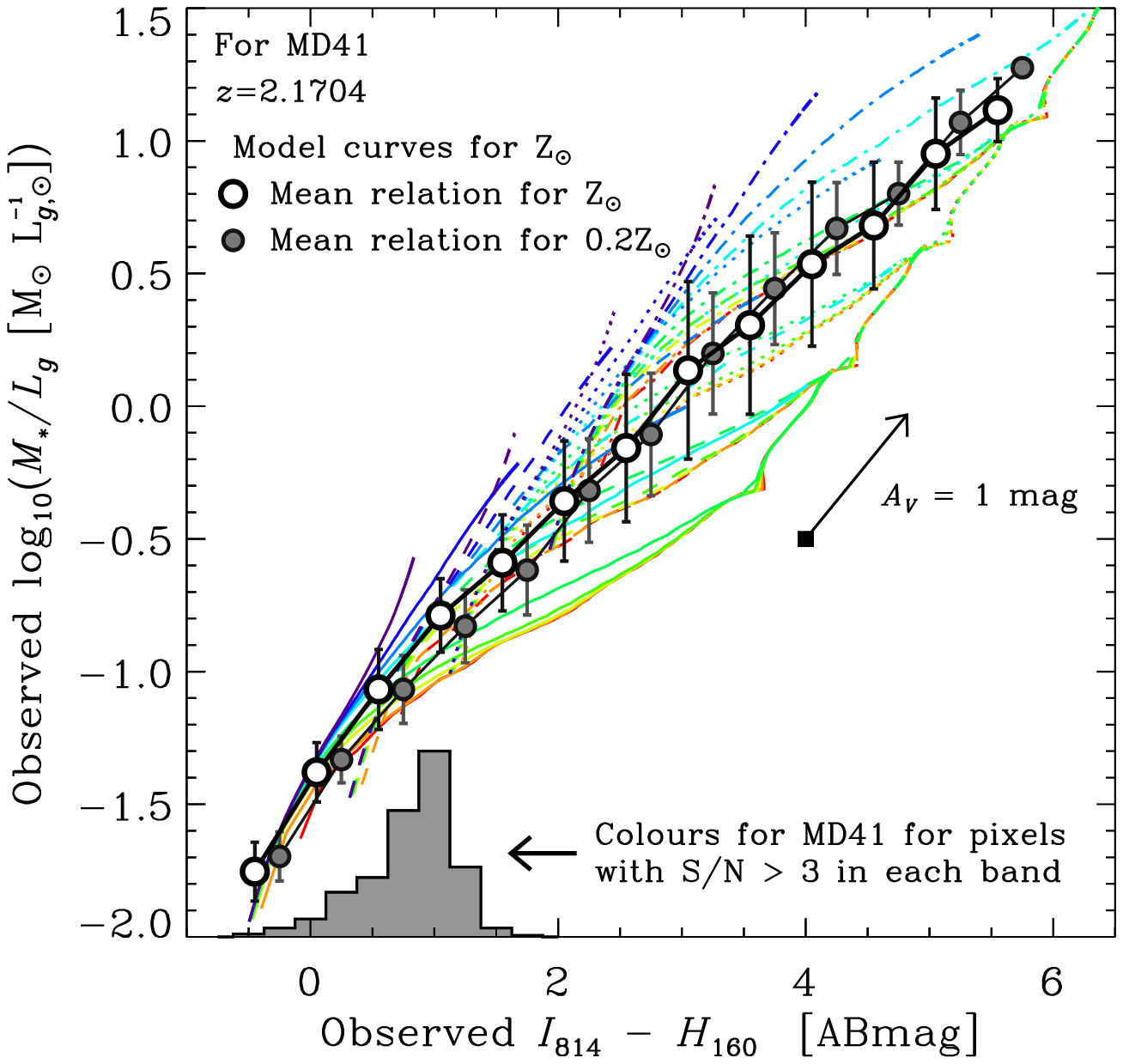}
\vspace{0.0cm}
\caption{
\small
Relationship between observed $i_{814} - H_{160}$ colors and
the ratio of stellar mass to rest-frame $g$-band luminosity
$M_{\star}/L_{g}^{\rm rest}$.
Both the colors and the $M_{\star}/L_{g}^{\rm rest}$ correspond to
{\em observed\/} quantities, uncorrected for the effects of dust
extinction.  The various curves plotted in colors are computed from
\citet{BC03} models with solar metallicity, a \citet{Chab03} IMF, and
the \citet{Cal00} reddening law.
Different colors are used for different star formation histories:
red to blue represent SFHs of increasing star formation timescale from
a single stellar population, a suite of exponentially declining SFRs
with $e$-folding timescales between 10~Myr and 1~Gyr, and constant SFR.
Different linestyles are also used for models computed with different
values for extinction: $A_{V} = 0$ (solid), 1 (dashed), 2 (dotted),
and 3~mag (dash-dotted).
Age increases along each model curve from blue to red $i_{814} - H_{160}$
colors and low to high $M_{\star}/L_{g}^{\rm rest}$ ratios.
As shown by the arrow, the effects of extinction are roughly parallel
to the locus of the various models.
The colors are synthesized from the model spectrum redshifted to the
$z = 2.1704$ of MD\,41, using the transmission curves for the ACS F814W
and NIC2 F160W filters.  The large white-filled circles and thick black
line shows the mean relationship derived from the solar metallicity models.
For comparison, the smaller grey-filled circles and thinner black line is
the relationship that is derived from models with $1/5$ solar metallicity.
The grey-shaded histogram shows the distribution of observed
$i_{814} - H_{160}$ colors of all individual pixels with $\rm S/N > 3$
in each band for MD\,41 (see \S~\ref{Sub-disc_md41}).
\label{fig-ihcol_lmlg}
}
\end{figure}

\end{document}